\documentclass[12pt,a4paper]{article}
\usepackage[a4paper]{geometry}

\usepackage{amsmath}
\usepackage{amsfonts}
\usepackage{amssymb}
\usepackage{dsfont}
\usepackage{graphicx}
\usepackage[nosort]{cite}
\usepackage{hyperref}
\usepackage{multirow}
\usepackage[margin=10pt,font=small,labelfont=bf]{caption}
\usepackage[margin=0pt,font=small,labelfont=normalfont,skip=22pt]{subcaption}
\usepackage{rotating}

\newcommand{\eq}[1]{\begin{equation}
                     \begin{split} #1 \end{split}
                     \end{equation}}

\newcommand{\ov}{\overline}
\newcommand{\op}{\hspace{1pt}}

\allowdisplaybreaks[2]
\numberwithin{equation}{section}

\begin{document}

\vspace*{-1.5cm}
\begin{flushright}
  {\small
  LMU-ASC 23/19\\
  MPP-2019-96
  }
\end{flushright}

\vspace{1.75cm}

\begin{center}
{\LARGE
Type IIB flux vacua and tadpole cancellation 
}
\end{center}

\vspace{0.4cm}

\begin{center}
  Philip Betzler$^{1,2}$, Erik Plauschinn$^1$
\end{center}

\vspace{0.3cm}

\begin{center} 
\textit{$^{1}$\op Arnold Sommerfeld Center for Theoretical Physics\\[1pt]
Ludwig-Maximilians-Universit\"at \\[1pt]
Theresienstra\ss e 37 \\[1pt]
80333 M\"unchen, Germany}
\\[1em]
\textit{$^{2}$\op Max-Planck-Institut f\"ur Physik\\[1pt]
F\"ohringer Ring 6 \\[1pt]
80805   M\"unchen, Germany}
\end{center} 

\vspace{0.8cm}


\begin{abstract}
\noindent
We consider flux vacua for type IIB orientifold compactifications 
and study their interplay with the tadpole-cancellation condition.
As a concrete example we focus on $\mathbb T^6/\mathbb Z_2\times \mathbb Z_2$,
for which we find that solutions to the F-term equations at 
weak coupling, large complex structure and large volume require large flux contributions.
Such contributions are however strongly disfavored by the tadpole-cancellation condition. 
We furthermore find that solutions which stabilize moduli in this perturbatively-controlled regime are only a very 
small fraction of all solutions, and that the space of solutions is not homogenous but shows
characteristic void structures and vacua concentrated on submanifolds.
\end{abstract}


\clearpage

\tableofcontents


\clearpage
\section{Introduction}

String theory is argued to be a consistent theory of quantum gravity including gauge interactions.
It is defined in ten space-time dimensions, and 
in order to make a connection  to four-dimensional physics  six spatial dimensions have to be compactified.
Such compactifications have to satisfy a number of consistency conditions, for instance, 
in the presence of D-branes and closed-string fluxes the tadpole-cancellation condition 
and the Freed-Witten anomaly-cancellation condition  have to be satisfied.
(For a review of the former see~\cite{Blumenhagen:2006ci} and for the latter see \cite{Freed:1999vc}.)
These conditions relate the closed- and open-string sectors to each other
and put strong restrictions on the allowed background configurations. More concretely, 
\begin{enumerate}

\item compactifications of string theory to four dimensions are typically performed on 
Calabi-Yau three-folds.
The resulting effective theory contains a number of massless scalar fields 
corresponding to deformations of the background, which should be absent due to experimental constraints. 
A way to achieve this for type II theories is to deform the background geometry by 
Neveu-Schwarz--Neveu-Schwarz (NS-NS) 
and Ramond-Ramond (R-R) fluxes, 
which generate a potential in the four-dimensional theory and 
provide mass-terms for the moduli fields. This procedure is called moduli stabilization,
and for early work in the type IIB context see for instance
\cite{Dasgupta:1999ss,Taylor:1999ii,Giddings:2001yu} and 
for work in the type IIA context see
\cite{Derendinger:2004jn,Villadoro:2005cu,DeWolfe:2005uu}.
However, especially in the type IIB setting some of the moduli cannot be stabilized by 
(geometric) fluxes. One therefore includes non-perturbative effects 
which lead to the KKLT \cite{Kachru:2003aw}
and large-volume \cite{Balasubramanian:2005zx} scenarios.
Moduli stabilization often results in anti-de-Sitter or Minkowski vacua, 
while it is difficult  to obtain de-Sitter solutions~\cite{Obied:2018sgi}.

\item A gauge-theory sector for type II theories can be engineered using D-branes.
D-branes filling four-dimensional space-time and wrapping submanifolds in 
the compact space have a gauge theory localized on their world-volume.
Chiral matter can be localized at the intersection loci of different D-branes in the 
compact space, and 
in this way four-dimensional gauge theories can be constructed in a geometric way (for
a review see for instance \cite{Blumenhagen:2006ci}).
However, when introducing D-branes one typically has to perform an orientifold projection
of the theory. The fixed-loci of this projection correspond to orientifold planes which 
generically have negative mass and negative charge.

\end{enumerate}
Moduli stabilization and the construction of a gauge-theory sector are two important aspects 
of connecting string theory to realistic four-dimensional physics. These tasks are 
often approached independently, however, as emphasized for instance in \cite{Blumenhagen:2007sm},
there is a complicated interplay between them. This interplay
can prevent moduli from being stabilized or can modify the stabilization procedure. 
Following this line of thought, the purpose of this paper is to study how  parts 1) and 2) are connected via the 
tadpole-cancellation condition (schematically) 
\eq{
  \label{tadpoles_master}
  \mbox{fluxes} = \mbox{D-branes} + \mbox{O-planes}\,.
}

We approach this question by analyzing how properties of the space of flux vacua depends
on the contribution of fluxes to the left-hand side of \eqref{tadpoles_master}.
We perform our analysis for the  type IIB orientifold of $\mathbb T^6/\mathbb Z_2\times \mathbb Z_2$,
and consider R-R three-form flux, geometric NS-NS $H$-flux as well as non-geometric NS-NS $Q$-flux. 
The geometric fluxes generically stabilize the complex-structure moduli and the 
axio-dilaton, while the non-geometric fluxes allow for stabilization of the K\"ahler moduli.
We then determine distributions for how the values of the stabilized moduli depend on the 
tadpole contribution of the fluxes.
Note that distributions of flux vacua have been discussed extensively in the literature before. 
For instance, for type IIB compactifications various aspects have been studied in
\cite{
Ashok:2003gk,
Douglas:2003um,
Douglas:2004kc,
Denef:2004ze,
Giryavets:2004zr,
Douglas:2004zg,
Conlon:2004ds,
DeWolfe:2004ns,
Denef:2004cf,
Eguchi:2005eh,
Shelton:2006fd,
MartinezPedrera:2012rs
}
and for type IIA related work can be found in
\cite{
DeWolfe:2005uu,
Dibitetto:2011gm
}.
In the context of M-theory similar questions have been discussed in 
\cite{
Acharya:2005ez
},
and for F-theory see
\cite{
Watari:2015ysa
}.
Recently also topological data analysis  has been used to investigate properties 
of flux vacua in
\cite{
Cirafici:2015pky,
Cole:2018emh
}.
The main results of our analysis can be summarized as follows:
\begin{itemize}

\item We observe that the space of flux vacua is not homogenous but shows
characteristic structures such as circular voids \cite{Denef:2004ze}. 
We show furthermore that solutions can be accumulated on submanifolds  in the space of solutions.

\item We find that flux configurations which stabilize moduli in a weak-coupling, large complex-structure
and/or large-volume regime are only a very small fraction of all possible configurations. The number
of reliable flux vacua is therefore much smaller than naively expected.

\item In order to stabilize moduli in a weak-coupling, large complex-structure
and large-volume regime, the flux contribution to the left-hand side of tadpole-can\-cellation 
condition \eqref{tadpoles_master} has to be larger than a certain threshold. 
The more reliable these vacua are required to be, the larger this threshold has to be.
However, the contribution of D-branes and orientifold planes to the right-hand side of 
\eqref{tadpoles_master} is typically small.
It is therefore difficult to perform moduli stabilization in a perturbatively-controlled regime 
and to satisfy the tadpole-cancellation condition.

\end{itemize}
Our findings for the structure of the space of flux vacua agree with for instance 
\cite{Denef:2004ze,Cole:2018emh} for the axio-dilaton, but we extend their
analysis by including the complex-structure moduli. 
Our observation concerning the difficulty of obtaining reliable flux vacua 
is consistent with for instance 
\cite{Halverson:2017vde}, who find that type IIB solutions at weak string-coupling 
are rare. Similarly, in \cite{Bena:2018fqc} the authors argue that in order to avoid
a certain runaway behaviour large fluxes have to be considered, 
also in \cite{Blaback:2018hdo} large fluxes are needed to obtain reliable solutions,
and related difficulties are encountered in 
\cite{CaboBizet:2019sku}.

\bigskip
This work is organized as follows: in section~\ref{sec_prelim} we review 
type IIB orientifold compactifications with geometric and
non-geometric fluxes, we discuss the corresponding tadpole-cancellation 
conditions, we specialize to the example of the 
$\mathbb T^6/\mathbb Z_2\times \mathbb Z_2$ orientifold and 
determine the relevant dualities. 
In section~\ref{sec_mod1} we study moduli stabilization for the axio-dilaton,
in section~\ref{sec_iso} we discuss the combined moduli stabilization of the  axio-dilaton and 
the complex-structure moduli, and in section~\ref{sec_stu} we stabilize all of the closed-string moduli
at tree-level. 
At the end of sections~\ref{sec_mod1}, \ref{sec_iso} and \ref{sec_stu} we have included 
brief summaries for each section, which may help the reader to get an overview of the main results.
Our conclusions can be found in section~\ref{sec_disc}.


\clearpage
\section{Flux compactifications }
\label{sec_prelim}

In this paper we are interested in compactifications of type IIB string theory
on Calabi-Yau orientifolds with geometric and non-geometric fluxes. 
In order to fix our notation, we start in sections~\ref{sec_pre_or_gen}
and~\ref{sec_tadpole}
by briefly reviewing orientifold compactifications and 
 tadpole-cancellation conditions
for general Calabi-Yau three-folds. 
In section~\ref{sec_pre_or_t6} we specialize to the example of
$\mathbb T^6/\mathbb Z_2\times \mathbb Z_2$, and 
in section~\ref{sec_dualities} we discuss  duality transformations for this background.


\subsection{Orientifold compactifications}
\label{sec_pre_or_gen}

Type IIB orientifold compactifications on Calabi-Yau three-folds give rise to a 
$\mathcal N=1$ supergravity theory in four dimensions. 
This theory  can be characterized in terms of a superpotential, K\"ahler potential 
and D-term potential, which we determine  in the following.


\subsubsection*{Calabi-Yau orientifolds}

We begin with type IIB string theory on $\mathbb R^{3,1}\times \mathcal X$, where $\mathcal X$ 
denotes a Calabi-Yau three-fold. The latter comes with a holomorphic three-form $\Omega\in H^{3,0}(\mathcal X)$ and a 
real K\"ahler form $J\in H^{1,1}(\mathcal X)$, and to perform an orientifold projection we impose 
 a holomorphic involution 
$\sigma$ on $\mathcal X$ which acts on $\Omega$ and $J$ as
\eq{
  \label{orient_choice}
  \sigma^*J=+J\,,\hspace{60pt}
  \sigma^*\op\Omega=-\Omega\,.
}
The involution 
 leaves the non-compact four-di\-men\-sional part invariant, and hence
the fixed loci of $\sigma$ correspond to orientifold three- and seven planes.
For the full orientifold projection $\sigma$ is 
combined with $\Omega_{\rm P} (-1)^{F_{\rm L}}$, where 
$\Omega_{\rm P}$ denotes the world-sheet parity operator and $F_{\rm L}$ is 
the left-moving fermion number.
The latter two act on the bosonic fields in the following way
\eq{
  \label{orient_signs}
  \arraycolsep2pt
  \begin{array}{lclclcl}
  \arraycolsep1.5pt
  \displaystyle \Omega_{\rm P}\, (-1)^{F_{\rm L}}\: g &=& + \:g \,,  & \hspace{60pt} &
  \displaystyle \Omega_{\rm P}\, (-1)^{F_{\rm L}}\: B &=& - B \,, \\[6pt]
  \displaystyle \Omega_{\rm P}\, (-1)^{F_{\rm L}}\: \phi &=& + \:\phi \,,  & \qquad &
  \displaystyle \Omega_{\rm P}\, (-1)^{F_{\rm L}}\: C_{2p} &=& (-1)^{p} 
    \,C_{2p} \,,
  \end{array}
}
with $g$ the metric, $B$ the Kalb-Ramond field, $\phi$ the dilaton and 
$C_{2p}$ the Ramond-Ramond potentials in type IIB. 
Since $\sigma$ is an involution,
the cohomology groups of the Calabi-Yau three-fold $\mathcal X$ split into even and odd 
eigenspaces as $H^{p,q}(\mathcal X) = H^{p,q}_+(\mathcal X) \oplus 
H^{p,q}_-(\mathcal X)$ \cite{Grimm:2004uq}, and for the discussion in this paper we assume that 
the corresponding Hodge numbers satisfy
\eq{
  \label{or_ass}
  h^{2,1}_+ = 0 \,, \hspace{50pt} 
  h^{1,1}_- = 0 \,.
}
However, more general cases can be considered as well. For the third de-Rham cohomology group 
$H^3_-(\mathcal X)$ we choose a symplectic basis as follows
\eq{
  \label{or_odd}
  \{\alpha_I ,\beta^I \} \,, \hspace{50pt}
  \int_{\mathcal X} \alpha_{I} \wedge \beta^J = \delta_I{}^J \,,
  \hspace{50pt}
  I,J = 0,\ldots, h^{2,1}_- \,,
}
with all other pairings vanishing. 
For the even Dolbeault cohomology groups 
we introduce bases
\eq{
  \label{or_even_2}
  \renewcommand{\arraystretch}{1.2}
  \arraycolsep2pt
  \begin{array}{rclcl}
  |g_{\mathcal X}|^{-1/2} d\mbox{vol}_{\mathcal X} &=&\omega_0& \in & H^{3,3}_+(\mathcal X) \,,
  \\
  &&\omega_{\mathsf A} &\in& H^{1,1}_+ (\mathcal X) \,,
  \\
  &&\sigma^{\mathsf A} &\in& H^{2,2}_+ (\mathcal X) \,,
  \\
  1 &=& \sigma^0 &\in & H^{0,0}_+ (\mathcal X) \,,
  \end{array}
  \hspace{50pt}
  \mathsf A = 1,\ldots, h^{1,1}_+\,,
}
where $|g_{\mathcal X}|$ denotes the determinant of the metric for $\mathcal X$.
These bases can be chosen such that they satisfy a condition analogous to \eqref{or_odd},
in particular, we can require
\eq{
  \label{or_even}
  \{\omega_{A} ,\sigma^{A} \} \,, \hspace{50pt}
  \int_{\mathcal X} \omega_{A} \wedge \sigma^{B} = \delta_{A}{}^{B} \,,
  \hspace{50pt}
  A,B = 0,\ldots, h^{1,1}_+ \,.
}


\subsubsection*{Moduli}

When compactifying string theory from ten to four dimensions, 
the deformations of the six-dimensional background become
dynamical  fields in the four-dimensional theory. 
These moduli fields 
are contained in the multiforms  \cite{Benmachiche:2006df}
\eq{
  \label{sg_n1_89919466}
  \arraycolsep2pt
   \begin{array}{lcl}
   \Phi^+ &=& \displaystyle  e^{-\phi}\, e^{B- i\op J} \,,\\[4pt]
   \Phi^- &=& \Omega \,,
   \end{array}
   \hspace{60pt}
   \Phi^+_{\rm c} = e^{B}\op \mathcal C_{\rm mod} + i \,\mbox{Re}\op \Phi^+ \,,
}
where the  sum over all  R-R potentials $\mathcal C = \sum_p C_p$ 
has been separated into a flux contribution and a moduli contribution 
as $\mathcal C=\mathcal C_{\rm flux} + \mathcal C_{\rm mod}$.
Complex scalar fields $\tau$ and $T_{\mathsf A}$ can be determined 
by expanding $\Phi^+_{\rm c}$ into its zero- and four-form components as
follows
\eq{
  \label{moduli_32784}
  \Phi^+_{\rm c} = \tau + T_{\mathsf A} \op \sigma^{\mathsf A} \equiv T_A \op \sigma^A\,,
}
where $T_0\equiv \tau$ is called the axio-dilaton and $T_{\mathsf A}$ contain the  K\"ahler moduli of 
$\mathcal X$. 
In general this expression also contains two-forms anti-invariant under $\sigma$, which  are however 
vanishing due to our choice \eqref{or_ass}.
The complex-structure moduli $U^i$ with $i=1,\ldots, h^{2,1}_-$ are contained in 
the holomorphic three-form $\Omega$.


\subsubsection*{Fluxes}

We furthermore consider non-vanishing fluxes for the internal space. These are the
R-R three-form flux
$F = d\op\mathcal C_{\rm flux}\bigr\rvert_3$ as well as
geometric and non-geometric fluxes in the NS-NS sector.
The latter are  the three-form flux $H$, the geometric flux $f$, and the non-geometric $Q$- and $R$-fluxes.
These fluxes  can be interpreted as operators acting on the differential forms as 
\eq{
  \label{fluxes_24247}
  \renewcommand{\arraystretch}{1.2}
  \arraycolsep3pt
  \begin{array}{l@{\hspace{7pt}}c@{\hspace{12pt}}lcl}
  H\,\wedge & :& \mbox{$p$-form} &\to& \mbox{$(p+3)$-form} \,, \\
  f\hspace{4.5pt}\circ & :& \mbox{$p$-form} &\to& \mbox{$(p+1)$-form} \,, \\
  Q\,\bullet & :& \mbox{$p$-form} &\to& \mbox{$(p-1)$-form} \,, \\  
  R\,\llcorner & :& \mbox{$p$-form} &\to& \mbox{$(p-3)$-form} \,,
  \end{array}
}
and can be conveniently summarized using a generalized derivative operator \cite{Shelton:2005cf}
\eq{
  \label{gen_der}
  \mathcal D = d + H \wedge - f\circ + Q\bullet - R \op\llcorner  \,.
}
The precise action of the fluxes on the cohomology will be specified below. 
We furthermore summarize the action of the combined world-sheet parity and 
left-moving fermion number on the fluxes as follows \cite{Shelton:2005cf,Blumenhagen:2015lta}
\eq{
  \label{sg_orient_82946}
  \arraycolsep2pt
  \begin{array}{lccl}
  \Omega_{\rm P}(-1)^{F_{\rm L}} \,H &=& -& H \,, \\[4pt]
  \Omega_{\rm P}(-1)^{F_{\rm L}} \,f &=& +& f \,, \\[4pt]
  \Omega_{\rm P}(-1)^{F_{\rm L}} \,Q &=& - &Q \,, \\[4pt]
  \Omega_{\rm P}(-1)^{F_{\rm L}} \,R &=& + &R \,.
  \end{array}
  \hspace{50pt}
  \Omega_{\rm P}(-1)^{F_{\rm L}} \,F = - F \,, 
}
For our assumption \eqref{or_ass} this implies that $f$ and $R$ are vanishing.
We also note that the R-R and NS-NS three-form fluxes have to 
satisfy  quantization conditions of the form (see e.g.~\cite{Grana:2005jc})
\eq{
  \label{flux_quant}
\int_{\Gamma} F \in \mathbb Z \,,
  \hspace{60pt}
\int_{\Gamma} H \in \mathbb Z \,,
}
where $\Gamma\in H_3(\mathcal X,\mathbb Z)$ is an arbitrary three-cycle on the 
Calabi-Yau three-fold $\mathcal X$.
For orbifolds and orientifolds this condition can be modified, and we come back 
to this point on  page~\pageref{page_subtle_fluxes} below. 
Furthermore, as will be explained in section~\ref{sec_dualities}, the NS-NS fluxes are related among
each other through T-duality transformations, and hence also the geometric 
$f$- and  the non-geometric $Q$- and $R$-fluxes should be appropriately quantized.


\subsubsection*{Supergravity data}

When compactifying type IIB string theory on orientifolds of Calabi-Yau three-folds,
the resulting four-dimensional effective theory can be described in terms of 
$\mathcal N=1$ supergravity \cite{Grimm:2004uq}. In particular, the K\"ahler potential takes the
following form
\eq{
  \label{kaehler_001}
  &\mathcal K = -\log\bigl[-i\op(\tau-\ov\tau)\bigr]  - 2\log \hat{\mathcal{V}}
  - \log\biggl[ -i\int_{\mathcal X} \Omega\wedge \ov\Omega \,\biggr]
\,,
}
where $\hat{\mathcal{V}}$ denotes the volume of the Calabi-Yau manifold in Einstein frame.
The superpotential is generated by the fluxes and can be expressed using the 
Mukai pairing $\langle \cdot, \cdot \rangle$ 
of the multiforms \eqref{sg_n1_89919466} and the generalized derivative \eqref{gen_der} 
in the following way \cite{Shelton:2005cf,Villadoro:2006ia,Shelton:2006fd}
\eq{
  \label{superpot_001}
 W = &\int_{\mathcal X} \bigl\langle \Phi^-, F - \mathcal D\op\Phi^+_{\rm c} \bigr\rangle
 \\[4pt]
  =&  \int_{\mathcal X}\Omega \wedge \Bigl[ F -\tau \op H 
  - \bigl(Q\bullet \sigma^{\mathsf A}\bigr) \op T_{\mathsf A}
  \Bigr].
}
In general, the fluxes \eqref{fluxes_24247} also generate a D-term potential which
can be expressed using 
the three-form part of $ \mathcal D (\mbox{Im}\op \Phi^+)$ \cite{Blumenhagen:2015lta} (see also \cite{Betzler:2017kme}). 
However,  due to \eqref{sg_orient_82946}
the latter belongs to the $\sigma$-even third cohomology and vanishes
when taking into account our requirements \eqref{or_ass}. In our setting therefore 
no D-term potential is generated.


\subsubsection*{Bianchi identities and tadpole-cancellation conditions}

Finally, the R-R and NS-NS fluxes have to satisfy a number of Bianchi identities. These can 
be expressed  using the generalized derivative $\mathcal D$ as follows
\eq{
  \label{bianchi_001}
  \mathcal D^2 = \mbox{NS-NS sources}\,,
  \hspace{50pt}
  \mathcal D F = \mbox{R-R sources} \,,
}
where NS-NS sources stand for NS5-branes, Kaluza-Klein monopoles or 
non-geo\-met\-ric $5^2_2$-branes
(see for instance \cite{Plauschinn:2018wbo} for a review and collection of references).
However, in this work we assume these to be absent and therefore 
require $\mathcal D^2=0$. 
The R-R sources stand for orientifold planes and D-branes, and 
the second condition in \eqref{bianchi_001} is also known 
as the tadpole cancellation condition. We discuss this condition
in more detail in the following section.


\subsection{Tadpole-cancellation condition}
\label{sec_tadpole}

The tadpole-cancellation condition is an important consistency conditions for 
type I string theories. It links the closed-string to the open-string sector and
puts  strong constraints on the allowed D-brane configurations 
(for a review see for instance \cite{Blumenhagen:2006ci}). 
From a conformal-field-theory point of view the tadpole-cancellation condition
ensures the absence of UV divergencies in one-loop amplitudes (see 
e.g. \cite{Kiritsis:2007zza,Blumenhagen:2009zz} for textbook reviews) and 
therefore plays an important role for  string theory being a consistent theory of 
gravity. 
From an effective-field-theory point of view, the tadpole-cancellation condition
is the integrated version of the equation of motion for the R-R potentials
and 
ensures the absence of certain anomalies in type II orientifold compactifications via
the generalized Green-Schwarz mechanism \cite{Sagnotti:1992qw}.
The tadpole-cancellation condition is thus an important consistency condition
for string compactifications.


\subsubsection*{Explicit expressions}

We now formulate the tadpole-cancellation condition 
for the setting of the previous section. 
The contribution of the R-R-sources can be described using 
the charges \cite{Minasian:1997mm,Aspinwall:2004jr}
\eq{  
  \mathcal Q_{{\rm D}p} =  \mbox{ch}\left( \mathcal F\right) \wedge
    \sqrt{ \frac{\hat{\mathcal A}(\mathcal R_T)}{\hat{\mathcal A}(\mathcal R_N)}} \wedge
    [\Gamma_{{\rm D}p}] \,,
    \hspace{30pt}
  \mathcal Q_{{\rm O}p} = Q_p \sqrt{ \frac{\mathcal{L}(\mathcal R_T/4)}{\mathcal{L}(\mathcal R_N/4)}} \wedge 
   [\Gamma_{{\rm O}p}] \,,
}
where $[\Gamma_{{\rm D}p}]$  and $[\Gamma_{{\rm O}p}]$ denote the Poincar\'e duals 
of the cycles wrapped by  D-branes and O-planes in $\mathbb R^{1,3}\times \mathcal X$.
The open-string gauge flux on the D-branes $\mathcal F$ appears in the Chern character, 
the tangential and normal part of the curvature two-form $\mathcal R$ appear
in the $\hat{\mathcal A}$-genus and the Hirzebruch polynomial $\mathcal L$,
and $Q_p=-2^{p-4}$ denotes the charge of an orientifold $p$-plane.
For more details we refer for instance to section 8.6 in \cite{Plauschinn:2018wbo}.
Denoting the orientifold image of a D$p$-brane with a prime, the 
Bianchi identity for the R-R fluxes then reads
\eq{
  \label{tp_038}
  \mathcal D F =   \sum_{{\rm D}p+ {\rm D}p'} \hspace{-3pt}\mathcal Q_{{\rm D}p}
  + \sum_{{\rm O}p}  \mathcal Q_{{\rm O}p} \,,
}
where the sum is over all D-branes and orientifold planes present in the background.
The Freed-Witten anomaly-cancellation condition \cite{Freed:1999vc} for D-branes takes the general 
form \cite{Blumenhagen:2015kja,Plauschinn:2018wbo}
\eq{
\mathcal D \mathcal Q_{{\rm D}p} = 0 \,, \hspace{50pt}
\mathcal D \mathcal Q_{{\rm O}p} = 0\,,
}
where we included the corresponding expression for an orientifold $p$-plane. Equation 
\eqref{tp_038} can therefore be interpreted as a relation in  $\mathcal D$-cohomology.

For the setting discussed in this paper, the orientifold projection 
satisfies \eqref{orient_choice} and therefore leads to spacetime filling O3- and O7-planes. 
Taking into account \eqref{or_ass} and that $F$ in \eqref{tp_038} is a 
three-form flux, we find the following explicit expressions
\eq{
  \label{tadpoles_001}
  \hspace{-30pt}
  Q\bullet F =& -2\op \sum_{{\rm D}7_a}  N_{{\rm D}7_a}   [\Gamma_{{\rm D}7_a}] + 8 
  \sum_{{\rm O}7_b} \,  [\Gamma_{{\rm O}7_b}] \,,
  \\[6pt]
  H\wedge F 
 =& - 2\left(  N_{{\rm D}3}- \frac{N_{{\rm O}3}}{4} \right) \omega_0
 \\
  &- \sum_{{\rm D}7_a}
    \mbox{tr}\bigl( \mathcal{F}_a\bigr)^2 \wedge  [\Gamma_{{\rm D}7_a}] 
  +2\left( \sum_{{\rm D}7_a}N_{{\rm D}7_a}\:\frac{\chi( \Gamma_{{\rm D}7_a}\bigr)}{24}  
  +\sum_{{\rm O}7_b} \frac{\chi\bigl( \Gamma_{{\rm O}7_b}\bigr)}{12}\right) \omega_0 \,.
  \hspace{-20pt}
}
Here, 
$N_{{\rm D}3}$ and $N_{{\rm O}3}$ denote the number of 
D$3$-branes and O$3$-planes, 
$N_{{\rm D}7_a}$ denotes the number of coincident D$7$-branes in a stack $a$,
$ [\Gamma_{{\rm D}7_a}]  = n_{a}^{\mathsf A} \omega_{\mathsf A}$ is 
the Poincar\'e dual of the cycle $\Gamma_{{\rm D}7_a}$ expanded in the basis 
\eqref{or_even_2},
$\mathcal F_a$ is the (quantized) two-form gauge flux on the stack of
D$7$-branes $a$ in the fundamental representation, and $\chi(\Gamma)$ denotes
the Euler number of the cycle $\Gamma$. The D-brane sums are over all 
D7-branes, and due to $h^{1,1}_-=0$ the orientifold images give a 
factor of two. 
For details on the derivation of these expressions see for instance
\cite{Plauschinn:2008yd}.


\subsubsection*{Orientifold contributions}

Let us now discuss the contribution of  orientifold planes to the 
right-hand sides in \eqref{tadpoles_001}. Typically  orientifold planes 
give a positive contribution while D-branes give a negative contribution. 
For some classes of models the orientifold-contributions can be estimated as follows:
\begin{itemize}

\item 
For orientifolds of $\mathbb T^6/\mathbb Z_M$ or
$\mathbb T^6/\mathbb Z_M\times \mathbb Z_N$
the numbers of O$3$- and O$7$-planes 
have been computed for instance in 
\cite{Lust:2006zh} for some examples. 
Here, the authors find that $N_{{\rm O}3},N_{{\rm O}7} \lesssim 60$ and 
the contribution of the Euler numbers to \eqref{tadpoles_001} 
are vanishing. 
The contribution of orientifold planes to the right-hand sides of  \eqref{tadpoles_001}
is therefore typically positive and of order $\mathcal O(10)$.

\item For del-Pezzo surfaces the possible orientifold projections 
have been classified in \cite{Blumenhagen:2008zz}. The number
of orientifold three- and seven-planes are of order $\mathcal O(10)$, 
and in some examples the Euler numbers of the four-cycles are 
of order $\mathcal O(100)$. 
Also here, the contribution of orientifold planes to the right-hand side of 
  \eqref{tadpoles_001} is positive and of order $\mathcal O(10)$.

\item In F-theory the geometry of Calabi-Yau four-folds $\mathcal Y$ encodes
the geometry of D$7$-branes and orientifold planes in Calabi-Yau three-folds $\mathcal X$. If a lift from 
type IIB orientifolds to F-theory is possible, one finds that
\eq{
  \frac{\chi(\mathcal Y)}{24} = \frac{N_{{\rm O}3}}{4} 
  + \sum_{{\rm D}7_a}N_{{\rm D}7_a}\op\frac{\chi( \Gamma_{{\rm D}7_a}\bigr)}{24}  
  +\sum_{{\rm O}7_b} \frac{\chi\bigl( \Gamma_{{\rm O}7_b}\bigr)}{12} \,,
}
where $\chi(\mathcal Y)$ denotes the Euler number of the Calabi-Yau four-fold $\mathcal Y$. 
In \cite{Candelas:1997eh,Lynker:1998pb} a manifold $\mathcal Y_{\rm max}$ was
identified with the largest known Euler number for a Calabi-Yau four-fold 
\eq{
  \chi(\mathcal Y_{\rm max}) = 1\, 820\, 448\,,
}
and more details for the present context can be found in \cite{Taylor:2015xtz}.
Hence, for this example the contribution to the D$3$-tadpole in \eqref{tadpoles_001} is of order
 $\mathcal O(10^5)$. 

\end{itemize}


\subsubsection*{D-brane contributions}

We furthermore note that the tadpole-cancellation conditions \eqref{tadpoles_001} are the integrated
versions of the R-R Bianchi identities \eqref{bianchi_001}. The former are therefore 
less restrictive than the latter, but for a proper string-theory solution also the Bianchi identities with 
localized sources have to be solved. 
When placing D-branes directly on top of orientifold planes solutions may be constructed more easily, 
but in general this a difficult task (see for instance \cite{Junghans:2013xza}).

However, we can make the following general argument: 
because D-branes have a non-vanishing mass, 
their probe approximation breaks down when too many 
D-branes are placed into  a compact space (away from the orientifold planes).
In this case the back-reaction of D-branes on the geometry has to be 
taken into account, and an extreme case for this mechanism is the formation of black holes. 
It would be desirable to make this more precise, but we can argue that  
for ignoring back-reaction effects the contribution of D-branes  to the right-hand sides in
 \eqref{tadpoles_001} should not be arbitrarily large.


\subsubsection*{Flux contributions}

Turning now to the flux contribution on the left-hand sides in \eqref{tadpoles_001},
we note that for vanishing $Q$-flux the $H\wedge F$-term typically has to be positive
in order to obtain physically-relevant solutions. 
Since the right-hand sides are bounded from above by the 
orientifold contributions, the flux contributions should not be larger than 
$\mathcal O(10)$ to $\mathcal O(10^5)$. 
In the presence of non-geometric $Q$-flux the left-hand sides in 
\eqref{tadpoles_001}  can  be negative -- but since also 
the D-brane contributions are bounded, again the flux contributions should not be
too large. 
This is an important point for our approach in this paper, which we summarize as\label{page_ma}
\begin{center}\fbox{
\begin{minipage}{0.8\textwidth}
In order to solve the tadpole-cancellation condition \eqref{tadpoles_001} 
and ignore the back-reaction of D-branes, the contribution of fluxes to 
the left-hand sides in 
\eqref{tadpoles_001} should not be too large. Depending on the 
setting, known bounds are of orders $\mathcal O(10)$ to $\mathcal O(10^5)$.
\end{minipage}}
\end{center}


\subsection{$\mathbb T^6/\mathbb Z_2\times \mathbb Z_2$ orientifold}
\label{sec_pre_or_t6}

Let us now turn to a specific example for a compactification space.
We consider the orbifold $\mathbb T^6/\mathbb Z_2\times \mathbb Z_2$ which provides 
a simple example of a Calabi-Yau three-fold with only few moduli.
For our purposes it is sufficient to stay in the orbifold limit and not blow-up the fixed-point singularities, that is 
we ignore the twisted sectors. 
This model has been extensively studied in the literature, and we refer for instance 
to \cite{Font:1988mk,Berkooz:1996dw,Antoniadis:1996vw,Lust:2005dy} for more details
in the present context.

For this model the contribution of orientifold planes to the tadpole-cancellation condition 
\eqref{tadpoles_001}
only allows for a small number of different flux choices. 
In order to be able to study general properties of the space of solutions, in the following we therefore 
ignore the precise form of the tadpole cancellation condition and allow for 
arbitrarily-large values of $H\wedge F$ and $Q\bullet F$.
We do however keep in mind that these tadpole contributions are bounded 
by the D-brane and orientifold contributions.


\subsubsection*{Compactification space}

We start from the following six-dimensional 
orbifold construction which has the properties of a  Calabi-Yau three-fold
\eq{
\label{orbifold_001}
\mathcal X = \frac{\mathbb T^2\times \mathbb T^2\times \mathbb T^2}{\mathbb Z_2\times\mathbb Z_2} \,.
}
On each of the two-tori we  introduce complex coordinates as
\eq{
  \label{coords_001}
  z^i = x^i + U^i \op y^i \,, \hspace{50pt} i = 1,2,3\,,
}
where $x^i$ and $y^i$ denote real coordinates 
with identifications $x^i\sim x^i + 1$ and $y^i\sim y^i+1$, 
$U^i$ denote the complex structures on each of the $\mathbb T^2$, and no summation is performed in 
\eqref{coords_001}.
The orbifold action is given by
\eq{
  \label{orbifold_action_01}
  \arraycolsep2pt
  \Theta: \left(\begin{array}{c} z_1 \\ z_2 \\ z_3 \end{array} \right)
  \rightarrow
  \left(\begin{array}{c} - z_1 \\ -z_2 \\ +z_3 \end{array} \right)  ,
  \hspace{50pt}
  \Theta':\left(\begin{array}{c} z_1 \\ z_2 \\ z_3 \end{array} \right)
  \rightarrow
  \left(\begin{array}{c} + z_1 \\ -z_2 \\ -z_3 \end{array} \right)  ,
}
where $\Theta$ and $\Theta'$ are the two generators of the orbifold group $\mathbb Z_2\times \mathbb Z_2$. 
In addition, we perform the following 
orientifold projection
\eq{
  \label{orientifold_action_01}
  \arraycolsep2pt
  \sigma: \left(\begin{array}{c} z_1 \\ z_2 \\ z_3 \end{array} \right)
  \rightarrow
  \left(\begin{array}{c} - z_1 \\ -z_2 \\ -z_3 \end{array} \right)  .
}


\subsubsection*{Cohomology}

Next, we turn to the cohomology of \eqref{orbifold_001}. We note that 
there are no one- or five-forms invariant under the 
orbifold action \eqref{orbifold_action_01}, and that 
the invariant three-forms are given by the following combinations
\eq{
  \label{basis_001}
  \arraycolsep2pt
  \begin{array}{lcl@{\hspace{50pt}}lcl}
  \alpha_0 &=& dx^1 \wedge dx^2 \wedge dx^3 \,,
  &
  \beta^0 &=& +dy^1 \wedge dy^2 \wedge dy^3 \,,
  \\[4pt]
  \alpha_1 &=& dy^1 \wedge dx^2 \wedge dx^3 \,,
  &
  \beta^1 &=& -dx^1 \wedge dy^2 \wedge dy^3 \,,
  \\[4pt]
  \alpha_2 &=& dx^1 \wedge dy^2 \wedge dx^3 \,,
  &
  \beta^2 &=& -dy^1 \wedge dx^2 \wedge dy^3 \,,
  \\[4pt]
  \alpha_3 &=& dx^1 \wedge dx^2 \wedge dy^3 \,,
  &
  \beta^3 &=& -dy^1 \wedge dy^2 \wedge dx^3 \,.
  \end{array}
}
Choosing the orientation of the six-dimensional space \eqref{orbifold_001}
such that we have $\int dx^1\wedge dx^2 \wedge dx^3 \wedge dy^1\wedge dy^2\wedge dy^3 = 1$, 
the three-forms in \eqref{basis_001} satisfy the intersection relation \eqref{or_odd}.
We can furthermore define a holomorphic three-form as
\eq{
  \Omega = dz^1 \wedge dz^2 \wedge dz^3\,,
}
which -- when  expanded in the  basis \eqref{basis_001} -- takes the form
\eq{
  \label{moduli_hol3}
  \arraycolsep2pt
  \begin{array}{rccrlcrlcrll}
  \Omega = \alpha_0 &+ & \bigl( & U^1 & \alpha_1 &+& U^2 &\alpha_2 &+& U^3 & \alpha_3  &\bigr)
  \\[5pt]
  &- &\bigl(& U^2\op U^3 & \beta^1 & +&  U^3\op U^1 &\beta^2 &+& U^1\op U^2& \beta^3 &\bigr) 
  + U^1\op U^2\op U^3 \op \beta^0 \,.
  \end{array}
}
Turning to the orientifold action \eqref{orientifold_action_01}, we see that all three-forms
\eqref{basis_001} are odd under $\sigma$ and therefore $h^{2,1}_- = 3$ and $h^{2,1}_+=0$.
We also note that  $\Omega$ is odd
under the orientifold action as required by \eqref{orient_choice}.

For the even cohomology we observe that the zero- and six-form cohomologies are 
even under the orbifold action \eqref{orbifold_action_01}. 
For the second cohomology we find the following invariant $(1,1)$-forms
\eq{
  \label{coho_01}
  \omega_{\mathsf A} &= \frac{i}{2 \,\mbox{Im}\op U^{\mathsf A}}\, dz^{\mathsf A}\wedge d\ov z^{\mathsf A} \,, 
  \hspace{50pt} \mathsf A= 1,2,3\,,
}
with no summation over $\mathsf A$,
and we define invariant $(2,2)$-forms as
\eq{
  \label{coho_02}
  \sigma^1 = -\omega_2 \wedge \omega_3 \,, \hspace{30pt}
  \sigma^2 = -\omega_3 \wedge \omega_1 \,, \hspace{30pt}
  \sigma^3 = -\omega_1 \wedge \omega_2 \,.
}
Note that these satisfy the relations shown in \eqref{or_even}.
For the orbifold \eqref{orbifold_001} we can now define a real K\"ahler form 
in the following way
\eq{
  J = t^1 \omega_1 + t^2 \omega_2 + t^3\omega_3 \,,
}
where the  $t^{\mathsf A}$ are the (real) K\"ahler moduli. 
The forms \eqref{coho_01} and \eqref{coho_02} are all even under the orientifold
projection \eqref{orientifold_action_01} and therefore $h^{1,1}_+=3$ and $h^{1,1}_-=0$. We
also note that $J$ is even under $\sigma$, in agreement with \eqref{orient_choice}.


\subsubsection*{Moduli}

With the explicit expressions for the even cohomologies discussed above, we can now 
determine the moduli fields contained in $\Phi_{\rm c}^+$ via equation \eqref{moduli_32784}. 
For the R-R zero- and four-form potentials (purely in the internal space) we use the following conventions
\eq{
  C_0 = c_0 \,, \hspace{50pt}
  C_4 \bigr\rvert_{\rm internal}= c_1 \op\sigma^1 + c_2\op \sigma^2 + c_3\op \sigma^3 \,,
}
and evaluating \eqref{moduli_32784} in the present situation leads to
\eq{
  \label{def_mod_003}
  T_0 = \tau = c_0 + i\op e^{-\phi} \,,
  \hspace{60pt}
  \arraycolsep2pt
  \begin{array}{lcl}
  T_1  &=& c_1 + i \op \mathsf t^2 \mathsf t^3 \,, \\[4pt]
  T_2  &=& c_2 + i \op \mathsf t^3 \mathsf t^1 \,, \\[4pt]
  T_3  &=& c_3 + i \op \mathsf t^1 \mathsf t^2 \,,
  \end{array}
}
with the Einstein-frame K\"ahler moduli defined as $\mathsf t^{\mathsf A} = e^{-\phi/2} t^{\mathsf A}$.
We also note that the R-R two-form potential $C_2$ is odd under the combined world-sheet parity 
and left-moving fermion number (cf. \eqref{orient_signs})
and should therefore be expanded in the $\sigma$-odd $(1,1)$-cohomology, which 
however vanishes. 
Finally, the complex-structure moduli $U^i$ are contained in $\Omega$ 
as can be seen from \eqref{moduli_hol3}.


\subsubsection*{Fluxes}
\label{page_fluxes}

Let us now turn to the fluxes. Using the basis of three-forms
\eqref{basis_001},
the NS-NS and R-R three-form fluxes can be  expanded in the following way
\eq{
  \label{flux_001}
  F = f^I \alpha_I + f_I \op\beta^I \,,
  \hspace{50pt}
  H = h^I \alpha_I+ h_I \op \beta^I \,,
}
where $I = 0,\ldots,3$. The expansion coefficients $f^I,f_I,h^I,h_I$ are quantized
due to the flux quantization conditions for $F$ and $H$ shown in \eqref{flux_quant}. 
For the remaining fluxes in the NS-NS sector we note that
due to \eqref{or_ass} and \eqref{sg_orient_82946}, the geometric $f$- and the
non-geometric $R$-flux vanish. 
The $Q$-flux is in general non-vanishing, and we specify it by
its action on the third and fourth cohomology as
\eq{
  \label{flux_002}
  \arraycolsep2pt
  \begin{array}{lcl}
  \displaystyle Q\bullet \alpha_I &=& -q_I{}^{\mathsf A}\op \omega_{\mathsf A} \,, 
  \\[6pt]
  \displaystyle Q\bullet \beta^I &=& +q^{I{\mathsf A}}\op \omega_{\mathsf A} \,,
  \end{array}
  \hspace{60pt}
  Q \bullet  \sigma^{\mathsf A}  =  q^{I\mathsf A} \alpha_I +q_I{}^{\mathsf A} \beta^I\,.
}
Here we have again $\mathsf A=1,2,3$ and the flux quanta are  integers. 
In order to shorten the notation for our subsequent discussion, we combine 
the $H$-flux with the $Q$-flux by defining
\eq{
  q_I{}^0 = h_I \,, \hspace{60pt}
  q^{I0} = h^I \,.
}

\label{page_subtle_fluxes}
Let us briefly discuss a
subtlety concerning the flux quantization condition \eqref{flux_quant}. 
It was first pointed out in \cite{Frey:2002hf} that on orbifold (or orientifold)
spaces besides bulk cycles inherited from the covering space,  twisted 
cycles of shorter length can exist. This implies that the
quantization condition of the fluxes shown above is slightly modified. 
For the present example of the type IIB $\mathbb T^6/\mathbb Z_2\times \mathbb Z_2$ 
orientifold this observation has been mentioned in 
\cite{Kachru:2002he,Shelton:2006fd}
and has been analyzed in detail for instance in \cite{Blumenhagen:2003vr,Cascales:2003zp}.
More concretely, for $\mathbb Z_2\times \mathbb Z_2$ orbifold actions with and without discrete torsion (see \cite{Vafa:1986wx})
one finds that fluxes on generic bulk cycles have to satisfy
\eq{\label{fluxes_TorsionConstraint}
  \arraycolsep10pt
  \begin{array}{rl}
  \mbox{without discrete torsion} & f^I,f_I,q^{IA},q_I{}^A \in8\op\mathbb Z \,,
  \\[4pt]
  \mbox{with discrete torsion} & f^I,f_I,q^{IA},q_I{}^A \in4\op\mathbb Z \,.
  \end{array}
}
As mentioned at the beginning of this subsection, in this paper we ignore the twisted sector which effectively 
implies that we consider models without discrete torsion \cite{Blumenhagen:2003vr}. Fluxes will therefore 
be quantized in multiples of eight. 
In the literature similar orientifolds have been studied
 \cite{Kachru:2002he,Ashok:2003gk,Denef:2004ze,DeWolfe:2004ns,Shelton:2006fd},
although with slightly different quantization conditions.


\subsubsection*{Bianchi identities}

Turning to the Bianchi identities \eqref{bianchi_001}, we  recall from \eqref{or_even_2}
that the collective basis for the even cohomology is denoted by $\omega_A$
and from \eqref{flux_001} and \eqref{flux_002} that the R-R and NS-NS flux quanta are given by
$f_I$, $f^I$, $q_I{}^A$ and $q^{IA}$.
For the left-hand side of the Bianchi identities we then
introduce the following
general notation 
\eq{
  \label{flux_020}
  \renewcommand{\arraystretch}{1.3}
  \arraycolsep2pt
  \begin{array}{lcl@{\hspace{1pt}}l@{\hspace{60pt}}lcl}
  \mathcal D F &=& \mathsf Q^{A} & \omega_A \,,
  & \mathsf Q^{A} &=&   f_I q^{IA} -f^I q_I{}^{A} \,,
  \\
  \mathcal D H &=& \mathsf Q^{0 B} & \omega_B \,,
  & \multirow{2}{*}{$\mathsf Q^{AB}$} &  \multirow{2}{*}{$=$}
  &  \multirow{2}{*}{$q_I{}^{A}q^{I{B}} -q^{I{A}}q_I{}^{B} \,, $}
  \\
  \mathcal D \op(Q\bullet \sigma^{\mathsf A}) &=& \mathsf Q^{\mathsf AB} & \omega_B \,,  
  \end{array}
}
where $A,B=0,\ldots, 3$. Note that these expressions can be combined 
into an anti-symmetric  five-by-five matrix of the form
\eq{
  \label{bianchi_010}
  \arraycolsep2pt
  \mathbf Q = \left( \begin{array}{cr}
  0 & +\mathsf Q^{ B} \\[4pt]
  - \mathsf Q^{A} & \mathsf Q^{AB}
  \end{array}\right),
  \hspace{50pt}
  A,B = 0,1,\ldots,3 \,.
}
The right-hand side of the Bianchi identities \eqref{bianchi_001} correspond to NS-NS and R-R sources, and 
schematically we have the relations
\eq{
  \renewcommand{\arraystretch}{1.2}
  \begin{array}{lcl}
  \mathsf Q^{0} & \longleftrightarrow & \mbox{O3-planes/D3-branes,} \\
  \mathsf Q^{\mathsf A} & \longleftrightarrow & \mbox{O7-planes/D7-branes,} \\
  \mathsf Q^{0 \mathsf A} & \longleftrightarrow & \mbox{NS5-branes,} \\
  \mathsf Q^{\mathsf A\mathsf B} & \longleftrightarrow & \mbox{$5^2_2$-branes,}
  \end{array}
}
where in particular $\mathsf Q^{A}$ for $A = 0,\ldots, 3$ are the contributions to the R-R tadpole cancellation conditions
\eqref{tadpoles_001}.
As mentioned above, in this paper we do not consider NS5-branes or non-geometric 
$5^2_2$-branes which leads to the requirement 
$\mathsf Q^{AB}=0$ for $A,B=0,\ldots,3$.


\subsubsection*{Supergravity data}

Let us finally determine the K\"ahler and superpotential for the $\mathbb T^6/\mathbb Z_2\times \mathbb Z_2$
orientifold com\-pacti\-fi\-cation.
Evaluating \eqref{kaehler_001} we find for the K\"ahler potential
\eq{
\label{kahler_001}
\mathcal K = 
-\sum_{A=0}^3 \log\Bigl[-i\op(T_{A}-\ov T_{A})\Bigr]
-\sum_{i=1}^3 \log\Bigl[-i\op(U{}^i-\ov U{}^i)\Bigr] \,,
}
up to an irrelevant constant term.
Turning to the superpotential \eqref{superpot_001}, the expansions of the fluxes  in \eqref{flux_001} and 
\eqref{flux_002}  give rise to 
\eq{\label{superpot_002}
  W = 
  \arraycolsep1.4pt
  \renewcommand{\arraystretch}{1.4}
  \begin{array}[t]{rllclll}
  &&f_0  &- & q_0{}^{{A}} & T_{{A}} &
  \\
  +U^i &\displaystyle \bigl(&  f_i& - & q_i{}^{{A}} & T_{{A}}  & \displaystyle \bigr) 
  \\
  +\tfrac{1}{2}\op\sigma_{ijk} \op U^iU^j& \displaystyle\bigl(&  f^k  &- & q^{k{A}} & T_{{A}} & \displaystyle \bigr)
  \\
  -\tfrac{1}{6}\op\sigma_{ijk}\op U^iU^jU^k & \displaystyle\bigl(& f^0 &- & q^{0{A}} & T_{{A}}&  \displaystyle \bigr) \,,
  \end{array}
}
where a summation 
over $A=0,\ldots,3$ and $i=1,2,3$ is understood. For ease of notation 
we also defined the symmetric symbol $\sigma_{ijk}$ which has the only non-vanishing components 
\eq{\label{sigma_symbol}
  \sigma_{123}=
  \sigma_{132}=
  \sigma_{231}=
  \sigma_{213}=
  \sigma_{312}=
  \sigma_{321}=      +1 \,.
}
The scalar F-term potential is determined in terms of 
the K\"ahler $\mathcal K$ and superpotential $W$ according to 
\eq{
  \label{f_term_pot}
  V_F = e^{\mathcal K} \left[ D_{\alpha} W \, \mathcal G^{\alpha \ov\beta} D_{\ov\beta} \ov W - 
  3\op |W|^2 \right] ,
}
where $\phi^{\alpha}$ collectively labels the complex scalar fields of the theory.
The K\"ahler metric is computed from the K\"ahler potential 
as $\mathcal G_{\alpha\ov\beta} = \partial_{\alpha}\partial_{\ov\beta}\op \mathcal K$, 
and the covariant derivative reads
\eq{
  \label{f_term_eq_001}
  D_{\alpha} W = \partial_{\alpha} W + \bigl( \partial_{\alpha} \mathcal K \bigr) W \,.
}
We also note that due to our assumption $h^{2,1}_+ = 0$ shown in \eqref{or_ass}, 
no D-term potential is generated by the fluxes.


\subsection{Dualities}
\label{sec_dualities}

We now want to discuss  dualities for the
orientifold of $\mathbb T^6/\mathbb Z_2\times \mathbb Z_2$ introduced in the previous section. 
We are interested in transformations
which leave the physical properties of a system invariant but which are not necessarily symmetries of the 
action. 
In particular, we note that an extremum of the F-term potential \eqref{f_term_pot} 
is reached for vanishing F-terms
\eq{
  \label{f_term_eq}
  0 = \partial_{\alpha} W + \bigl( \partial_{\alpha} \mathcal K \bigr) W \,,
}
and in our subsequent analysis we are interested in duality transformations which map 
solutions of \eqref{f_term_eq} to new solutions.


\subsubsection*{Overall sign-change}

Let us start by noting that the F-term potential \eqref{f_term_pot} 
as well as the F-term equations \eqref{f_term_eq} are invariant 
under changing the sign of all fluxes \cite{Shelton:2006fd}
\eq{
  \label{dual_010}
  \bigl(f_I, f^I, q_I{}^A, q^{IA}\bigr)
   \;\longrightarrow \;
   \bigl(-f_I, -f^I, -q_I{}^A, -q^{IA}\bigr) \,.
}
This $\mathbb Z_2$ transformation maps $W\to -W$, which indeed leaves 
the scalar potential 
\eqref{f_term_pot}, the equations \eqref{f_term_eq} 
and  the 
 tadpole contributions \eqref{flux_020}
invariant.


\subsubsection*{$SL(2,\mathbb Z)$ for complex-structure moduli $U^i$}

Next, we consider the group of large diffeomorphisms for each of the two-tori in 
\eqref{orbifold_001} \cite{Shelton:2006fd}. 
For a single $\mathbb T^2$ this group is $SL(2,\mathbb Z)$, which is generated by 
$T$- and $S$-transformations of the form 
\eq{
  T: \, U^i \to U^i + 1 \,, \hspace{50pt}
  S: \, U^i \to -1/U^i \,,
}
with $i=1,2,3$.
In order for the F-term equation \eqref{f_term_eq} to stay invariant under $T$-transformations, the fluxes have to transform
in the following way 
\eq{\label{CS-transformations_T}
  U^i \to U^i + b^i \,,
  \hspace{60pt}
  \arraycolsep2pt
  \renewcommand{\arraystretch}{1.2}
  \begin{array}{lclcl}
  g_0 &\to& g_0 &-& b^i g_i +\frac{1}{2}\op\sigma_{ijk}\op b^{i}b^{j}g^{k}+\frac{1}{6}\op\sigma_{ijk}\op b^{i}b^{j}b^{k}\,, \\
  g_i &\to & g_i &-& \sigma_{ijk}\op b^j g^k-\frac{1}{2}\op \sigma_{ijk}\op b^{j}b^{k}g^{0} \,, \\
  g^i &\to &g^i &+& b^i  g^0\,,\\
  g^0 &\to& \multicolumn{2}{l}{g^0 \,,}
  \end{array}
}
where $(g_I,g^I)$ stands collectively for $(f_I,f^I)$ and $(q_I{}^A,q^{IA})$,
and  $\sigma_{ijk}$ was defined in \eqref{sigma_symbol}. 
Under $S$-transformations of the complex structure moduli, the fluxes transform as
follows 
\eq{
  \label{dual_012}
  U^1\to - 1/U^1 \,,
  \hspace{60pt}
  \arraycolsep2pt
  \renewcommand{\arraystretch}{1.1}
  \begin{array}{lcl}
  g_0 &\to& - g_1 \,, \\
  g_1 &\to & +g_0\,,\\
  g_2 &\to&  -g^3 \,,\\
  g_3 &\to & -g^2 \,,
  \end{array}
  \hspace{20pt}
  \begin{array}{lcl}
  g^0 &\to& - g^1 \,, \\
  g^1 &\to & +g^0\,,\\
  g^2 &\to&  +g^3 \,,\\
  g^3 &\to & +g^2 \,,
  \end{array}
}
and similarly for $U^2$ and $U^3$. Note that for the fluxes this is not a $\mathbb Z_2$ but a $\mathbb Z_4$ 
action, which is however reduced to $\mathbb Z_2$ using \eqref{dual_010}.
We also note that for a simultaneous $S$-transformation of all three complex-structure moduli, the transformation reads
\eq{
    U^i\to - 1/U^i \,,
  \hspace{60pt}
  \arraycolsep2pt
  \renewcommand{\arraystretch}{1.1}
  \begin{array}{lcl}
  g_I &\to& + g^I \,, \\
  g^I &\to & -g_I\,.
  \end{array}
}
Furthermore, all Bianchi identities and tadpole contributions $\mathsf Q^A$ and $\mathsf Q^{AB}$ 
are invariant under these transformations.


\subsubsection*{T-duality}

We  now turn to T-duality transformations. It is well-known that  performing an odd number 
of T-dualities for type IIB string theory results in the type  IIA theory and 
vice versa, and applying two or six T-dualities to type IIB string theory with O3-/O7-planes
results in type IIB with O5-/O9-planes. 
For T-duality to map the present setting of type IIB with O3-/O7-planes to itself, we therefore have to
perform four collective T-duality transformations.

Let us now consider more closely the $\mathbb T^6/\mathbb Z_2\times \mathbb Z_2$ orientifold 
with O3-/O7-planes.
Using the Buscher rules \cite{Buscher:1987sk,Buscher:1987qj}, a collective T-duality transformation \cite{Plauschinn:2014nha}
say along the first and second $\mathbb T^2$ results in the following transformation 
of the moduli
\eq{
  \label{dual_011}
  \mbox{T-duality along $z^1,z^2\op$:}
  \hspace{20pt}
  \left\{
  \arraycolsep3pt
  \begin{array}{@{\hspace{6pt}}lcl@{\hspace{20pt}}lcl} 
   \tau &\to& T_3 \,, \\
   T_1 &\to&  T_2 \,, &   U^1 &\to&   -1/U^1 \,, \\
   T_2 &\to&  T_1 \,, & U^2 &\to&   -1/U^2 \,,  \\
   T_3 &\to&  \tau \,.
  \end{array}
  \right.
}
In \eqref{dual_011} we have only shown how the moduli transform, 
but also the fluxes transform in a non-trivial way under T-duality. 
However, \eqref{dual_011} contains an S-transform of the 
complex-structure moduli $U^1$ and $U^2$.
To better show the underlying structure, let us undo the 
$U^i$ transformation in  \eqref{dual_011} using \eqref{dual_012}. We then obtain
the transformation
\eq{
  \label{dual_13}
  \arraycolsep3pt
  \begin{array}[t]{lcl}
  \tau &\to& \displaystyle T_3 \,, \\
  T_1 &\to& \displaystyle T_2 \,, \\
  T_2 &\to& \displaystyle T_1 \,, \\
  T_3 &\to& \displaystyle \tau \,,
  \end{array} 
  \hspace{70pt}
  \begin{array}[t]{lcl}
  q_I{}^1 &\leftrightarrow& q_I{}^2 \,, \\
  q_I{}^3 &\leftrightarrow& h^I \,, \\
  q^{I1} &\leftrightarrow& q^{I2} \,, \\
  q^{I3} &\leftrightarrow& h^I \,,
  \end{array}
}
which by a slight abuse of notation we will refer to as T-duality in the following. 
Similar transformations are obtained  for T-duality along the second \& third and first \& third 
two-torus.\footnote{We also mention that the transformation of the moduli under T-duality 
shown in \eqref{dual_13} was to be expected: 
for type IIB orientifolds the 
R-R zero- and four-form potentials $C_0$ and $C_4$ are
the real parts of $\tau$ and $T_{\mathsf A}$, respectively.
Under a single T-duality 
the R-R potentials transform as $C_p\to C_{p\pm1}$ \cite{Hassan:1999bv}, 
where the upper/lower sign is for a transformation transversal/longitudinal
to $C_p$. 
For a collective T-duality along four directions we therefore 
map $C_0\to C_4$ and some components of 
$C_4\to C_0$. This agrees with \eqref{dual_13}.}

We furthermore  observe that the F-term equations \eqref{f_term_eq} are invariant under 
a permutation of the K\"ahler moduli $T_{\mathsf A}$ and fluxes 
$(q_I{}^{\mathsf A},q^{I\mathsf A})$. This is just a re-labelling of indices and 
corresponds to the permutation group $\mathcal S_3$. 
Using now the T-duality action \eqref{dual_13} together with the 
permutation of K\"ahler moduli, we see that $\mathcal S_3$ is enhanced to 
$\mathcal S_4$ acting on $T_A=(T_0,T_{\mathsf A})$ and
fluxes $(q_I{}^{A},q^{IA})$.
Indeed, for $S_A{}^B\in \mathcal S_4$ the superpotential \eqref{superpot_002} is invariant under
\eq{
  \label{perm_dual}
  T_A \to S_A{}^B\op T_B \,, \hspace{50pt}
  \arraycolsep2pt
  \begin{array}{lcl}
  q_I{}^A &\to& \displaystyle q_I{}^B (S^{-1})_B{}^A \,,
  \\[6pt]
  q^{IA} &\to& \displaystyle q^{IB} (S^{-1})_B{}^A \,,
  \end{array}
}
and the flux contribution to the Bianchi identities shown in
\eqref{flux_020} transform as
\eq{\
  \arraycolsep3pt
  \mathbf Q  \to \mathbf S^{-T}  \mathbf Q \, \mathbf{S}^{-1} \,,
  \hspace{60pt}
  \mathbf S = \left( \begin{array}{cc} 1 & 0 \\ 0 & S_A{}^B \end{array}\right)\,.
}
We emphasize that four collective T-duality transformations 
for type II orientifold compactification are  permutations 
of moduli and fluxes. They do not correspond to transformations which invert $T_{\mathsf A}$.


\subsubsection*{S-duality}

We finally consider the $SL(2,\mathbb Z)$ duality  of type IIB string theory. 
For vanishing $Q$-flux its action on the axio-dilaton and the 
$F$- and $H$-flux takes the following form
\eq{\label{S-duality}
  \arraycolsep3pt
  \tau \to \frac{a \op\tau + b}{c\op\tau + d} \,,
  \hspace{30pt}
  \binom{F}{H} \to \left( \begin{array}{cc} a & b \\ c & d \end{array}\right) \binom{F}{H} \,,
  \hspace{30pt}
  \left( \begin{array}{cc} a & b \\ c & d \end{array}\right) \in SL(2,\mathbb Z)\,,
}
where $a,b,c,d\in\mathbb Z$. 
In particular, this transformation leaves the F-term equations \eqref{f_term_eq} invariant.
However, for non-vanishing $Q$-flux only part of this duality survives.
\begin{itemize}

\item For constant shifts of the axio-dilaton and the K\"ahler moduli $T_{\mathsf A}$, 
the K\"ahler potential \eqref{kahler_001} as well as the superpotential 
\eqref{superpot_002}
are invariant under 
\eq{
  \label{gen_s_dual_001}
  T_A \to T_A + b_A \,, \hspace{60pt}
  \arraycolsep2pt
  \renewcommand{\arraystretch}{1.2}
  \begin{array}{lclcl}
  f_I &\to& f_I &+& q_I{}^A b_A \,, \\
  f^I &\to& f^I &+& q^{IA} b_A \,, \\
  q_I{}^A &\to& \multicolumn{3}{l}{ q_I{}^A \,,}\\
  q^{IA} &\to & \multicolumn{3}{l}{q^{IA} \,,}
  \end{array}
}
where the parameter $b_A$ has to be quantized as $b_{A}\in\mathbb Z$.
This corresponds to a gauge transformation of the R-R zero- and four-form potentials.

\item For an $S$-transformation $\tau\to - 1/\tau$ in the presence of non-geometric
fluxes, the F-term equations \eqref{f_term_eq} are in general not invariant. 
To restore the $SL(2,\mathbb Z)$ duality the authors of
\cite{Aldazabal:2006up} introduced additional non-geometric $P$-fluxes as the 
counterpart of the $Q$-fluxes.
In this paper we do not consider such $P$-fluxes, but refer for instance 
to 
\cite{Aldazabal:2008zza,Guarino:2008ik,
Chatzistavrakidis:2013jqa,Sakatani:2014hba,Blumenhagen:2015kja,Bergshoeff:2015cba,Shukla:2015rua,Lombardo:2016swq}
for more details on this topic.

\end{itemize}


\clearpage
\section{Moduli stabilization I}
\label{sec_mod1}

As a first example for  moduli stabilization on 
the $\mathbb T^6/\mathbb Z_2\times \mathbb Z_2$ orientifold
we consider a specific choice of fluxes which stabilizes the 
axio-dilaton $\tau$, fixes
the complex-structure moduli $U^i$ to a symmetric point
but leaves the K\"ahler moduli $T_{\mathsf A}$ unstabilized. 
This setting has been studied for instance in 
\cite{Ashok:2003gk,Denef:2004ze,DeWolfe:2004ns},
and here we use it as a toy model for the more involved settings 
in the subsequent sections.


\subsection{Setting}

We start by specifying the superpotential \eqref{superpot_002}. We consider a  configuration with vanishing non-geometric fluxes \eqref{flux_002},
\eq{
q_I{}^{\mathsf A} = 0 \,, \hspace{40pt} q^{I\mathsf A} = 0 \,,
}
which implies that the K\"ahler moduli $T_{\mathsf A}$ do not appear in the potential and hence are not stabilized. 
The remaining R-R and NS-NS fluxes \eqref{flux_001} are chosen as follows
\eq{\label{conditions_rigidCalabiYau}
  \arraycolsep2pt
  \renewcommand{\arraystretch}{1.3}
  \begin{array}{lcl@{\hspace{50pt}}lclclcl@{\hspace{50pt}}l}
  f^0 &=& 3 \op\tilde f^0 \,,
  &
  f_1 &=& f_2 &=& f_3 &=& -\tilde f^0 \,,
  &\tilde f^0\in8\mathbb Z \,,
  \\
  f_0 &=& 3\op \tilde f_0 \,,
  &
  f^1 &=&f^2 &=& f^3 &=&+ \tilde f_0 \,,
  &
  \tilde f_0\in8\mathbb Z \,,
  \\
   h^0 &=& 3 \op \tilde h^0 \,,
  &
  h_1 &=&h_2 &=& h_3 &=& -\tilde h^0 \,,
  &
  \tilde h^0\in8\mathbb Z\,,
  \\
  h_0 &=& 3 \op\tilde h_0 \,,
  &
  h^1 &=& h^2 &=& h^3 &=&+\tilde h_0 \,,
  &
  \tilde h_0\in8\mathbb Z \,,
  \end{array}
}
where $\tilde h^0$ and $\tilde h_0$ should not be zero simultaneously. 
Since the superpotential $W$ is  independent of the K\"ahler moduli
the F-term equations \eqref{f_term_eq}  take the 
 simple form
\eq{\label{F-Term_Equations_DouglasPlot}
  \arraycolsep2pt
  \begin{array}{lclcrcr@{\hspace{3pt}}l}
 0 &=&  F_{T_{\mathsf A}}  &=& & & \partial_{T_{\mathsf A}} \mathcal K & W \,, \\[4pt] 
 0 &=&   F_{U^{i}}  &=& \partial_{U^{i}} W &+& \partial_{U^{i}} \mathcal K & W  \,,\\[4pt]
 0 &=&  F_{\tau} &=& \partial_\tau W &+& \partial_{\tau}  \mathcal K & W \,, \\
\end{array}
\hspace{40pt} \Rightarrow \hspace{40pt}
\arraycolsep2pt
\begin{array}{lcl}
0 & = & W \,, \\[4pt]
0 &=& \partial_{U^i} W  \,,\\[4pt]
0 &=& \partial_\tau W \,.
\end{array}
}
Ignoring unphysical values for $U^i$ and $\tau$ with negative or vanishing imaginary part, 
we obtain the following solution to \eqref{F-Term_Equations_DouglasPlot} 
\eq{
  \label{sol_axio_001}
  U^{1}=U^{2}=U^{3}= i  \,, 
  \hspace{40pt}
  \tau = \frac{\tilde{f}_{0}-i\tilde{f}^ {0}}{\tilde{h}_{0}-i\tilde{h}^ {0}}\,.
}
The fluxes in \eqref{conditions_rigidCalabiYau} are not arbitrary but 
are subject to the Bianchi identities \eqref{flux_020}.
Since all non-geometric  $Q$-fluxes vanish, the only nontrivial condition  is the D3-tadpole 
contribution
\eq{\label{tadpole_rigid}
  \mathsf Q^{0} = 12\Bigl(\tilde{f}_{0}\tilde{h}^{0} -\tilde{f}^{0}\tilde{h}_{0}\Bigr) >0 \,,
}
where the requirement of $\mathsf Q^0$ being positive is related to having $\mbox{Im}\op \tau >0$. 
Note that due to the quantization condition for the fluxes in \eqref{conditions_rigidCalabiYau}, 
$\mathsf Q^0$ is a multiple of $768$. 


\subsection{Finite number of solutions for fixed $\mathsf Q^{0}$}
\label{sec_t_mod_finite}

Next, we briefly review the arguments of \cite{Denef:2004ze,DeWolfe:2004ns}
showing that the number of physically-distinct solutions \eqref{sol_axio_001}
is finite for finite $\mathsf Q^0$. 
We restrict the values of the axio-dilaton $\tau$ to the fundamental domain
of the corresponding $SL(2,\mathbb Z)$ duality \eqref{S-duality}
\eq{\label{FundamentalDomain_AxioDilaton}
\mathcal F_{\tau}=\left\{ 
-\frac{1}{2}\leq\tau_{1}\leq0,\, |\tau|^2 \geq 1 
\; \cup \;
0<\tau_{1}< +\frac12, \,|\tau|^2 > 1 
\right\}.
}
For ease of notation we express the axio-dilaton in terms of its real and imaginary part as
$\tau = \tau_{1}+ i\tau_{2}$,
and for the solution \eqref{sol_axio_001} to the F-term equations we have
\eq{\label{DouglasPlot_tau_solution}
  \tau = \frac{\tilde{h}_{0}\op \tilde{f}_{0} + \tilde{h}^{0}\op \tilde{f}^{0}}{(\tilde{h}_{0})^2+ (\tilde{h}^{0})^2} + \frac{i}{12}\, \frac{\mathsf Q^{0}}{(\tilde{h}_{0})^2 + (\tilde{h}^{0})^2} \,.
}
For a fixed positive value of $\mathsf Q^0$ the imaginary part of 
$\tau$ is bounded from above as $\tau_2 \leq \frac{\mathsf Q^0}{768}$, since $\tilde{h}_{0}$ and $\tilde{h}^{0}$ are integer multiples of eight 
which 
cannot be zero simultaneously.
We now argue along the following lines:
\begin{itemize}

\item The tadpole contribution $\mathsf Q^0$ is invariant under the $SL(2,\mathbb Z)$ transformations
\eqref{S-duality}. Using then a $T$-transformations acting on the axio-dilaton as $\tau \to \tau + b$ with
$b\in\mathbb Z$, we can bring $\tau_1$ into the region $-\frac12 \leq \tau_1<+\frac12$. 
This $T$-transformation is a duality transformation, and therefore we have the equivalence
\eq{
  \tilde{f}_{0} \sim \tilde{f}_{0}+ b\, \tilde{h}_{0} \,,\hspace{50pt}
  \tilde{f}^{0} \sim \tilde{f}^{0} + b\, \tilde{h}^{0}\,.
}
Choosing without loss of generality $\tilde{h}_{0}\neq0$, 
for fixed $\tilde h_0$ there are only finitely-many  inequivalent values for $\tilde{f}_{0}$ given by
\eq{
  \label{fin_92}
  \tilde{f}_{0} = 0, \ldots, \tilde{h}_{0}-1 \,.
}

\item Next, using an $S$-transformation $\tau\to -1/\tau$ (possibly together with additional $T$-transformations)
we can bring $\tau$ into the fundamental domain $\mathcal F_{\tau}$. 
In $\mathcal F_{\tau}$ a lower bound for the imaginary part $\tau_2$ is obtained by considering 
$\tau_1=-\frac12$ for which 
$\tau_2 \geq \sqrt{3}/2$.
Using \eqref{DouglasPlot_tau_solution} we then find
\eq{
  0< (\tilde{h}_{0})^2 + (\tilde{h}^{0})^2 \leq \frac{\mathsf Q^{0}}{6\sqrt{3}} \,,
}
which leaves only finitely many possibilities for the integers $\tilde{h}_{0},\tilde{h}^{0}$.
Together with \eqref{fin_92}, this implies also a finite number of choices for $\tilde f_0$.

\item The remaining flux $\tilde{f}^{0}$ is now determined via the tadpole contribution $\mathsf Q^{0}$ 
shown in \eqref{tadpole_rigid}.

\end{itemize}
In summary, for a fixed positive value of the D3-tadpole contribution $\mathsf Q^{0}$,
the F-term equations \eqref{F-Term_Equations_DouglasPlot} have only a finite number of 
of physically-distinct solutions  for $\tau$.


\subsection{Space of solutions}

In \cite{Denef:2004ze,DeWolfe:2004ns} it was shown that 
the solutions \eqref{sol_axio_001} mapped to the fundamental domain for $\tau$
are not distributed homogeneously. In particular, the space of solutions  contains voids
with large degeneracies in their centers. 
In this section we review these findings and provide some new
results on the dependence of these distributions on the D3-tadpole contribution $\mathsf Q^0$. 
Our data has been obtained using a computer algorithm to generate all phy\-si\-cally-distinct
flux vacua for a given upper bound on the D3-brane tadpole contribution.


\subsubsection*{Distribution of solutions}

As we argued above, for a fixed value of  $\mathsf Q^0$, the number 
of physically inequivalent solutions for the axio-dilaton $\tau$ is finite. 
Using the $SL(2,\mathbb Z)$ duality \eqref{S-duality} we can 
map these solutions to the fundamental domain \eqref{FundamentalDomain_AxioDilaton},
and we have shown the corresponding space of solutions in figures~\ref{DouglasPlot_Plane_Global}
and \ref{DouglasPlot_Plane}.
\begin{itemize}

\item For figure~\ref{DouglasPlot_Plane_Global} we have included all flux configurations
for which the tadpole contribution satisfies $0< \frac{\mathsf Q^0}{768}\leq 300$, and  
in order to have a symmetric plot we have added points on the boundary 
of the fundamental domain at $\tau_1 = +\frac12$.
We see that the space of solutions for \eqref{sol_axio_001} is bounded as 
$\tau_2 \leq 300$, and that solutions are located on lines with fixed $\tau_1$.

\item In figure~\ref{DouglasPlot_Plane} we show a zoom of figure~\ref{DouglasPlot_Plane_Global}
for a small range of $\tau_2$. Here we see a characteristic structure of voids 
\cite{Denef:2004ze} with accumulated points in their centers. 
The large voids are encircled by smaller ones, and we note that 
the higher density of points near $|\tau|^2 =1$ is not a physical property 
as we have not taken into account the metric on moduli space.

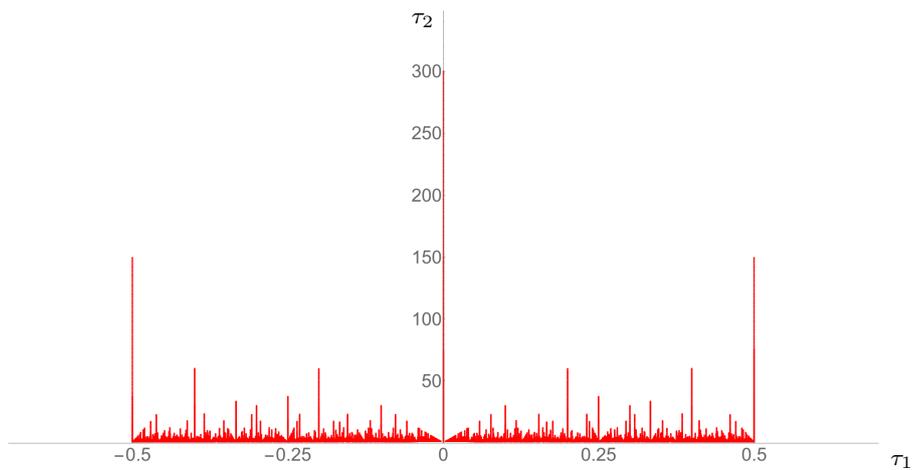
\begin{figure}[p]
\centering
\resizebox{0.8\textwidth}{!}{%
\includegraphics[width=300pt]{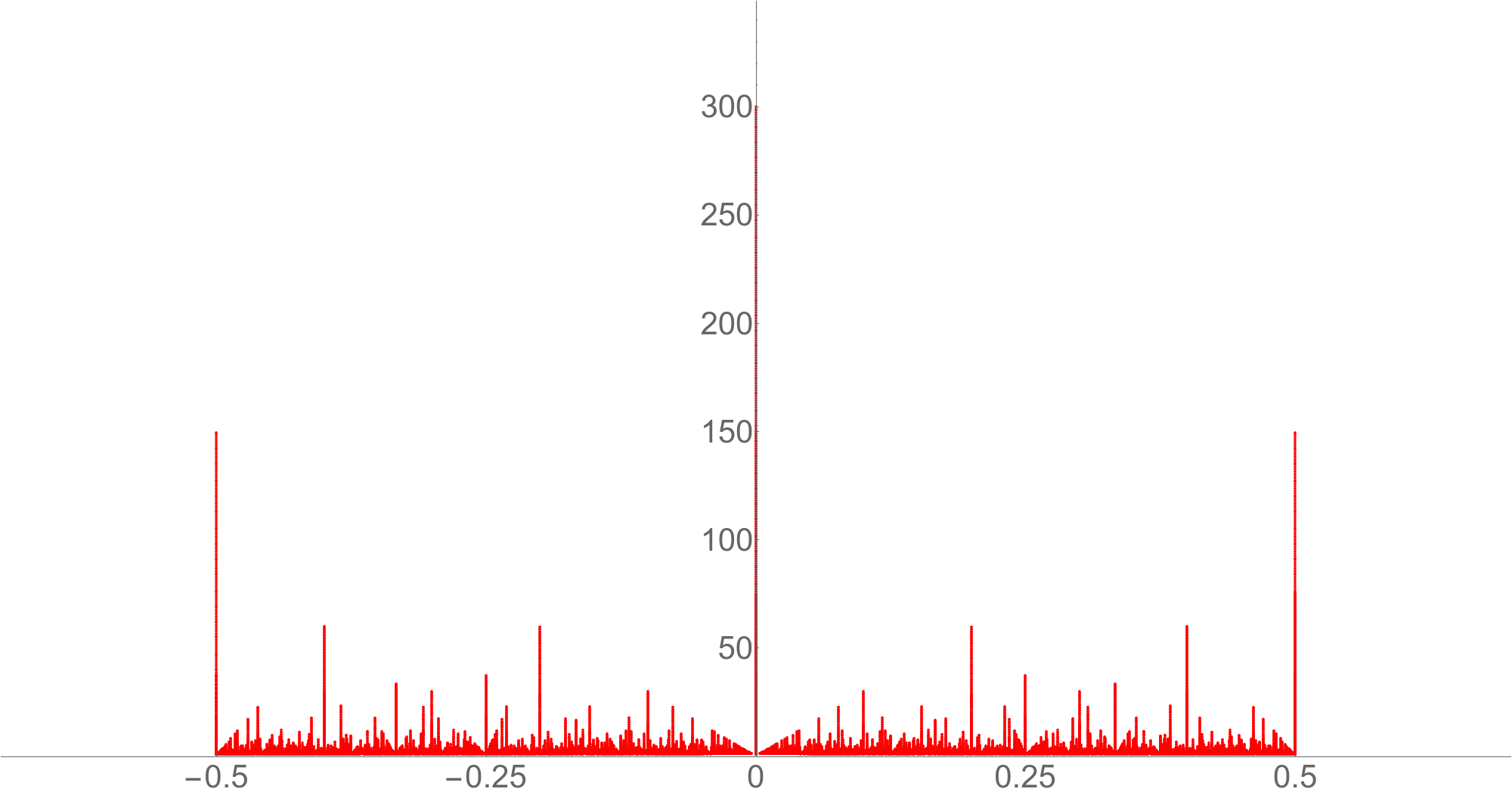}
\begin{picture}(0,0)
\put(0,0){\scriptsize $\tau_1$}
\put(-165,153){\scriptsize $\tau_2$}
\end{picture}
}
\caption{Space of solutions for the axio-dilaton $\tau$
with fluxes  \eqref{conditions_rigidCalabiYau}, mapped to the fundamental domain $\mathcal F_{\tau}$. 
All solutions satisfy the bound 
\raisebox{0pt}[0pt][0pt]{$\frac{\mathsf Q^{0}}{768}  \leq \frac{\mathsf Q_{\rm max}^{0}}{768}=300$}.} 
\label{DouglasPlot_Plane_Global}
\end{figure}

\begin{figure}[p]
\centering
\vskip1em
\resizebox{0.4\textwidth}{!}{%
\includegraphics[width=150pt]{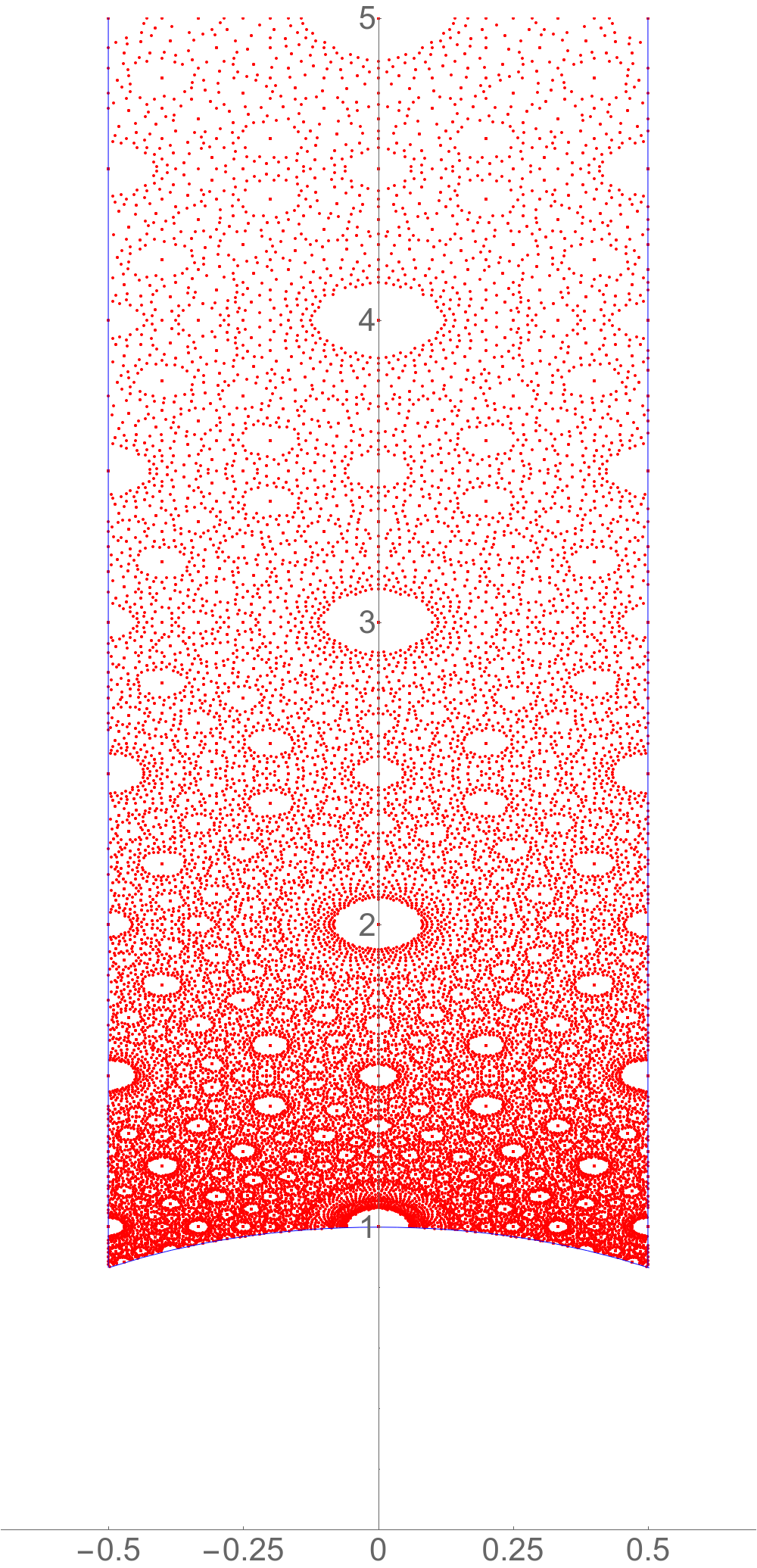}
\begin{picture}(0,0)
\put(0,0){\scriptsize $\tau_1$}
\put(-91,312){\scriptsize $\tau_2$}
\end{picture}
}
\caption{Zoom of figure~\ref{DouglasPlot_Plane_Global}
for a $0\leq \tau_2\leq 5$.} 
\label{DouglasPlot_Plane}
\end{figure}
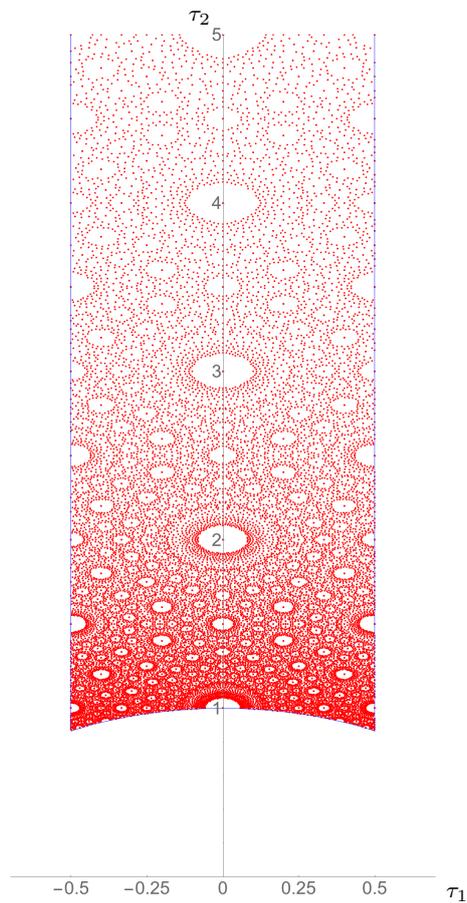

\end{itemize}
Let us next note that the moduli space of the axio-dilaton $\tau$ is hyperbolic. Indeed from the 
K\"ahler potential \eqref{kahler_001} we can derive the corresponding K\"ahler metric with components
\eq{\label{mod_met_008}
\mathcal G_{\tau_{1}\tau_{1}}= \mathcal G_{\tau_{2}\tau_{2}}=\frac{1}{4{\tau_{2}}^{2}}\,,\hspace{50pt}
\mathcal G_{\tau_{1}\tau_{2}}=0\,.
}
A convenient way to visualize this hyperbolic space is by mapping  the  Poincar\'e half-plane to the
Poincar\'e disk via the conformal transformation
\eq{\label{conf_map_001}
\left(\tau_{1},\tau_{2}\right) \to 
\left(\tilde \tau_{1},\tilde \tau_{2}\right) = \left(\frac{2\op\tau_{1}}{\tau_{1}^{2}+\left(1+\tau_{2}\right)^{2}}\,,\,
\frac{\tau_{1}^{2}+\tau_{2}^{2}-1}{\tau_{1}^{2}+\left(1+\tau_{2}\right)^{2}}\right).
}
The space of solutions for the axio-dilaton mapped to the Poincar\'e disk is then shown in figure~\ref{DouglasPlot_Disk},
which is  the mapping of figure~\ref{DouglasPlot_Plane_Global} under \eqref{conf_map_001}.
\begin{itemize}

\item In figure~\ref{DouglasPlot_Disk} the characteristic structure of voids is visible.
In this plot effects of the moduli-space metric are incorporated.

\begin{figure}[p]
\centering
\resizebox{0.8\textwidth}{!}{%
\includegraphics[width=300pt]{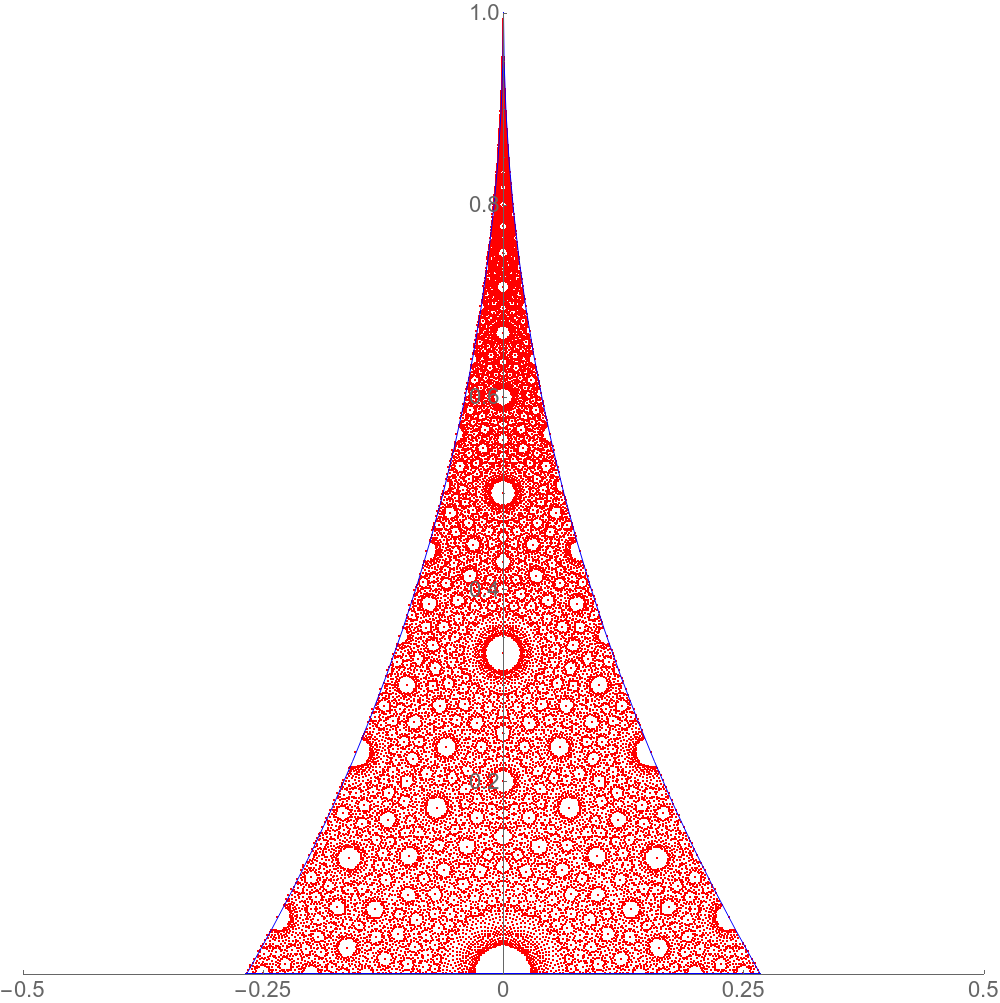}
\begin{picture}(0,0)
\put(0,0){\scriptsize $\tilde\tau_1$}
\put(-174,295){\scriptsize $\tilde\tau_2$}
\end{picture}
}
\caption{
Space of solutions for the axio-dilaton $\tau$
with fluxes of the form \eqref{conditions_rigidCalabiYau}, restricted to the fundamental domain and
mapped to the Poincar\'e disk.
All solutions satisfy the bound \raisebox{0pt}[0pt][0pt]{$\frac{\mathsf Q^{0}}{768}  \leq 300$}.} 
\label{DouglasPlot_Disk}
\end{figure}
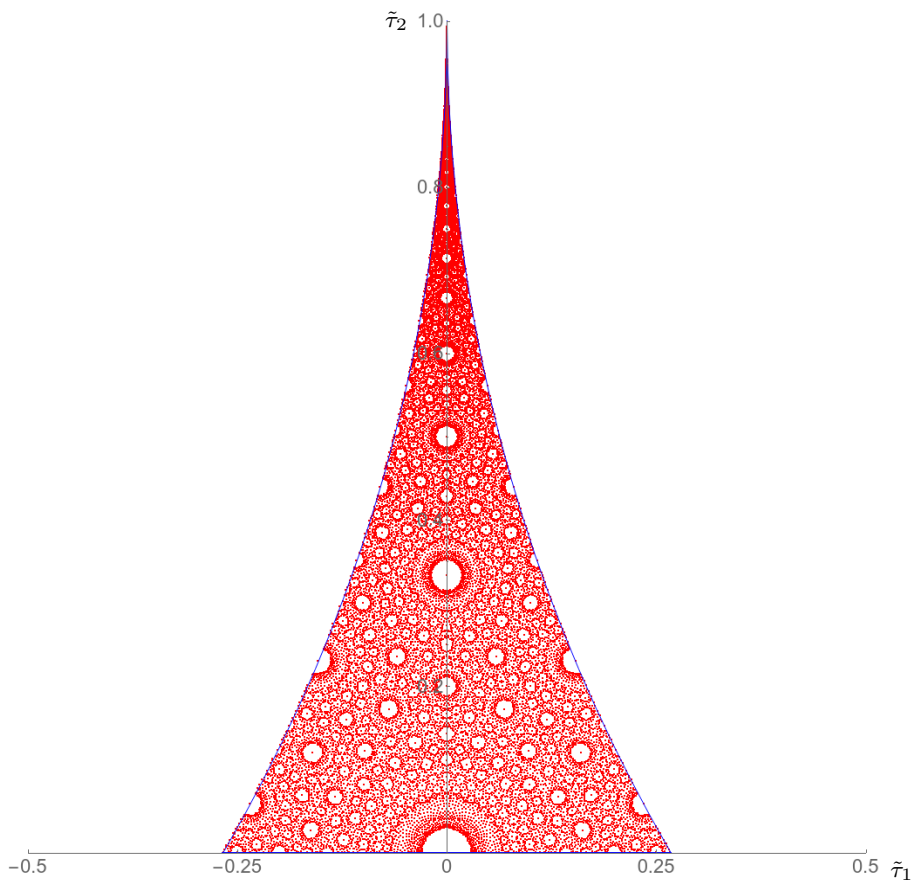
\begin{figure}[p]
\centering
\vspace*{10pt}
\includegraphics[width=300pt]{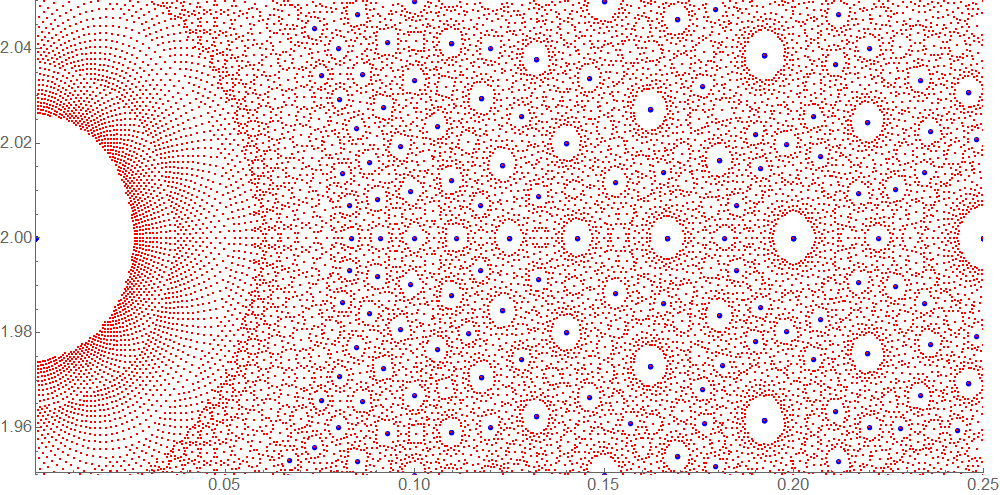}
\begin{picture}(0,0)
\put(2,0){\scriptsize $\tau_1$}
\put(-310,145){\scriptsize $\tau_2$}
\end{picture}
\caption{
Space of solutions for the axio-dilaton $\tau$ with fluxes of the form \eqref{conditions_rigidCalabiYau}
near  $\tau=2i$ on the Poincar\'e plane for
\raisebox{0pt}[0pt][0pt]{$\frac{\mathsf Q^{0}}{768}  \leq 300$}  (blue) and
\raisebox{0pt}[0pt][0pt]{$\frac{\mathsf Q^{0}}{768}  \leq 3000$} (red).} 
\label{DouglasPlot_Q3000_Zoom}
\end{figure}
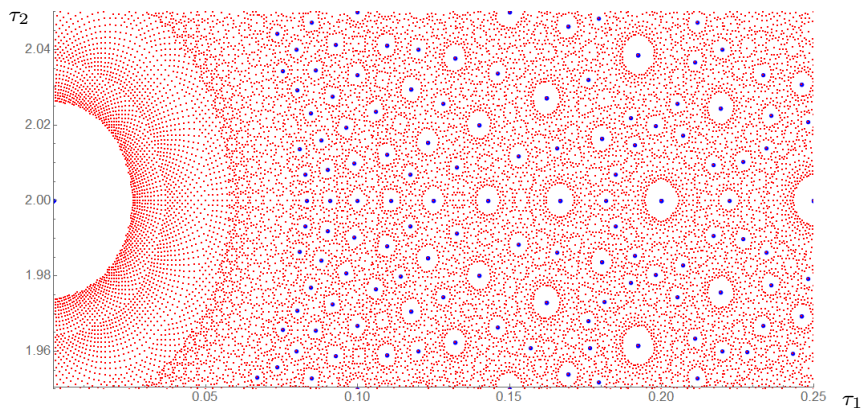

\end{itemize}


\subsubsection*{Analysis of voids}

The number of physically-distinct solutions for the axio-dilaton
is finite for fixed tadpole-contribution $\mathsf Q^0$. 
The number of solutions $\mathsf N$ with $\mathsf Q^0\leq \mathsf Q^0_{\rm max}$  can 
be determined numerically, which leads to the 
following scaling behaviour 
\eq{
  \label{res_00033}
  \mathsf N \approx 0.823 
  \left[\frac{\mathsf Q_{\rm max}^0}{768}\right]^{2}
}
for large $\mathsf Q_{\rm max}^0$.
We now want to study how the voids change depending on $\mathsf N$, 
or, equivalently, depending on $\mathsf Q^0_{\rm max}$. 
In particular, we are interested how the size of the voids depends on $\mathsf Q^0_{\rm max}$. 
Qualitatively, this behaviour is illustrated in figure~\ref{DouglasPlot_Q3000_Zoom}:
\begin{itemize}

\item In figure~\ref{DouglasPlot_Q3000_Zoom} the space of solutions 
for the axio-dilaton around the point $\tau = 2i$ is shown. 
The blue points correspond to solutions which satisfy 
\raisebox{0pt}[0pt][0pt]{$\frac{\mathsf Q^{0}}{768}  \leq 300$}, and the red points  correspond
to solutions with 
\raisebox{0pt}[0pt][0pt]{$\frac{\mathsf Q^{0}}{768}  \leq 3000$}.
For larger $\mathsf Q^0_{\rm max}$ the void around $\tau = 2i$ therefore becomes 
smaller, and finer void structures appear.
These results are in agreement with the topological data analysis in \cite{Cole:2018emh}.

\end{itemize}
Let us denote the origin of a void by $\tau_{\rm void}$, and define 
its size by the distance to the nearest solution $\tau_{\rm sol}$ (not located at $\tau_{\rm void}$). 
The geodesic distance $d$ is measured using the metric \eqref{mod_met_008} 
on the axio-dilaton moduli space, for which we have
\eq{\label{geodesic_distance_DouglasPlot}
d\left(\tau\,,\widetilde{\tau}\right)=\frac{1}{2}\op \textrm{arccosh}\left[1+\frac{\left(\widetilde{\tau_{1}}-\tau_{1}\right)^{2}
+\left(\widetilde{\tau_{2}}-\tau_{2}\right)^{2}}{2\op\widetilde{\tau_{2}}\op\tau_{2}}\right].
}
As we can see for instance from figure~\ref{DouglasPlot_Disk}, in the 
proper distance the voids can be approximated by a circle whose 
radius we define as
\eq{
  R_{\rm void} =  \min_{\tau_{\rm sol}\neq \tau_{\rm void}} d(\tau_{\rm void},\tau_{\rm sol}) \,.
}
The scaling behaviour of $R_{\rm void}$ with $\mathsf Q^0_{\rm max}$ has been obtained 
for instance in \cite{Denef:2004ze,DeWolfe:2004ns} 
as $R^{2}_{\rm void}\sim1/\mathsf Q^{0}_{\rm max}$, and below we have 
determined the prefactors for some families of voids numerically.
For voids located in the fundamental domain on 
the Poincar\'e plane we have the following relation between 
the radius of the void $R_{\rm void}$, the tadpole contribution $\mathsf Q^0_{\rm max}$ 
and the number of solutions located at the center of the void $n_{\rm void}$
\eq{
\label{DouglasPlot_RadiusExplicit}
\arraycolsep2pt
\begin{array}{lcl}
R_{\rm void}^{2} &\approx&\displaystyle  \frac{1}{C\op\tau_{2{\rm void}} } \left[\frac{768}{\mathsf Q_{\rm max}^{0}}\right]
\\[18pt]
n_{\rm void} &\approx& \displaystyle \frac{2\pi}{C\op\tau_{2{\rm void}}} \left[{\frac{\mathsf Q_{\rm max}^{0}}{768}}\right]
\end{array}
\hspace{60pt}
\arraycolsep15pt
\begin{array}{c|c||c}
\tau_{1{\rm void}} & \tau_{2{\rm void}} & C
\\
\hline\hline
0 & n & 4 \\
0 & n+0.5 & 16 \\
\pm0.2 & n+0.4 & 20 \\
\pm0.2 & n+0.6 & 20
\end{array}
}
where $n\in\mathbb Z_+$. The constant $C$ depends on the family of voids under consideration
and can be read-off from the table in \eqref{DouglasPlot_RadiusExplicit} for several examples. 
Note also that the number of solutions located at the center of the void 
divided by the area of the void takes the simple form 
\eq{
  \frac{n_{\rm void}}{2\pi R_{\rm void}^2}  \approx \left[  \frac{\mathsf Q_{\rm max}^{0}}{768} \right]^2 \,.
}


\subsubsection*{Solutions at small coupling}

The imaginary part of the axio-dilaton $\tau$
is bounded from above by the D3-tadpole contribution $\mathsf Q^0$,
which via \eqref{def_mod_003} implies a restriction on the string coupling $g_{\rm s}$ as
\eq{
  \tau_2 = e^{-\phi} = \frac{1}{g_{\rm s}} \leq \frac{\mathsf Q^0}{768} 
  \hspace{40pt}\Rightarrow \hspace{40pt}
  g_{\rm s} \geq \frac{768}{\mathsf Q^0} \,.
}
Recall that in our conventions $\mathsf Q^0$ is a multiple of $768$.
In the following we determine the number of physically-distinct solutions $\mathsf N_{\mathsf c}$
which satisfy $\tau_2\geq \mathsf c$ for some cutoff $\mathsf c >0$ so that we have
\eq{
 \frac{768}{\mathsf Q^0_{\rm max}} \leq g_{\rm s} \leq \frac{1}{\mathsf c} \,.
}
Note that in order to ignore string-loop corrections and corrections from world-sheet instantons, 
we need to stabilize the axio-dilaton at small $g_{\rm s}$. This implies that 
$\mathsf Q^0$ and $\mathsf c$ should be sufficiently large. 
Using then the exact data for the space of solutions, we can obtain fits for $\mathsf N_{\mathsf c}$ 
for values of  $\mathsf Q^0_{\rm max}$ of the 
order \raisebox{0pt}[0pt][0pt]{$\frac{\mathsf Q_{\rm max}^0}{768}= \mathcal O(10^3)$}.
In particular, with the scaling of the total number of solutions $\mathsf N$ shown in \eqref{res_00033} 
we have
\eq{
 \frac{\mathsf Q^0_{\rm max}}{768} \gg 1
 \hspace{60pt}
 \renewcommand{\arraystretch}{1.3}
 \arraycolsep10pt
 \begin{array}{c||cc}
 \mathsf c & \mathsf N_{\mathsf c} & \mathsf N_{\mathsf c}/\mathsf N 
 \\
 \hline\hline
 2 & 0.393 \bigl[\frac{\mathsf Q_{\rm max}^0}{768}\bigr]^{2} & 0.478
 \\
 5 & 0.157 \bigl[\frac{\mathsf Q_{\rm max}^0}{768}\bigr]^{2} & 0.191
 \\
 10 & 0.078 \bigl[\frac{\mathsf Q_{\rm max}^0}{768}\bigr]^{2} & 0.095
 \\
 20 & 0.039 \bigl[\frac{\mathsf Q_{\rm max}^0}{768}\bigr]^{2} & 0.047
 \end{array}
}
We observe that in this limit the percentage of solutions with $g_{\rm s}\ll 1$ 
is small and independent of $\mathsf Q^0_{\rm max}$. For instance, only about 5\%
of the solutions  have a string coupling satisfying $g_{\rm s} \leq 0.05$.
\begin{figure}[t!]
\centering
\includegraphics[width=200pt]{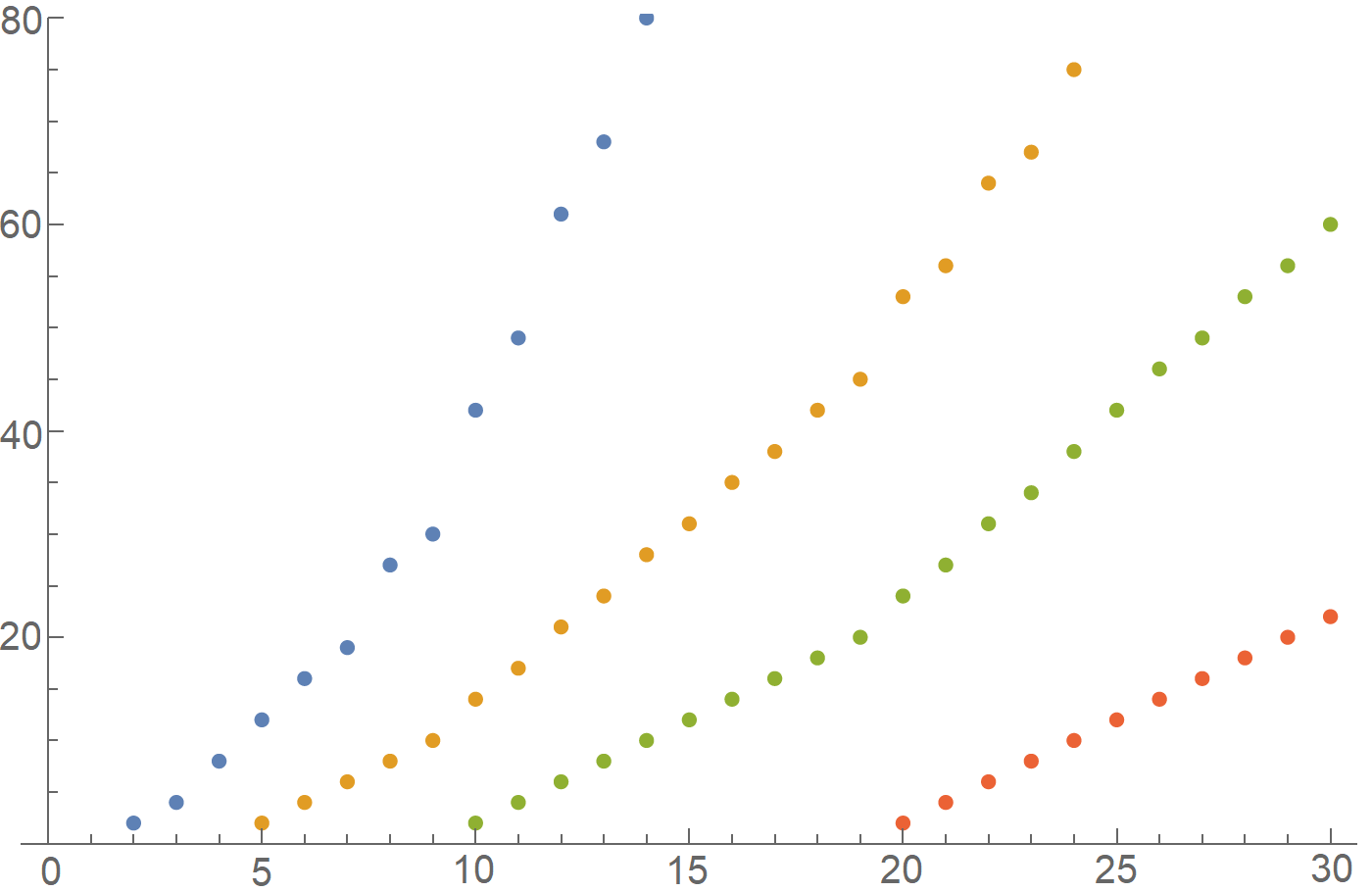}
\begin{picture}(0,0)
\put(0,4){\scriptsize $\frac{\mathsf Q_{\rm max}^0}{768}$}
\put(-216,125){\scriptsize $\mathsf N_{\mathsf c}$}
\end{picture}
\caption{
Number of solutions $\mathsf N_{\mathsf c}$ 
(for the axio-dilaton $\tau$ with fluxes of the form \eqref{conditions_rigidCalabiYau})
which satisfy
\raisebox{0pt}[0pt][0pt]{$\mathsf c \leq \tau_2 \leq \frac{\mathsf Q_{\rm max}^0}{768}$} for 
$\mathsf c = 2,5,10,20$ in colors blue, orange, green and red, respectively.} 
\label{plot04}
\end{figure}
However, the region of small tadpole contributions $\frac{\mathsf Q^0}{768}= \mathcal O(1)$ is 
more interesting. Here the number of solutions $\mathsf N_{\mathsf c}$ does not 
follow a simple quadratic behavior, and the precise numbers are
shown in figure~\ref{plot04}.
We see that for a particular $\mathsf c$ in $g_{\rm s}\leq 1/\mathsf c$, 
the D3-tadpole contributions 
\raisebox{0pt}[0pt][0pt]{$\frac{\mathsf Q^0}{768}$}
has to be larger than some threshold. 
Furthermore, above this threshold 
the number of solutions is not large but only $\mathcal O(10)$.


\subsection{Summary}

Let us summarize the results and observations of this section for 
moduli stabilization of the axio-dilaton with the choice of fluxes given in equation \eqref{sol_axio_001}:
\begin{itemize}

\item As already known before,  for a fixed D3-brane tadpole contribution 
$\mathsf Q^0$, the number of physically-distinct solutions to the F-term equations for the axio-dilaton $\tau$ 
is finite due to the corresponding $SL(2,\mathbb Z)$ duality \cite{Denef:2004ze,DeWolfe:2004ns}.

\item The solutions for the axio-dilaton in the fundamental domain are not 
distributed homogeneously, but show characteristic void structures as 
illustrated in figures~\ref{DouglasPlot_Plane} and \ref{DouglasPlot_Disk}.

\item With increasing upper bound $\mathsf Q^0_{\rm max}$ on the tadpole contribution, the area of 
these voids shrinks and the number of solutions located at the center $n_{\rm void}$ increases
as shown in \eqref{DouglasPlot_RadiusExplicit}.
The precise behaviour for the radius $R_{\rm void}$ and $n_{\rm void}$ 
is proportional to a constant depending on the location 
of the void, however, the ratio
$n_{\rm void}/2\pi R_{\rm void}^2$ is universal.

\item The string coupling is bounded from below by the tadpole contribution as 
\raisebox{0pt}[0pt][0pt]{$\frac{768}{\mathsf Q^0}\leq g_{\rm s}$}.
In order to ignore string corrections and trust 
the solutions  \eqref{sol_axio_001}, we have to demand 
$g_{\rm s}\ll 1$ which implies 
\raisebox{0pt}[0pt][0pt]{$\frac{\mathsf Q^0}{768}\gg1$}.
This is in contrast to our discussion of the tadpole-cancellation condition on page~\pageref{page_ma} which 
requires \raisebox{0pt}[0pt][0pt]{$\frac{\mathsf Q^0}{768}$} to be small, 
and illustrates the difficulty of obtaining reliable solutions to the F-term equations.

\item We have furthermore analyzed the number of physically-distinct solutions
satisfying $g_{\rm s}\leq 1/\mathsf c$.
Requiring a small string coupling of for instance $g_{\rm s}\leq 1/10$, we find that only about $10\%$ of the 
solutions satisfy this condition. 
If we require $g_{\rm s}$ to be smaller, then the corresponding fraction of solutions is smaller.

\end{itemize}


\clearpage
\section{Moduli stabilization II}
\label{sec_iso}

In this section we extend our previous discussion by including the complex-struc\-ture moduli $U^i$.
We choose flux configurations which stabilize the axio-dilaton and fix the complex-structure moduli at an isotropic minimum with $U^1=U^2=U^3$. 
Such vacua have previously been studied for instance in \cite{Kachru:2002he,DeWolfe:2004ns}.


\subsection{Setting}
\label{sec_tu_set}

We start again by specifying the superpotential \eqref{superpot_002} and set to zero the non-geometric fluxes 
\eqref{flux_002}
\eq{\label{IsotropyConditions2}
q_I{}^{\mathsf A} = 0 \,, \hspace{40pt} q^{I\mathsf A} = 0 \,,
}
and for the R-R and remaining NS-NS fluxes \eqref{flux_001} we choose the following restricted setting
\eq{\label{IsotropyConditions}
  \arraycolsep2pt
  \begin{array}{lclcl@{\hspace{40pt}}lclcl}
  f_1 &=& f_2 &=& f_3\,,& h_1 &=&h_2 &=& h_3\,, 
  \\[6pt]
  f^1 &=&f^2 &=& f^3\,,& h^1 &=& h^2 &=& h^3\,,
  \end{array}
  \hspace{40pt}
  f^I, f_I, h^I, h_I \in 8\mathbb Z\,.
} 
Since the superpotential is independent of
the K\"ahler moduli $T_{\mathsf{A}}$, 
the F-term equations \eqref{f_term_eq} 
simplify as in  \eqref{F-Term_Equations_DouglasPlot} and we obtain
\eq{\label{F-Term_Equations_IsotropicTorus}
\arraycolsep2pt
0  =  W \,, \hspace{40pt}
0 = \partial_{U^i} W  \,,  \hspace{40pt}
0 = \partial_\tau W \,.
} 
Due to the isotropic choice of fluxes in \eqref{IsotropyConditions}, the complex-structure moduli
$U^i$ are stabilized such that 
\eq{
U^{1}=U^{2}=U^{3}=:U\,,
}
and the F-term equations \eqref{F-Term_Equations_IsotropicTorus} reduce to 
\begin{align}
\label{IsotropicTorus_EoMIsotropic1-2}
\arraycolsep2pt
\begin{array}{lcl@{\op}lcl@{\op}lcl@{\op}lcl@{}}
- f_{0}&-&3\op U&f_{1}&-&3\op(U)^{2}&f^{1}&+&(U)^{3}&f^{0} & =&0 \,,
\\[4pt]
-h_{0}&-&3\op U&h_{1}&-&3\op(U)^{2}&h^{1}&+&(U)^{3}&h^{0} & =&0 \,,
\end{array}&
\\[4pt]
\label{IsotropicTorus_EoMIsotropic2-2}
(f_{1}-\tau h_{1})+2\op U(f^{1}-\tau h^{1})-(U)^{2}(f^{0}-\tau h^{0})=0\,.&
\end{align}
The R-R and NS-NS fluxes in  \eqref{IsotropyConditions}  are furthermore 
subject to the Bianchi identities \eqref{flux_020}, and due to the vanishing $Q$-fluxes the only nontrivial
relation is again given by the D$3$-brane tadpole contribution
\eq{\label{IsotropicTorus_Tadpole2}
\mathsf Q^{0} = f_{0}\op h^{0}-f^{0}\op h_{0}+3\left(f_{1}\op h^{1}-f^{1}\op h_{1}\right)>0\,.
}
Note that due to the quantization condition for the fluxes, the tadpole contribution $\mathsf Q^0$ is
an integer multiple of $64$. 
However, as it has been explained in footnote 10 of \cite{Kachru:2002he}, 
in order to obtain physically-viable solutions  $\mathsf Q^0$ 
receives an additional factor of three. The tadpole contribution is therefore 
always a multiple of $192$, which is also what we see explicitly in our data.


\subsection{Finite number of solutions for fixed $\mathsf Q^{0}$}

The two equations for the complex-structure modulus shown in \eqref{IsotropicTorus_EoMIsotropic1-2} 
define an overdetermined cubic system for $U$, which in general does not 
allow for a  solution in closed form.
Since the coefficients in \eqref{IsotropicTorus_EoMIsotropic1-2} are real, one can bring these
equations into the form
\eq{
   (U- u_0)\op (U-u_1)\op (U-\bar u_1) =0 \,,
  \hspace{40pt}
  u_0\in \mathbb R\,, \; u_1\in\mathbb C \,,
}
where $u_0,u_1,\bar u_1$ denote the solutions. Physically-acceptable solutions 
have to satisfy $\mbox{Im}\op U>0$, and therefore the F-term equations
\eqref{IsotropicTorus_EoMIsotropic1-2} have at most one solution for $U$ 
of interest to us.
The equation \eqref{IsotropicTorus_EoMIsotropic2-2} can be 
solved for the axio-dilaton as
\eq{
  \label{rel_0039}
  \tau = \frac{f_1 + 2\op U  f^1 - (U)^2 \op f^0}{h_1 + 2\op U  h^1 - (U)^2 \op h^0} \,,
}
which however depends on $U$. 
More details on these solutions can be found in appendix~\ref{app_proof_tu}, where 
we follow the discussion of \cite{Kachru:2002he,DeWolfe:2004ns}.
As reviewed in section~\ref{sec_dualities},
in the absence of non-geometric $Q$-fluxes the axio-dilaton and the complex-structure 
moduli enjoy $SL(2,\mathbb Z)$ dualities. These can be used 
to bring $\tau$ and $U$ into their fundamental domains
\eq{\label{FundamentalDomain_IsotropicTorus}
\arraycolsep2pt
\begin{array}{lclclrlclllrcl}
\mathcal{F}_{\tau}
&=&
\displaystyle\biggl\{ -\frac{1}{2}\leq 
&\tau_{1}&
\leq0,\,& |\tau|^2& \geq 1 \;
& \cup &\;
0<&\tau_{1}&\displaystyle < +\frac12, \, & |\tau|^2 &>& 1 
\biggr\}\,, 
\\[14pt]
\mathcal{F}_{U}
&=&
\displaystyle\biggl\{ -\frac{1}{2}\leq 
&U_{1}&
\leq0,\,& |U|^2& \geq 1 
& \cup &\;
0<&U_{1}&\displaystyle < +\frac12, \, & |U|^2 &>& 1 
\biggr\}\,,
\end{array}
}
where we again split $\tau$ and $U$ into their real and imaginary parts
as 
$\tau = \tau_{1}+ i\op\tau_{2}$ and $U= U_{1}+ i\op U_{2}$.
We furthermore note that the two $SL(2,\mathbb Z)$ dualities 
leave the D3-tadpole contribution $\mathsf Q^0$ invariant. 
Now, as shown by \cite{Kachru:2002he,DeWolfe:2004ns} and reviewed
in appendix~\ref{app_proof_tu}, the dualities can be used to 
show that the number of physically-distinct vacua in the fundamental domain
is finite for fixed $\mathsf Q^0$. 
In the following we explore how the properties of the space of solutions for 
$\tau$ and $U$ depend on $\mathsf Q^0$.


\subsection{Space of solutions}

In this section we study the space of solutions to the F-term equations \eqref{f_term_eq}
for the combined axio-dilaton and 
complex-structure-modulus system. 
Since for the axio-dilaton system we found two-dimensional circular voids in the two-dimensional 
moduli space, it is natural to expect four-dimensional spherical voids in the four-dimensional 
moduli space. However, we can not confirm this expectation. 
Our data has again been obtained using a computer algorithm, which  generated all phy\-si\-cally-distinct
flux vacua for a given upper bound on the D3-brane tadpole contribution $\mathsf Q^0$.


\subsubsection*{Distribution of solutions}

In \cite{Kachru:2002he,DeWolfe:2004ns} (as well as in appendix~\ref{app_proof_tu}) it is shown that 
for fixed $\mathsf Q^0$ the number of physically-distinct solutions is finite. 
We have determined all solutions for the setting described in section~\ref{sec_tu_set} 
numerically, and have visualized them in the following figures.
\begin{itemize}

\item In figure~\ref{IsotropicTorus_Plane_Global} we have shown the solutions 
for the fluxes of the form \eqref{IsotropyConditions} projected onto the $\tau$ and 
onto the $U$-plane \cite{DeWolfe:2004ns}. 
All solutions satisfy the bound on the tadpole contribution $\frac{\mathsf Q^{0}}{192}  \leq 1000$, and 
in order to have a symmetric plot we included points on the 
boundary of the fundamental domains. 
These plots are similar to the one in figure~\ref{DouglasPlot_Plane_Global}.
When comparing figures~\ref{tu_2001} and \ref{tu_2002}, we note that
for the same $\mathsf Q^0$ the maximum values for $\tau_2$ and $U_2$ differ significantly. 
Furthermore, we note that the number of different values for $\tau_1$ is much larger
than for $U_1$.

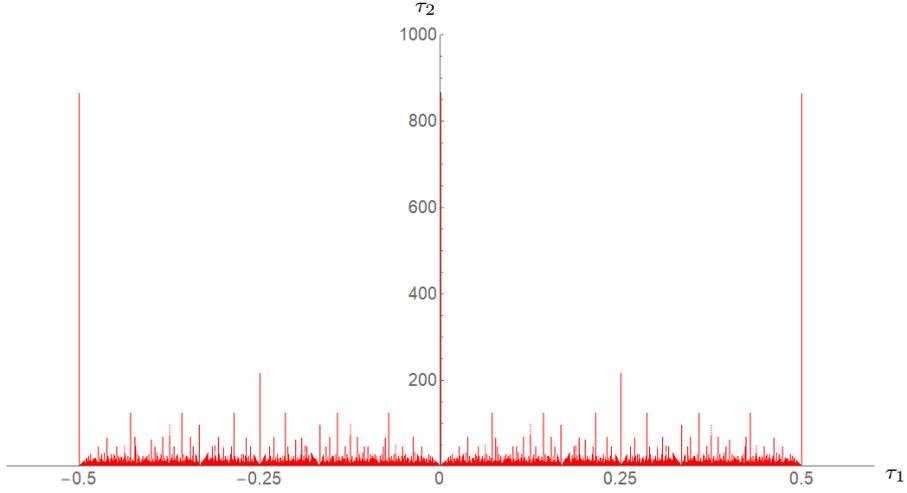
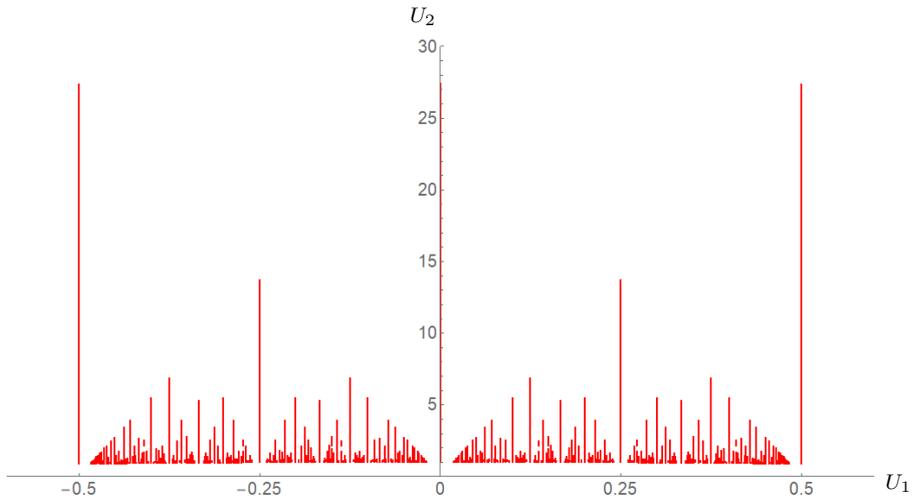
\begin{figure}[p]
\centering
\vspace*{20pt}
\begin{subfigure}{0.95\textwidth}
\centering
\includegraphics[width=325pt]{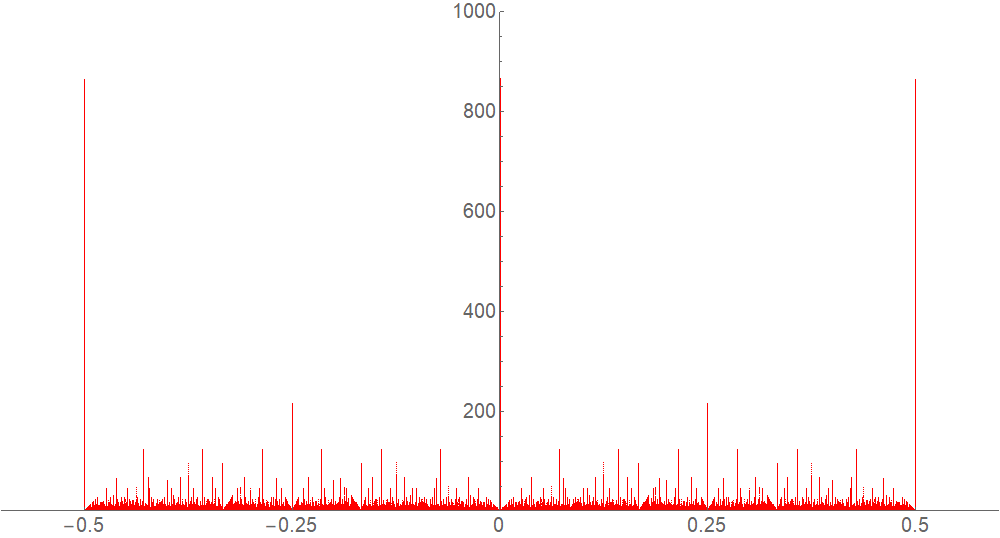}
\begin{picture}(0,0)
\put(0,4){\scriptsize $\tau_1$}
\put(-176,180){\scriptsize $\tau_2$}
\end{picture}

\vspace{-15pt}
\caption{Projection of solutions onto the $\tau$-plane.\label{tu_2001}}
\end{subfigure}

\vspace*{20pt}

\begin{subfigure}{0.95\textwidth}
\centering
\vspace*{20pt}
\includegraphics[width=325pt]{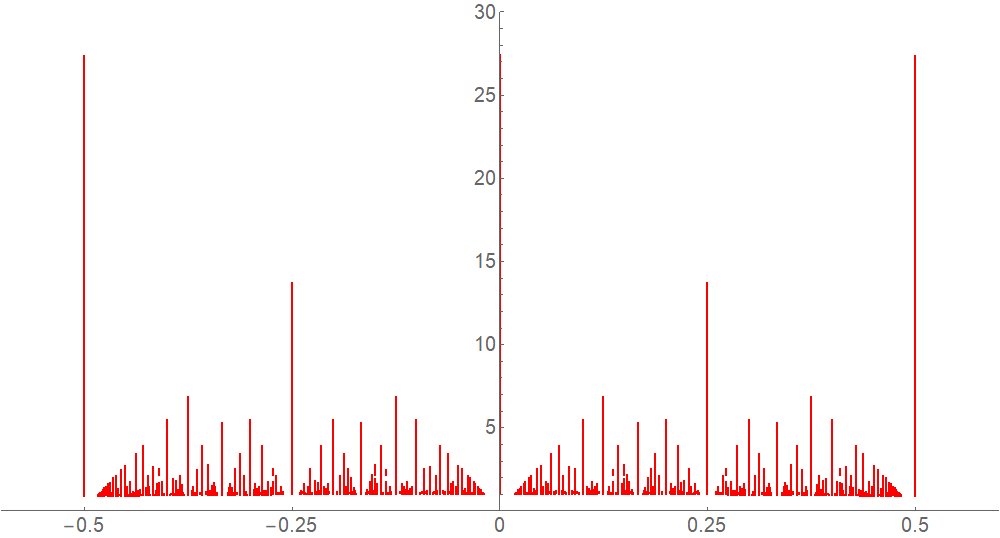}
\begin{picture}(0,0)
\put(0,4){\scriptsize $U_1$}
\put(-178,180){\scriptsize $U_2$}
\end{picture}

\vspace{-15pt}
\caption{Projection of solutions onto the $U$-plane.\label{tu_2002}}
\end{subfigure}

\caption{
Space of solutions 
for the setting described in section~\ref{sec_tu_set}, mapped to the fundamental 
domains $\mathcal F_{\tau}$ and $\mathcal F_U$ and projected 
onto the $\tau$- and $U$-plane.
All solutions satisfy the bound \raisebox{0pt}[0pt][0pt]{$\frac{\mathsf Q^{0}}{192}  \leq 1000$}.
\label{IsotropicTorus_Plane_Global}
}
\end{figure}

\item In figure~\ref{IsotropicTorus_Sections} we show sections through the 
four-dimensional space of solutions for $\tau_2\leq 2$, characterized by 
different values of the complex-structure modulus. 
All solutions satisfy $\frac{\mathsf Q^{0}}{192}  \leq 1000$,
and these plots show void structures similar as in figure~\ref{DouglasPlot_Plane}.
We note however that 
although the location of the voids stays the same when going from 
$U=i$ to $U=2i$ and similarly from $U=\sqrt{2} i$ to $U=2\sqrt{2}i$,
the density of solutions decreases. This appears to be a general 
feature which we observe in the data. 

\begin{figure}[p]
\centering
\begin{subfigure}{0.45\textwidth}
\centering
\vspace{10pt}%
\resizebox{\textwidth}{!}{%
\includegraphics[width=200pt]{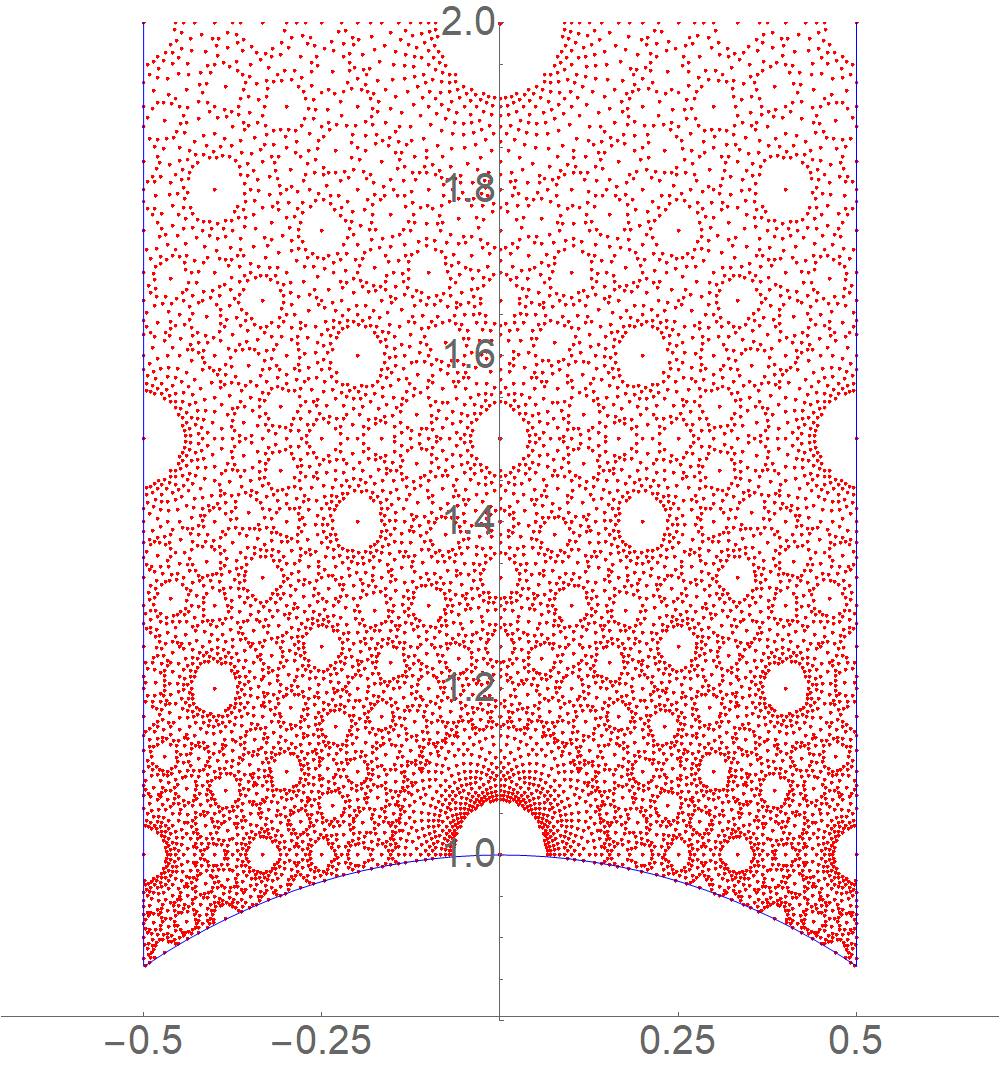}%
\begin{picture}(0,0)
\put(0,2){\scriptsize  $\tau_1$}
\put(-114,218){\scriptsize $\tau_2$}
\end{picture}%
}%
\vspace*{-14pt}%
\caption{$U=i$}
\end{subfigure}
\hspace{20pt}
\begin{subfigure}{0.45\textwidth}
\centering
\vspace{10pt}%
\resizebox{\textwidth}{!}{%
\includegraphics[width=200pt]{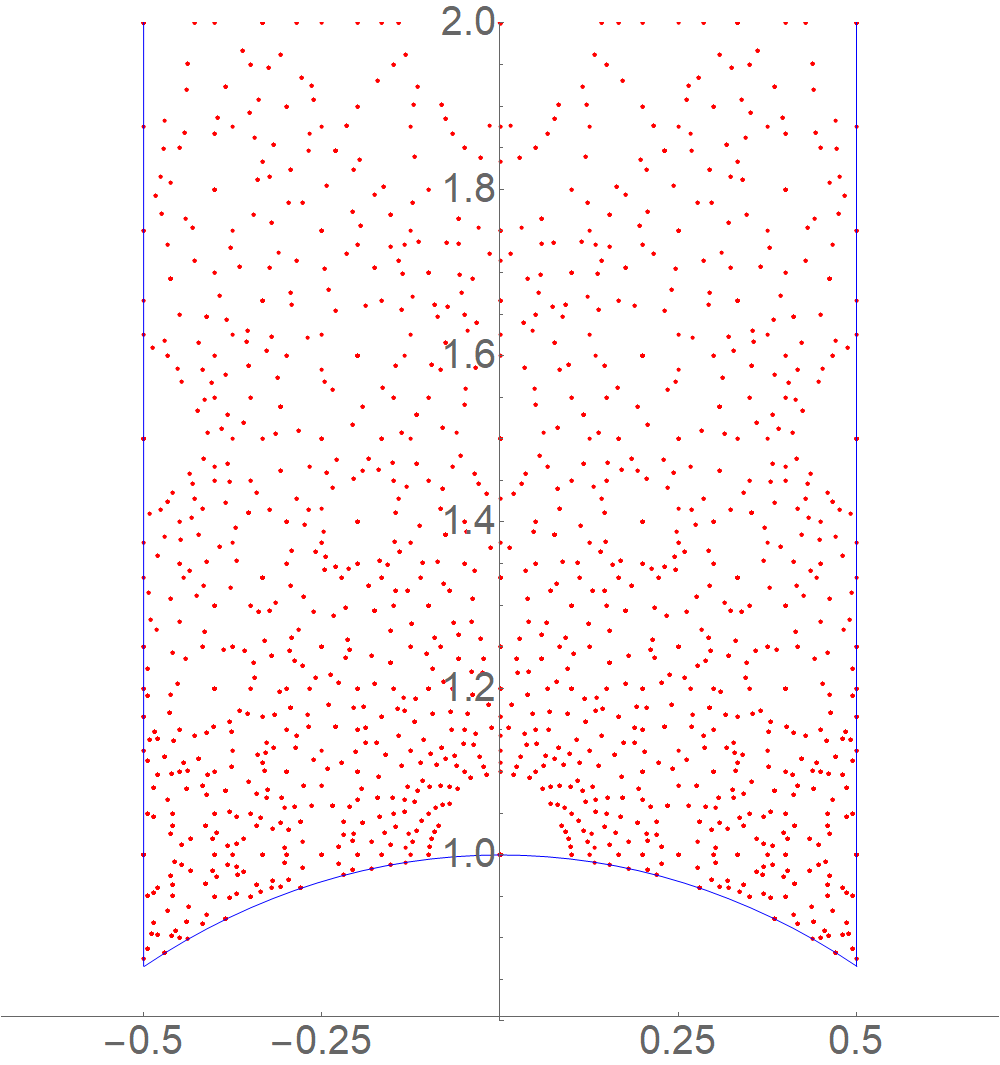}%
\begin{picture}(0,0)
\put(0,2){\scriptsize  $\tau_1$}
\put(-114,218){\scriptsize $\tau_2$}
\end{picture}%
}%
\vspace*{-14pt}%
\caption{$U=2i$}
\end{subfigure}%
\hspace{6pt}

\vspace{15pt}

\begin{subfigure}{0.45\textwidth}
\centering
\vspace{10pt}%
\resizebox{\textwidth}{!}{%
\includegraphics[width=200pt]{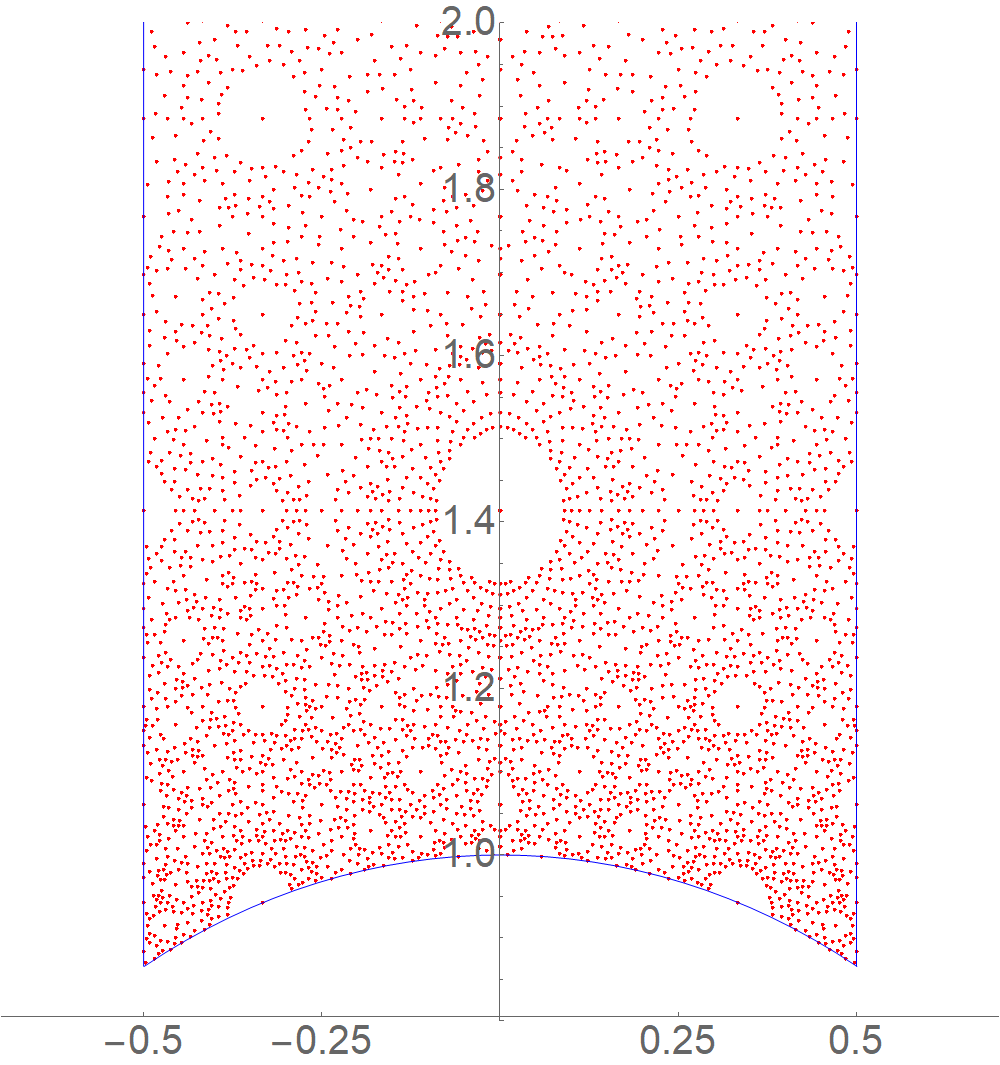}%
\begin{picture}(0,0)
\put(0,2){\scriptsize  $\tau_1$}
\put(-114,218){\scriptsize $\tau_2$}
\end{picture}%
}%
\vspace*{-14pt}%
\caption{$U=\sqrt{2}i$}
\end{subfigure}
\hspace{20pt}
\begin{subfigure}{0.45\textwidth}
\centering
\vspace{10pt}%
\resizebox{\textwidth}{!}{%
\includegraphics[width=200pt]{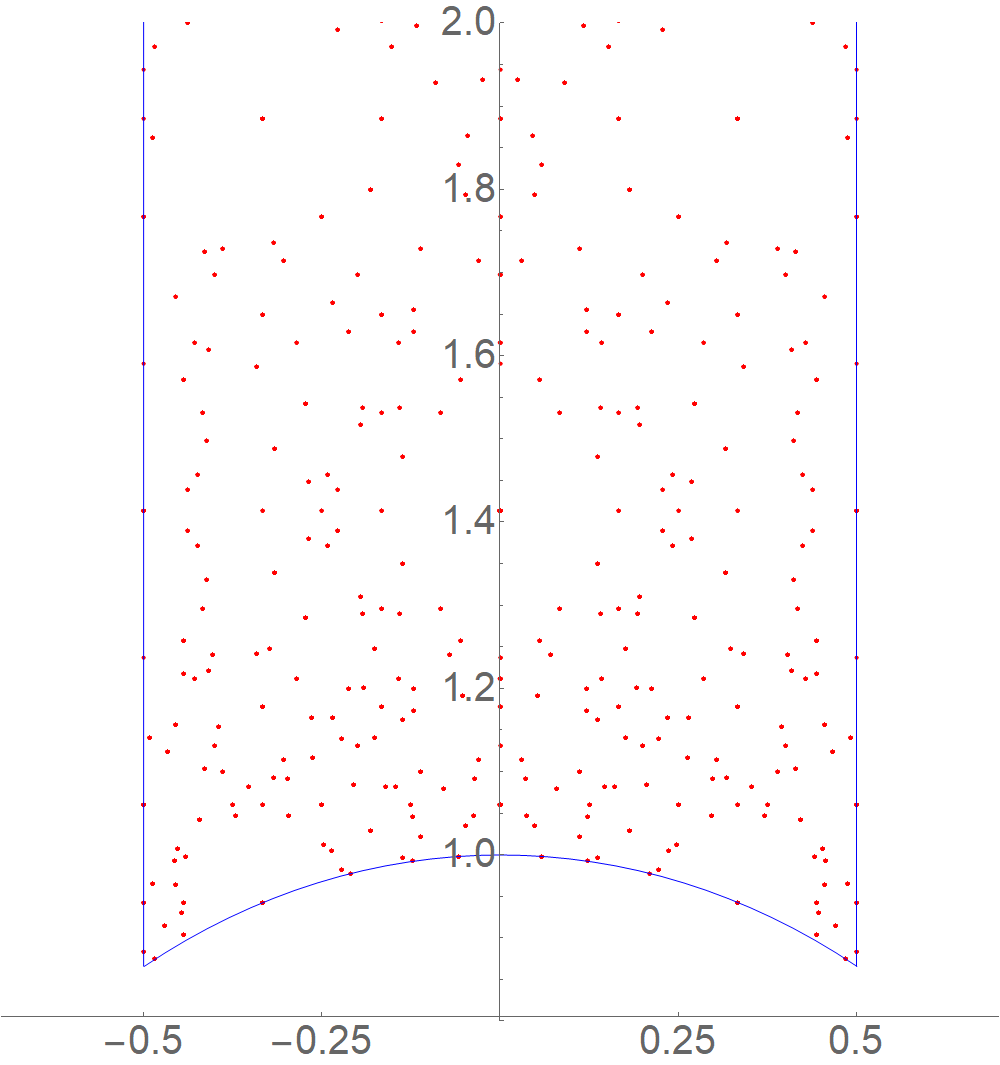}%
\begin{picture}(0,0)
\put(0,2){\scriptsize  $\tau_1$}
\put(-114,218){\scriptsize $\tau_2$}
\end{picture}%
}%
\vspace*{-14pt}%
\caption{$U=2\sqrt{2}i$}
\end{subfigure}%
\hspace{6pt}

\caption{
Section through the four-dimensional space of solutions
for the setting described in section~\ref{sec_tu_set}.
The solutions have been mapped to the fundamental domains,
and the sections are for fixed complex-structure modulus at  
$U=i$,  $U=2i$,   $U=\sqrt{2}i$ and  $U=2\sqrt{2}i$. All solutions satisfy the bound 
$\frac{\mathsf Q^{0}}{192}  \leq 1000$. } 
\label{IsotropicTorus_Sections}
\end{figure}

\item In figure~\ref{IsotropicTorus_3DPlots} we have shown three-dimensional sections of
the four-dimensional space of solutions for $U_1=0$. 
All solutions have been mapped to the fundamental domain.
Figures~\ref{fig_tu_01} and \ref{fig_tu_02} show two different points of view, which 
illustrate that the three-dimensional 
section of the space of solutions is not homogenous. Solutions are  accumulated on planes 
for particular values of $U_2$, while the space between these planes 
is only sparsely populated. 
This is in agreement with our observations in figures~\ref{IsotropicTorus_Sections}, which 
also show that the density of solutions varies. 

The lines in figures~\ref{fig_tu_01} and \ref{fig_tu_02} connect voids for different values 
of $U_2$ and are described by the 
following equations for $t\in\mathbb R_+$
\eq{
  \label{tu_lines_004}
  \arraycolsep2pt
  \begin{array}{l@{\hspace{15pt}}l@{\hspace{30pt}}lcl@{\op}r@{\op}r@{\op}l@{\op}l@{\op}l}
  \mbox{orange} & l_1 & (\tau_1,\tau_2,U_1,U_2) &=& ( & 0 &,\, 1+t & ,\, 0 & ,\, 1+ \frac{1}{1}\op t &)\,,
  \\[4pt]
  \mbox{red} &l_2 & (\tau_1,\tau_2,U_1,U_2) &=& ( & 0 &,\, 2+t & ,\, 0 & ,\, 1+ \frac{1}{2}\op t &)\,,
  \\[4pt]
  \mbox{purple} &l_3 & (\tau_1,\tau_2,U_1,U_2) &=& ( & 0 &,\, 3+t & ,\, 0 & ,\, 1+ \frac{1}{3}\op t &)\,,
  \\[4pt]
  \mbox{green} &l_4 & (\tau_1,\tau_2,U_1,U_2) &=& ( & -\frac{1}{2} &,\, \frac{3}{2}+t & ,\, 0 & ,\, 1+ \frac{2}{3}\op t &)\,.
  \end{array}
}

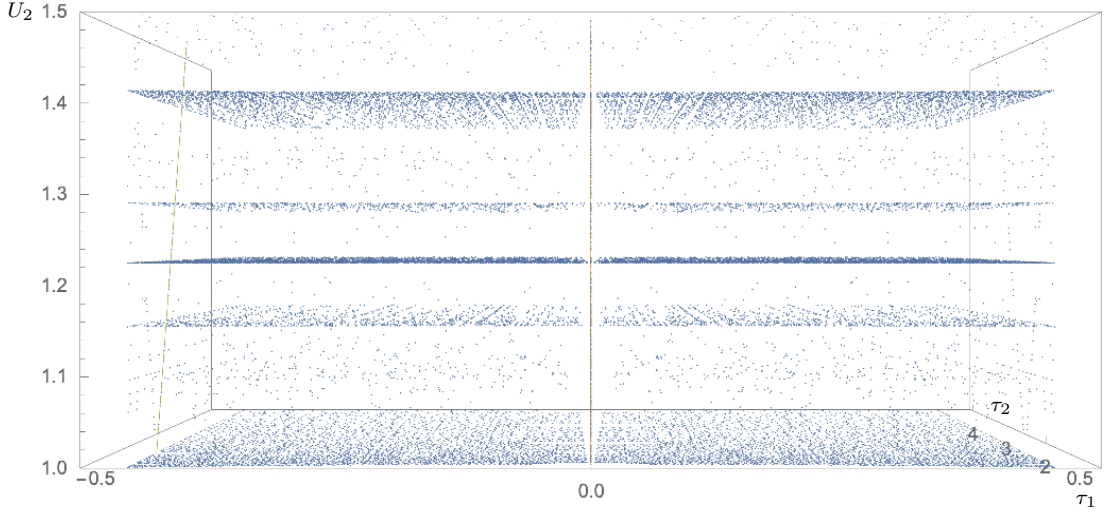
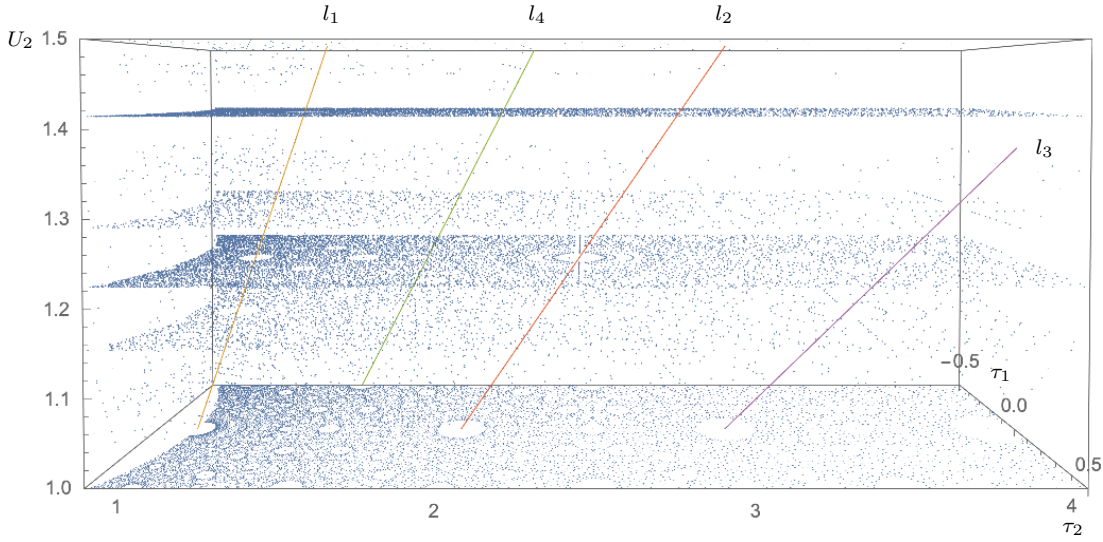
\begin{figure}[p]
\centering

\begin{subfigure}{0.95\textwidth}
\includegraphics[width=\textwidth]{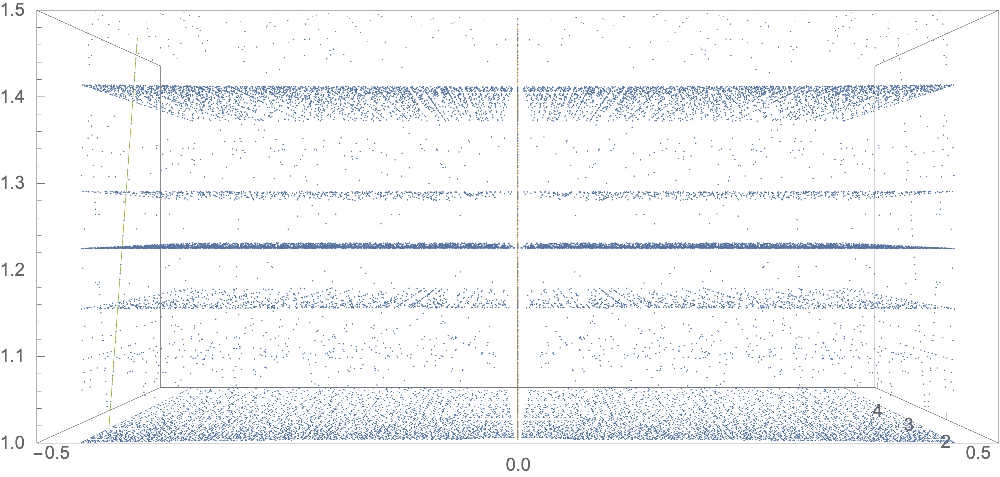}%
\begin{picture}(0,0)
\put(-10,0){\scriptsize $\tau_1$}
\put(-42,35){\scriptsize$\tau_2$}
\put(-410,183.5){\scriptsize $U_2$}
\end{picture}%
\vspace*{-8pt}%
\caption{Point of view along the $\tau_2$-direction.\label{fig_tu_01}}
\end{subfigure}

\vspace{30pt}

\begin{subfigure}{0.95\textwidth}
\includegraphics[width=\textwidth]{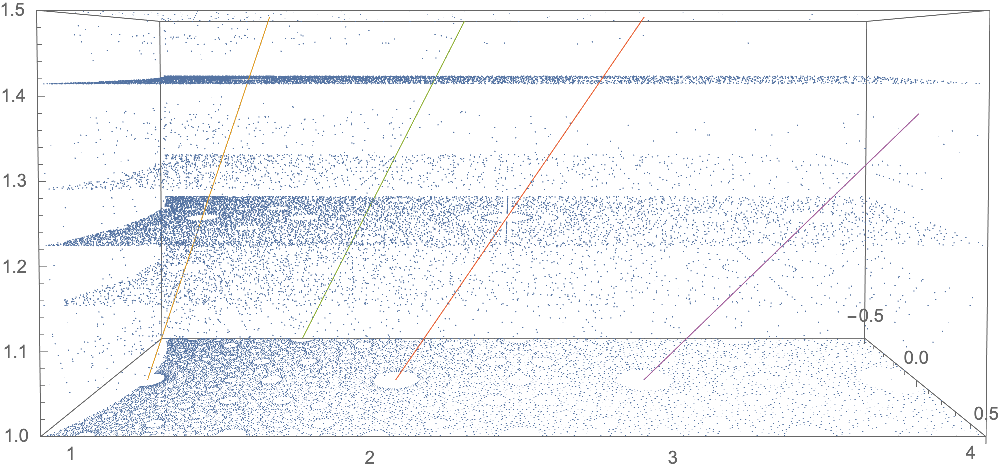}%
\begin{picture}(0,0)
\put(-15,-3){\scriptsize $\tau_2$}
\put(-42,55){\scriptsize $\tau_1$}
\put(-410,180.5){\scriptsize $U_2$}
\put(-292,190){\scriptsize $l_1$}
\put(-215,190){\scriptsize $l_4$}
\put(-145,190){\scriptsize $l_2$}
\put(-25,139.5){\scriptsize $l_3$}
\end{picture}%
\vspace*{-8pt}%
\caption{Point of view along the $\tau_1$-direction.\label{fig_tu_02}}
\end{subfigure}

\vspace{10pt}

\caption{
Section through the four-dimensional space of solutions with $U_1=0$
for the setting described in section~\ref{sec_tu_set}.
All solutions satisfy the bound \raisebox{0pt}[0pt][0pt]{$\frac{\mathsf Q^{0}}{192}  \leq 1000$}
and have been mapped to the fundamental domains.
The lines in \ref{fig_tu_01} and \ref{fig_tu_02} connect voids for 
different values of $U_2$ and are described by the expressions  
in equation \eqref{tu_lines_004}.
\label{IsotropicTorus_3DPlots}}
\end{figure}

\item In figure~\ref{IsotropicTorus_3DPlots_2} we have shown the same
three-dimensional section of the space of solutions as in 
figure~\ref{IsotropicTorus_3DPlots}. The point of 
view in figure~\ref{fig_10001} is along the line $l_1$ (orange)
of \eqref{tu_lines_004} and 
the point of view in figure~\ref{fig_10002} is along
the line $l_2$ (red). 
In these three-dimensional sections of the four-dimensional space of solutions
we  therefore
have a cylindrical void centered around the lines in 
\eqref{tu_lines_004}.

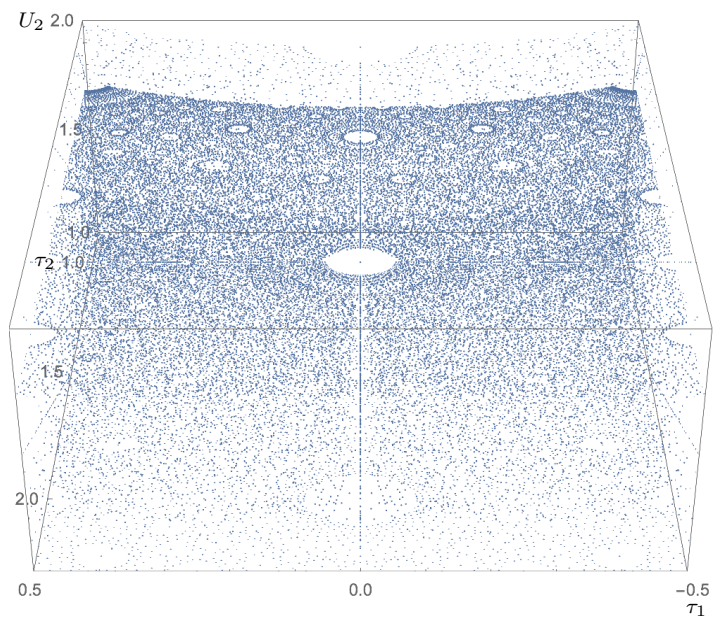
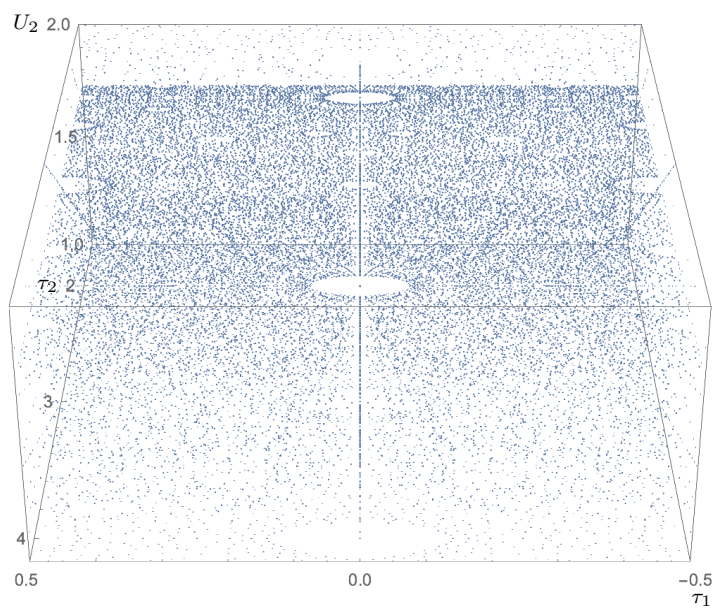
\begin{figure}[p]
\centering

\begin{subfigure}{0.63\textwidth}
\includegraphics[width=\textwidth]{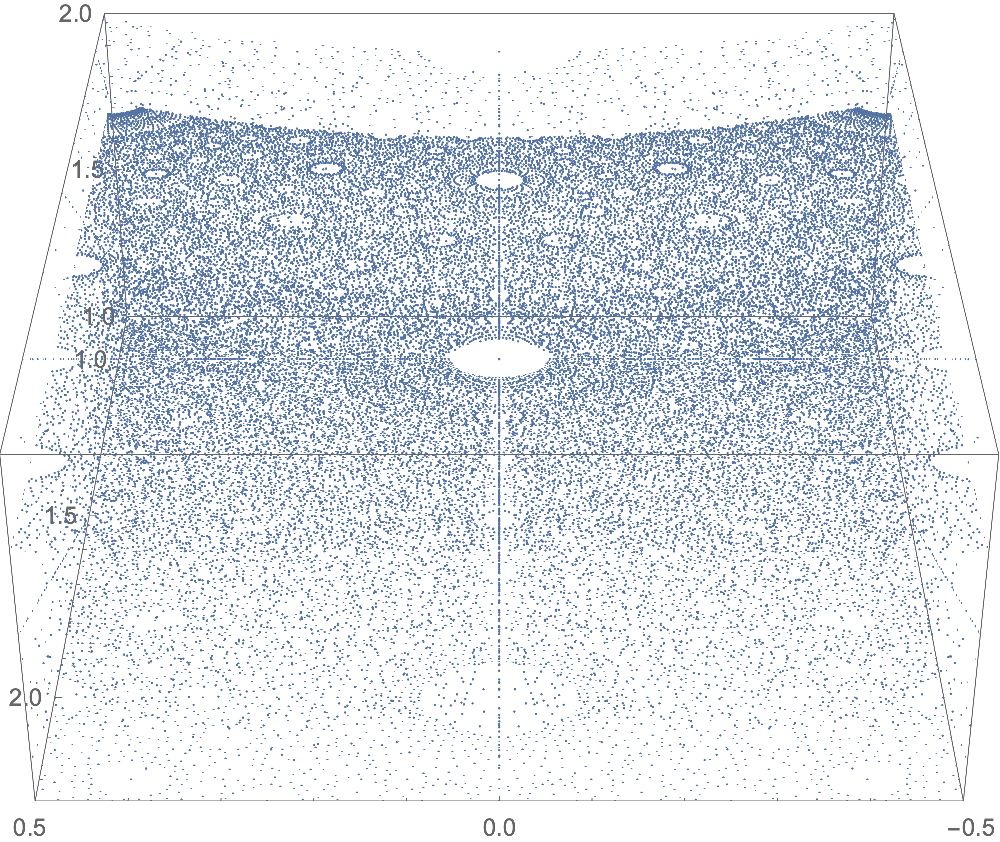}%
\begin{picture}(0,0)
\put(-10,-5){\scriptsize $\tau_1$}
\put(-254,126){\scriptsize$\tau_2$}
\put(-260,216){\scriptsize $U_2$}
\end{picture}%
\vspace*{-8pt}%
\caption{View along the line $l_1$ (orange) in figures~\ref{IsotropicTorus_3DPlots}. \label{fig_10001}}
\end{subfigure}

\vspace{10pt}

\begin{subfigure}{0.63\textwidth}
\includegraphics[width=\textwidth]{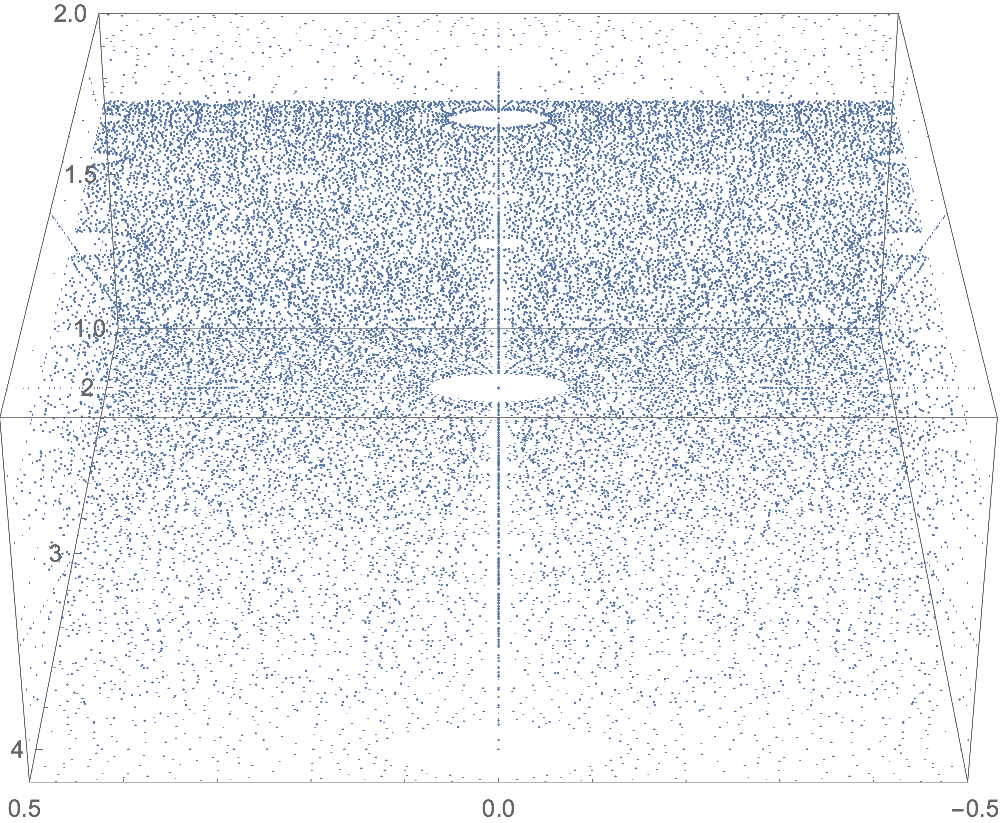}%
\begin{picture}(0,0)
\put(-8,-5){\scriptsize $\tau_1$}
\put(-253,113.5){\scriptsize$\tau_2$}
\put(-262,211.5){\scriptsize $U_2$}
\end{picture}%
\vspace*{-8pt}%
\caption{View along the  line $l_2$ (red) in figures~\ref{IsotropicTorus_3DPlots}.\label{fig_10002}}
\end{subfigure}

\vspace{5pt}

\caption{
Section through the four-dimensional space of solutions with $U_1=0$
for the setting described in section~\ref{sec_tu_set}.
All solutions satisfy the bound \raisebox{0pt}[0pt][0pt]{$\frac{\mathsf Q^{0}}{192}  \leq 1000$}
and have been mapped to the fundamental domains.
The points of view are along the line $l_1$ (figure~\ref{fig_10001}) and 
line $l_2$ (figure~\ref{fig_10002}) in figures~\ref{IsotropicTorus_3DPlots},
which show a void structure around $l_1$ and $l_2$.
\label{IsotropicTorus_3DPlots_2}}
\end{figure}

\end{itemize}


\subsubsection*{Solutions at small coupling and large complex structure}

\begin{figure}[t!]
\centering
\includegraphics[width=100pt]{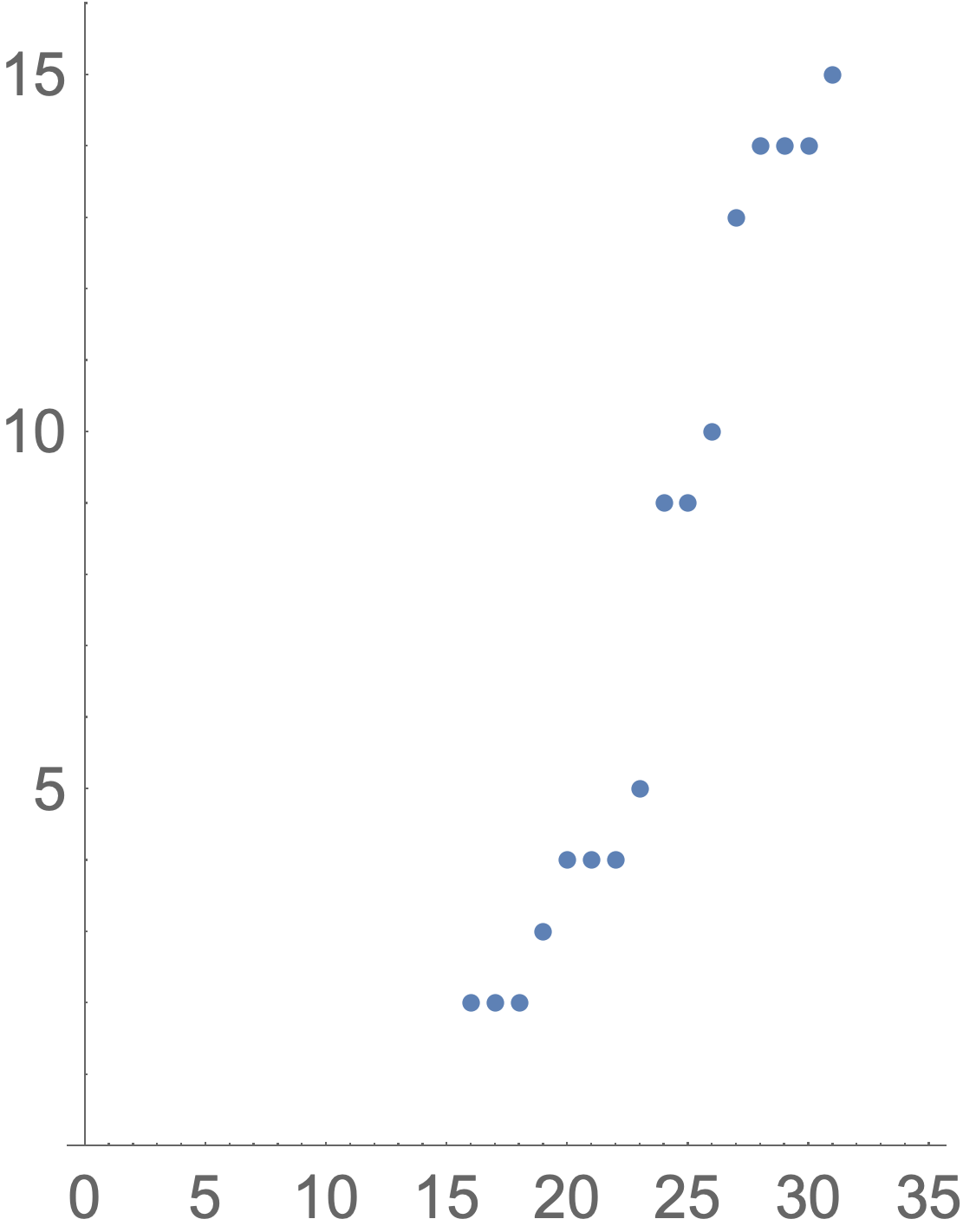}\hspace{15pt}
\includegraphics[width=100pt]{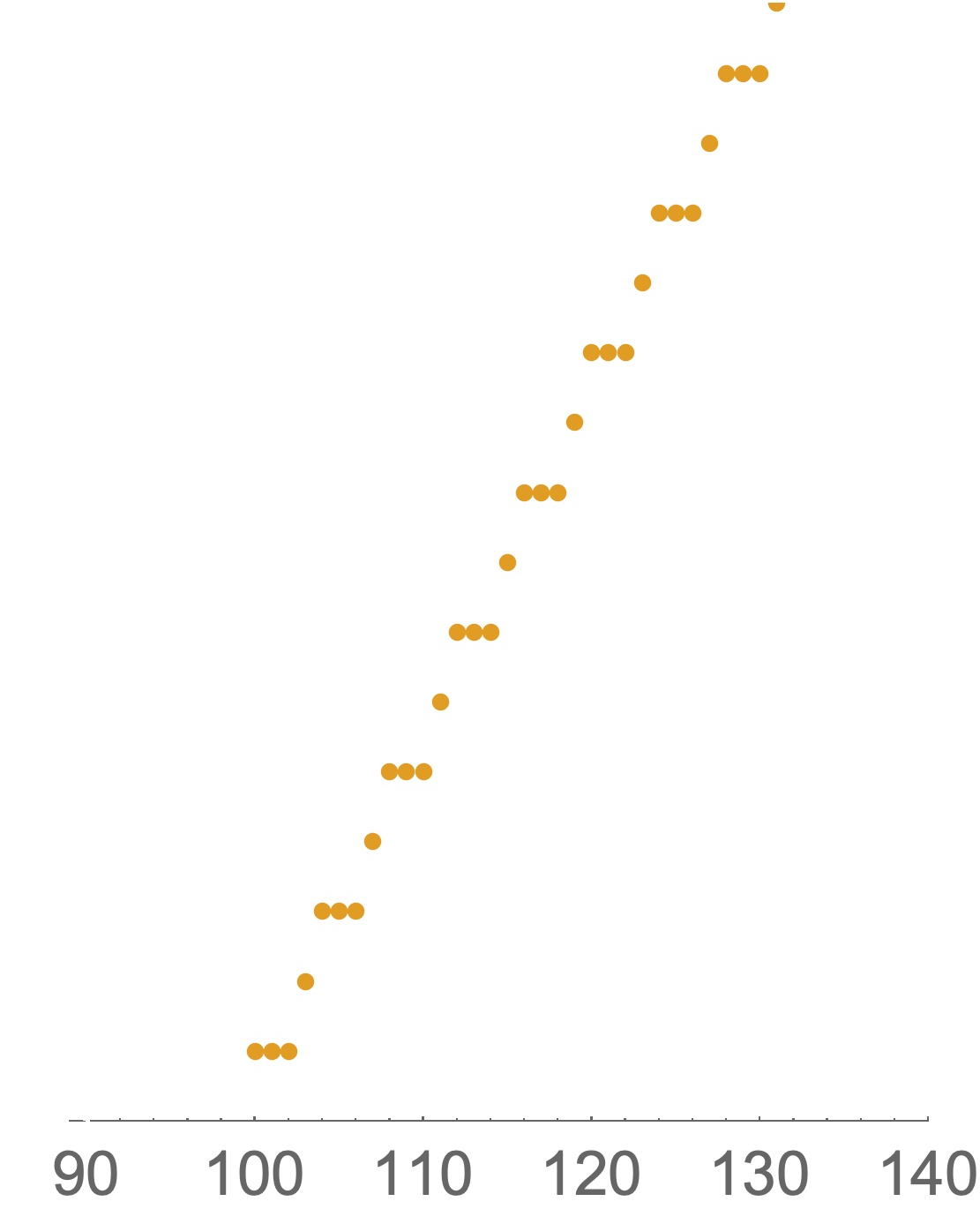}\hspace{15pt}
\includegraphics[width=100pt]{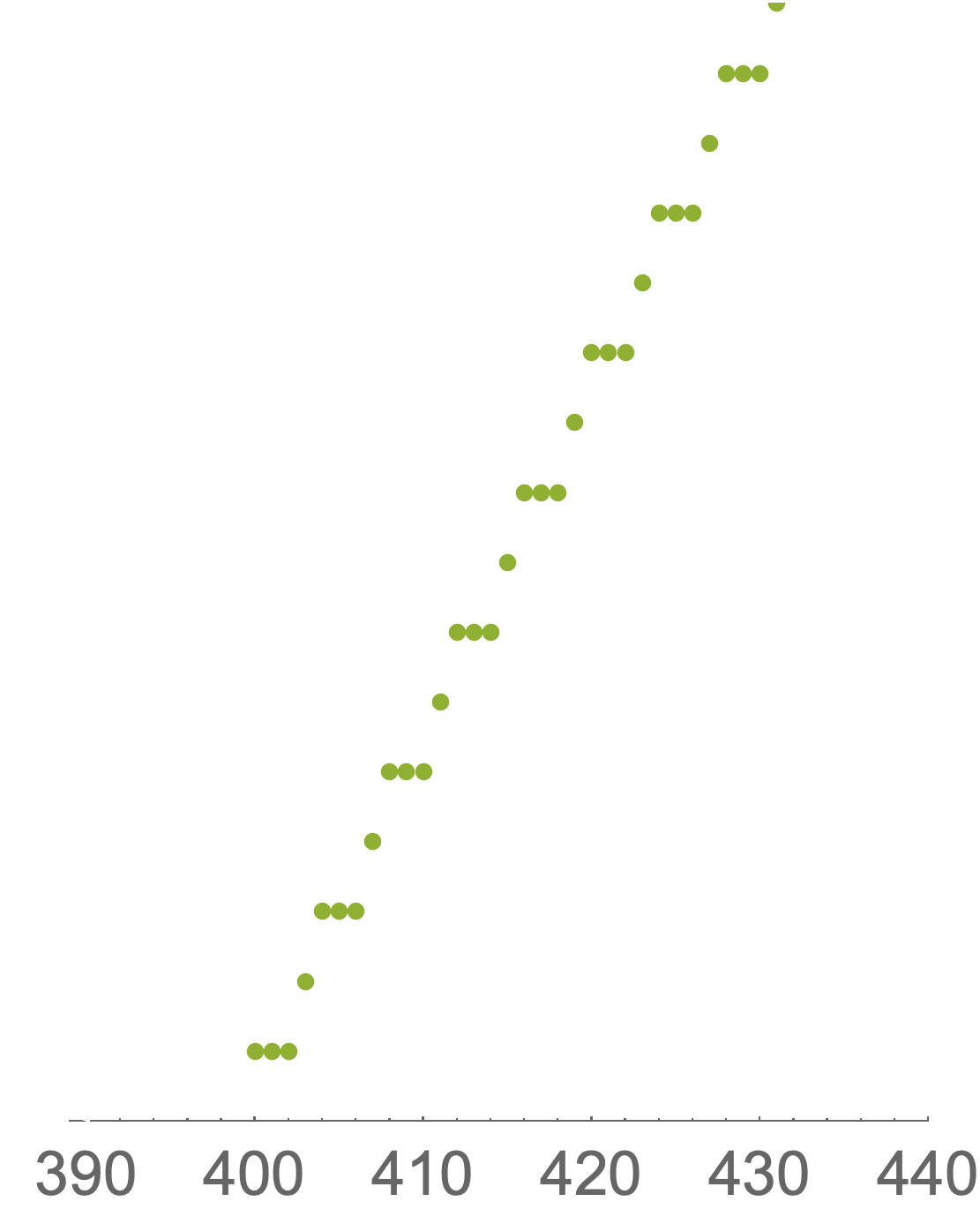}%
\begin{picture}(0,0)
\put(-350,123){\scriptsize $\mathsf N_{\mathsf c}$}
\put(4,4.5){\scriptsize $\frac{\mathsf Q_{\rm max}^0}{192}$}
\end{picture}%
\caption{
Number of solutions $\mathsf N_{\mathsf c}$ 
as a function of \raisebox{0pt}[0pt][0pt]{$\frac{\mathsf Q_{\rm max}^0}{192}$}
(for the setting described in section~\ref{sec_tu_set})
which satisfy $\tau_2,U_2\geq \mathsf c$ 
for $\mathsf c = 2,5,10$ in colors blue, orange and green, respectively.
Note that different ranges for \raisebox{0pt}[0pt][0pt]{$\frac{\mathsf Q_{\rm max}^0}{192}$}
on the horizontal axis.
} 
\label{plot10}
\end{figure}
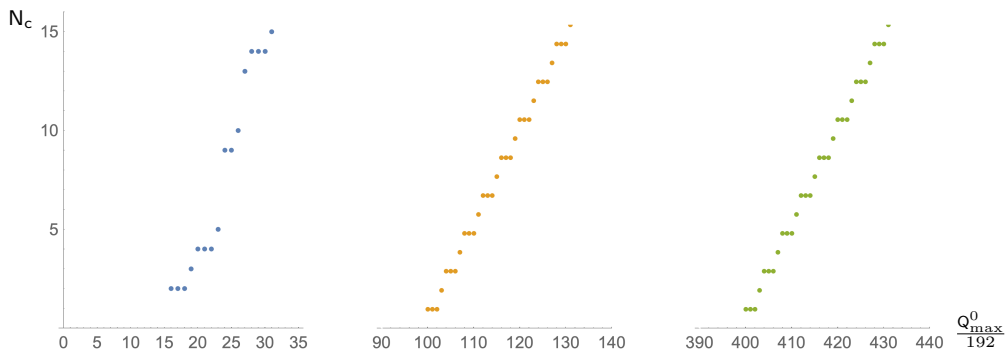

We now consider the number $\mathsf N$ of physically-distinct solutions for the combined axio-dilaton and 
complex-structure moduli system defined in section~\ref{sec_tu_set}.
This number is finite for fixed D3-tadpole contribution $\mathsf Q^0$, and since
we have the numerical data  we can determine this number explicitly. 
For large $\mathsf Q^0$ the dependence takes the form
\eq{
  \label{rel_9347775}
  \mathsf N \approx 1.2501
  \left[\frac{\mathsf Q^0}{192}\right]^{2} .
}
We next note that in the fundamental domains, 
the imaginary parts of the axio-dilaton and complex-structure moduli
satisfy a lower bound similarly as in the previous example. 
An upper bound can be obtained from the numerical data, which 
can be expressed as\,\footnote{
More precisely, with
\raisebox{0pt}[0pt][0pt]{$x=\frac{\mathsf Q^0}{192}$} 
the bound on $U_2$ can be expressed as
$U_2\leq\sqrt{C\op x}$, where the constant $C$ takes values
\raisebox{0pt}[0pt][0pt]{$C=3/4$} for \raisebox{0pt}[0pt][0pt]{$x=0,1\mod 4$},
\raisebox{0pt}[0pt][0pt]{$C=3/8$} for \raisebox{0pt}[0pt][0pt]{$x=2\mod 8$},
\raisebox{0pt}[0pt][0pt]{$C=1/4$} for \raisebox{0pt}[0pt][0pt]{$x=3,7\mod 8$}, and
\raisebox{0pt}[0pt][0pt]{$C=1/8$} for \raisebox{0pt}[0pt][0pt]{$x=6\mod 8$}.
}
\eq{
  \label{rel_0038}
  \frac{\sqrt{3}}{2}\leq \tau_2  \leq \frac{\sqrt{3}}{2}\left[ \frac{\mathsf Q^0}{192} \right],
  \hspace{80pt}
  \frac{\sqrt{3}}{2}\leq U_2  \leq \frac{\sqrt{3}}{2}\left[ \frac{\mathsf Q^0}{192}   \right]^{1/2}.
}
Note that in our conventions the tadpole contribution $\mathsf Q^0$ is a multiple of $192$.   
However, as we have seen in \eqref{rel_0039}, the solution for the axio-dilaton 
depends on the complex-structure modulus. Although this dependence is difficult to 
analyze analytically, the numerical data gives the following bound on the solutions 
\eq{
  \tau_2 \, U_2 \leq \frac{3}{4}\op \frac{\mathsf Q^0}{192} \,.
}
This  bound is stronger than in \eqref{rel_0038}, and it implies that for 
fixed $\mathsf Q^0$ the imaginary parts of $\tau$ and $U$ cannot be made 
simultaneously large. 
In particular, in order to have solutions at small coupling $g_{\rm s} = \frac{1}{\tau_2}\ll 1$ and 
large complex structure $U_2\gg 1$, the tadpole contribution has to be sufficiently large. 
Let us make this more precise and determine numerically the number of solutions $\mathsf N_{\mathsf c}$ 
with $\mathsf Q^0\leq \mathsf Q^0_{\rm max}$
for which 
\eq{
  g_{\rm s} \leq \frac{1}{\mathsf c} \hspace{50pt} \mbox{and} \hspace{50pt}
  U_2 \geq \mathsf c \,.
}
In the limit of large \raisebox{0pt}[0pt][0pt]{$\frac{\mathsf Q_{\rm max}^0}{192}$} 
we obtained the following approximations 
\eq{
 \frac{\mathsf Q^0_{\rm max}}{192} \gg 1
 \hspace{40pt}
 \renewcommand{\arraystretch}{1.3}
 \arraycolsep10pt
 \begin{array}{c||r@{\hspace{2pt}}rc}
 \mathsf c & \multicolumn{2}{c}{\mathsf N_{\mathsf c}} & \mathsf N_{\mathsf c}/\mathsf N 
 \\
 \hline\hline
 2 & 0.0553 \bigl[\frac{\mathsf Q_{\rm max}^0}{192}\bigr]^{2} &-3.4617\bigl[\frac{\mathsf Q_{\rm max}^0}{192}\bigr] &  0.041
 \\
 5 & 0.0047 \bigl[\frac{\mathsf Q_{\rm max}^0}{192}\bigr]^{2} &-0.7627\bigl[\frac{\mathsf Q_{\rm max}^0}{192}\bigr]   & 0.004
 \\
 10 & 0.0009 \bigl[\frac{\mathsf Q_{\rm max}^0}{192}\bigr]^{2} &-0.3569\bigl[\frac{\mathsf Q_{\rm max}^0}{192}\bigr]   & 0.001
 \end{array}
}
These approximations do not describe the data very well, but are sufficient for our purposes here. 
In particular, we see that at leading order $\mathsf N_{\mathsf c}$ depends 
quadratically on $\mathsf Q^0_{\rm max}$ and that the ratios $ \mathsf N_{\mathsf c}/\mathsf N$
are rather small. Thus, only a small percentage of the solutions to the F-term equations 
are in a perturbatively-controlled 
regime.
More interesting is the limit of small 
\raisebox{0pt}[0pt][0pt]{$\frac{\mathsf Q^0}{192}$}, which we have
illustrated in figure~\ref{plot10}.
We see that for $\mathsf c=2$ (blue) 
there are  solutions starting at 
\raisebox{0pt}[0pt][0pt]{$\frac{\mathsf Q^0}{192}=16$}.
For $\mathsf c=5$ (orange) we find 
 solutions starting at 
\raisebox{0pt}[0pt][0pt]{$\frac{\mathsf Q^0}{192}=100$},
and for $\mathsf c=10$ (green) 
solutions can be obtained starting at
\raisebox{0pt}[0pt][0pt]{$\frac{\mathsf Q^0}{192}=400$}.
The main conclusion we want to draw from this analysis is that for 
solutions at weak coupling $g_{\rm s}\ll 1$ and large complex structure $U_2\gg1$, 
the D3-tadpole contribution \raisebox{0pt}[0pt][0pt]{$\frac{\mathsf Q^0}{192}$} has
to be large. As discussed on page~\pageref{page_ma}, this is in tension with the tadpole cancellation condition.


\subsection{Summary}

Let us summarize the results obtained in this section for the space of solutions
of the combined axio-dilaton and complex-structure-moduli  system
with fluxes characterized by the setting described in section~\ref{sec_tu_set}:
\begin{itemize}

\item As known before, for a fixed D3-brane tadpole contribution $\mathsf Q^0$ the number
of physically-distinct solutions to the F-term equations \eqref{f_term_eq} 
for the axio-dilaton and complex-structure moduli is finite. 
This is again due to the $SL(2,\mathbb Z)$ dualities for the axio-dilaton and the complex-structure 
moduli. 

\item The values of the fixed moduli (mapped to their fundamental domains) are not 
distributed homogeneously in the space of solutions. As shown in figures~\ref{IsotropicTorus_3DPlots_2}, 
the solutions are accumulated on submanifolds in 
the four-dimensional space with few points in between. 
We also find void structures in the space of solutions, which are however not spherical 
but take a cylindrical form in three-dimensional sections (cf. figures~\ref{IsotropicTorus_3DPlots_2}).

\item The imaginary parts of the axio-dilaton and the complex-structure moduli 
$\tau_2$ and $U_2$ are bounded from above and below as shown in equation \eqref{rel_0038}.
In our data we find however the stronger bound on the product 
\raisebox{0pt}[0pt][0pt]{$ \tau_2 \, U_2 \leq \frac{3}{4}\op \frac{\mathsf Q^0}{192}$},
which implies that in the weak-coupling and large-complex-structure regime 
the tadpole contribution \raisebox{0pt}[0pt][0pt]{$\frac{\mathsf Q^0}{192}$} has to be large.
This is again in contrast to our arguments regarding the tadpole-cancellation condition 
on page~\pageref{page_ma} which 
requires \raisebox{0pt}[0pt][0pt]{$\frac{\mathsf Q^0}{192}$} to be small, 
and illustrates the tension between the closed- and open-string sectors for 
obtaining reliable solutions.

\item We have furthermore shown that the fraction of reliable flux solutions within 
all solutions for fixed tadpole  \raisebox{0pt}[0pt][0pt]{$\frac{\mathsf Q^0}{192}$} is
only of orders $\mathcal O(10^{-3})$, which is a  reduction as compared 
to the setting of section~\ref{sec_mod1}.

\end{itemize}


\clearpage
\section{Moduli stabilization III}
\label{sec_stu}

We now generalize the setting from section~\ref{sec_iso} by including non-geometric $Q$-fluxes. 
The fluxes are restricted such that the complex-structure and K\"ahler moduli are fixed 
isotropically as  $T_1 = T_2 = T_3:=T$ and $U^1=U^2=U^3:=U$,
which reduces the system to the three complex moduli fields $\tau$, $U$ and $T$.
Such vacua have previously been analyzed for instance in \cite{Shelton:2006fd}.


\subsection{Setting}
\label{sec_stu_steting}

We specify the superpotential  \eqref{superpot_002} 
by imposing the following restrictions on the R-R and NS-NS fluxes
\eqref{flux_001} and \eqref{flux_002}
\eq{\label{IsotropyConditions_Nongeometric}
  \arraycolsep2pt
  \begin{array}{lclcl@{\hspace{50pt}}clclclcllclcl}
  f^1 &=&f^2 &=& f^3\,, &   h^1 &=& h^2 &=& h^3\,,
  \\
  f_1 &=& f_2 &=& f_3\,, &   h_1 &=&h_2 &=& h_3\,,
  \\[6pt]
  q^{01} &=& q^{02} &=& q^{03} \,,
  \\
  q_0{}^1 &=& q_0{}^2 &=& q_0{}^3\,,
  \\[6pt]
  q^{11} &=& q^{22} &=& q^{33} \,,
  &
  q^{12} &=& q^{13} &=& q^{21} &=& q^{23} &=& q^{31} &=& q^{32} &=:& \tilde q^{11}\,,
  \\
  q_1{}^1 &=& q_2{}^2 &=& q_3{}^3\,,  
  &
  q_1{}^2 &=& q_1{}^3 &=& q_2{}^1 &=& q_2{}^3 &=& q_3{}^1 &=& q_3{}^2 &=:& \tilde q_1{}^1 \,,
  \end{array}
} 
which  leaves four independent R-R $F$-flux components, four independent $H$-flux components
and six independent $Q$-flux components. 
As discussed around equation \eqref{fluxes_TorsionConstraint},  these fluxes are subject to the quantization 
conditions
\eq{
\label{stu_flux_quant}
 f^0,\, f^1,\, f_0,\, f_1,\hspace{10pt} 
 h^0,\, h^1,\, h_0,\,h_1,\hspace{10pt} 
 q^{01},\, q^{11},\, \tilde q^{11},\, q_0{}^1,\, q_1{}^1,\, \tilde q_1{}^1
  \in 8\mathbb Z\,.
}
Together with the K\"ahler potential \eqref{kahler_001}, the F-term equations  \eqref{f_term_eq} 
can then be determined explicitly. Since in the present situation the superpotential $W$ depends on the K\"ahler moduli 
$T_{\mathsf A}$, the condition $W=0$ is in general not obtained and the 
resulting system of equations is more involved.
However, provided that solutions to the F-term equations with non-vanishing imaginary parts exist, 
then for the fluxes \eqref{IsotropyConditions_Nongeometric} 
the moduli are stabilized isotropically 
\eq{
  U^{1}=U^{2}=U^{3}=:U\,,
  \hspace{50pt}
  T_1=T_2=T_3=:T\,.
}
A necessary condition to achieve this stabilization is that $q_1{}^1 \neq \tilde q_1{}^1$ and $q^{11}\neq \tilde q^{11}$.
The system of seven F-term equations for $\tau$, $U^i$, $T_{\mathsf A}$ 
then reduces to the following three equations
\begin{subequations}
\label{eom_STU}
\begin{align}
&
\scalebox{0.87}{$\displaystyle
\arraycolsep1pt
\begin{array}{lcrlclcllcllcllcllcllll}
0 &=&&\bigl[ & f_0 &-& h_0 &\ov\tau &-&  q_0{}^1 &3\op T &\bigr] &+& 3\op U &\bigl[ &f_1 &-& h_1 &\ov\tau &-& ( 2 \op\tilde q_1{}^1 + q_1{}^1 )& T &\bigr]
\\[4pt]
&&- (U)^3  & \bigl[ & f^0 &-& h^0 &\ov\tau &- & q^{01} &3\op T &\bigr]& +& 3 (U)^2 &\bigl[ &f^1 &-& h^1 &\ov\tau &-& ( 2\op\tilde q^{11} + q^{11} ) &T &\bigr] \,,
\end{array}
$}
\\[10pt]
&
\scalebox{0.87}{$\displaystyle
\arraycolsep1pt
\begin{array}{lcrlclcllcllcllcllcllll}
0 &=&&\bigl[ & f_0 &-& h_0 &\tau &-& q_0{}^1 &(\ov T+ 2\op T) &\bigr] &+& 
3 \op U &\bigl[ &f_1 &-& h_1 &\tau &-& ( 2\op  \tilde q_1{}^1 + q_1{}^1 )& \tfrac{1}{3}(\ov T+2 \op T) &\bigr]
\\[4pt]
&&- (U)^3  & \bigl[ & f^0 &-& h^0 &\tau &- &q^{01} &(\ov T+2\op T) &\bigr]& +& 3 (U)^2 &\bigl[ &f^1 &-& h^1 &\tau 
&-& ( 2\op\tilde q^{11} + q^{11} ) &\tfrac{1}{3}(\ov T +2\op T)&\bigr] \,,
\end{array}
$}
\displaybreak
\\[10pt]
&
\scalebox{0.87}{$\displaystyle
\arraycolsep1pt
\begin{array}{lcrlclcllcllcllcllcllll}
0 &=&&\bigl[ & f_0 &-& h_0 &\tau &-&  q_0{}^1 & 3\op T &\bigr] &+& (\ov U+2\op U) &\bigl[ &f_1 &-& h_1 &\tau &-& ( 2 \op\tilde q_1{}^1 + q_1{}^1 )& T &\bigr]
\\[4pt]
&&- \ov U (U)^2  & \bigl[ & f^0 &-& h^0 &\tau &- &q^{01} &3\op T &\bigr]& +& (2\op  \ov U U + (U)^2) &\bigl[ &f^1 &-& h^1 &\tau &-& ( 2\op\tilde q^{11} + q^{11} ) &T &\bigr] \,.
\end{array}
$}
\end{align}
\end{subequations}
The R-R and NS-NS fluxes are furthermore subject to the 
Bianchi identities \eqref{flux_020}, and using the restrictions \eqref{IsotropyConditions_Nongeometric}
we find for the tadpole contributions
\eq{\label{Bianchi_STU}
  \arraycolsep2pt
  \begin{array}{lclclcl@{\op}lcl@{\op}lclclcl}
  {\mathsf Q}^{0}  &= &
   f_0\op h^0 &-& f^0\op h_0 &
  +& \displaystyle 3\bigl( & f_1 \op h^1 & -& f^1\op h_1 & \multicolumn{2}{@{\op}l}{\displaystyle \bigr) ,}
  \\[6pt]
  {\mathsf Q}^{1}  &= &
   f_0\op q^{01} &-& f^0\op q_0{}^1 &
  +& \displaystyle 2\bigl( & f_1 \op \tilde q^{11} & -& f^1\op \tilde q_1{}^1 & \displaystyle \bigr) 
  &
  +& f_1 \op q^{11} & -& f^1\op q_1{}^1 \,,
  \\[3pt]
  {\mathsf Q}^{01}  &= &
   h_0\op q^{01} &-& h^0\op q_0{}^1 &
  +& \displaystyle 2\bigl( & h_1 \op \tilde q^{11} & -& h^1\op \tilde q_1{}^1 & \displaystyle \bigr) 
  &
  +& h_1 \op q^{11} & -& h^1\op q_1{}^1 &\overset{!}{=}& 0\,.  
 \end{array}
}
Finally, as we discussed in section~\ref{sec_dualities}, 
the present setting is duality invariant under $SL\left(2,\mathbb{Z}\right)$ transformations
of the complex-structure modulus $U$
whereas the $SL(2,\mathbb Z)$ duality \eqref{S-duality} of the axio-dilaton is broken 
to constant shifts \eqref{gen_s_dual_001}
due to the non-vanishing $Q$-flux. 
Furthermore, T-duality \eqref{perm_dual} is in general broken because of the 
isotropic choice of fluxes in \eqref{IsotropyConditions_Nongeometric}.


\subsection{Infinite number of solutions for fixed $\mathsf Q^{A}$}

In contrast to the settings of sections~\ref{sec_mod1} and~\ref{sec_iso}, for non-vanishing 
$Q$-flux the number of solutions for fixed tadpole contributions $\mathsf Q^A$ is 
infinite \cite{Shelton:2006fd}. 
This can be  illustrated with the following example from  \cite{Shelton:2006fd}:
the D7-brane tadpole contribution is fixed as $\mathsf Q^{1}=0$
and the fluxes are chosen as follows
\eq{\label{fluxes_STU_restricted}
\arraycolsep2pt
\begin{array}{lcl@{\hspace{40pt}}lcl@{\hspace{40pt}}lcl}
f^0 &=& 0 \,, & f^1 &=& 0 \,, 
\\
f_0 &=& \frac{\mathsf Q^0}{b} \,, & f_1 &=& 0 \,,
\\[8pt]
h^0 &=& b \,, & h^1 &=& \hphantom{+}b \,, 
\\
h_0 &=& b \,, & h_1 &=& -b \,,
\\[8pt]
q^{01} &=& 0 \,, & q^{11} &=& -m \,, & \tilde q^{11} &=& 0 \,, 
\\
q_0{}^1 &=& -m-n \,, & q_1{}^1 &=& \hphantom{-}n \,, & \tilde q_1{}^1 &=& m\,,
\end{array}
}
where $m,n\in 8\mathbb Z$ and $b\in8\mathbb Z$ is restricted such that $f_{0}\in8\mathbb Z$.
To obtain non-trivial solutions we require $m,n,b\neq0$,
and the above choice of fluxes always satisfies the Bianchi identities \eqref{Bianchi_STU}. A solution to the 
equations of motion \eqref{eom_STU} 
is given by
\eq{\label{vacua_STU_restricted}
\tau=\frac{m^3\op \mathsf Q^{0}}{8\op b^2\op n^3}\left(-4\pm i\right),
\hspace{40pt}
U=\frac{m+n}{m}\pm\frac{n}{m}\,i\,,
\hspace{40pt}
T=\frac{m\op \mathsf Q^{0}}{4\op b\op n^2}\left(2\pm i\right),
}
where the sign takes the same value for all three moduli. 
In order for the imaginary parts to be positive we require that this sign is chosen 
appropriately and that $\mathsf Q^0>0$ and $b\, n>0$.

Note that \eqref{vacua_STU_restricted} describes is an infinite set of vacua since
$m,n$ are not bounded, which
is in contrast to the situations studied in
sections~\ref{sec_mod1} and~\ref{sec_iso}.
However, in order to trust these solutions we have 
to require that ${\rm Im}\op \tau,\, {\rm Im}\op U,\,{\rm Im}\op T >1$ which
 translates into the conditions
\eq{
  \label{rel_884949}
  1<\left( \pm \frac{n}{m} \right)^3 < \frac{\mathsf Q^0}{8\op b^2} \,, \hspace{60pt}
  1<\pm \frac{n}{m}  < \frac{\mathsf Q^0}{4\op b\op n}   \,.
}
For a fixed $\mathsf Q^0$ there is only a finite number of choices for $(m,n,b)$ which satisfy
\eqref{rel_884949}, and therefore the number of reliable solutions for the particular flux choice \eqref{fluxes_STU_restricted} 
is finite for fixed $\mathsf Q^0$.
We remark however that duality transformations can change 
the form of \eqref{fluxes_STU_restricted}, and therefore a similar analysis has to be performed for 
the transformed flux choices. 
We do not know whether this leads to a finite number of reliable flux solutions.


\subsection{Space of solutions}

Since the number of physically-distinct solutions for fixed tadpole contributions $\mathsf Q^A$ is
in general infinite, for the present setting we cannot construct a complete data set of flux vacua
for fixed tadpole contribution. 
However, we can generate flux vacua using Monte-Carlo sampling.


\subsubsection*{Data set}

Our data set of flux vacua for the setting described in section~\ref{sec_stu_steting} has
been obtained in the following way:
\begin{itemize}

\item We restrict the contributions to the D3- and D7-brane tadpoles
$\mathsf Q^0$ and $\mathsf Q^1$ shown in equations \eqref{Bianchi_STU} as 
\eq{
  \label{stu-cutoff}
  \left\lvert \frac{\mathsf Q^0}{64} \right\rvert \leq 1000\, ,
  \hspace{40pt}
  \left\lvert \frac{\mathsf Q^1}{64} \right\rvert \leq 1000\,, 
  \hspace{40pt}
  \mathsf Q^{01}=0\,.
}  
Note that due to the flux-quantization condition \eqref{stu_flux_quant} 
the $\mathsf Q^A$ are always a multiple of $64$, and that 
$\mathsf Q^0$ and $\mathsf Q^1$ can be negative while still leading 
to positive imaginary parts for $\tau$, $U$, $T$.

\item The  fluxes in \eqref{IsotropyConditions_Nongeometric} are 
chosen randomly with a uniform distribution. The restriction on the value
of the fluxes reads
\eq{
  \label{stu-cutoff_2}
  \left\lvert \frac{\mbox{flux quantum}}{8} \right\rvert \leq 100\,.
}

\item  We have generated $6.8\cdot 10^6$ flux configurations for which 
1) all moduli $\tau$, $U$, $T$ are fixed, 2) all imaginary parts of 
the moduli fields are strictly positive, and 3) 
the vacua are physically distinct (i.e. not related by 
$SL(2,\mathbb Z)$ transformations of the complex-structure moduli
nor by $T$-transformations of the axio-dilaton or K\"ahler moduli).

\item For these flux contributions all moduli are fixed, however, 
not all of these extrema are stable. The number of 
vacua with all moduli fixed and without tachyonic or flat directions 
is $1.5\cdot 10^6$.

\end{itemize}


\subsubsection*{Solutions at small coupling, large complex structure and large volume}

Since we do not have a complete set of solutions for fixed tadpole contributions 
$\mathsf Q^0$ and $\mathsf Q^1$, the same analysis as for the previous cases cannot 
be performed. However, 
for our data set we have determined the number of solutions $\mathsf N_{\mathsf c}$ for which 
$\tau_2={\rm Im}\op \tau$,
$U_2={\rm Im}\op U$ and
$T_2={\rm Im}\op T$ satisfy
\eq{
  \label{stu_994}
  \tau_2,\,U_2,\, T_2 \,\geq \mathsf c\,.
}
For sufficiently large $\mathsf c$, this corresponds to the weak-coupling, large-complex-structure and large-volume
regime.
We furthermore denote by $|\mathsf Q^A/64|_{\rm min}$ 
the lowest value of the expression $|\mathsf Q^A/64|=\sqrt{(\mathsf Q^0)^2+ 3 (\mathsf Q^1)^2}/64$  in the 
set of vacua determined by \eqref{stu_994}. For configurations which fix all moduli but may contain tachyonic directions 
we find: 
\eq{
 \begin{minipage}{80pt}\fbox{all}\end{minipage}
 \renewcommand{\arraystretch}{1.1}
 \arraycolsep12pt
 \begin{array}{c||cc||c}
 \mathsf c & \mathsf N_{\mathsf c} & \mathsf N_{\mathsf c}/\mathsf N_0 & |\mathsf Q^A/64|_{\rm min} 
 \\
 \hline\hline
 0 & 6.84\cdot10^{6} & 1 & 1 
 \\
 1 & 3.06 \cdot 10^5 & 4.5 \cdot 10^{-2} & 17.7 
 \\
 2 & 1.08 \cdot 10^4 & 1.6 \cdot 10^{-3} & 42.1 
 \\
 3 & 94 & 1.4 \cdot 10^{-5} & 129.8 
 \\
 5 & 3 & 4.4 \cdot 10^{-7} & 811.8 
 \end{array}
}
From here we see that vacua with large imaginary parts $\tau_2, U_2,T_2$ are extremely rare but not 
excluded.  For larger $\mathsf c$ -- that is for more reliable solutions -- the tadpole contributions have to be larger. 
For $\mathsf c\geq 6$ we did not find any solutions, which we expect to be due to the cutoffs
shown in \eqref{stu-cutoff} and \eqref{stu-cutoff_2}. 
These observations are in line with the results of the previous sections.
For the stabilized vacua (without flat or tachyonic directions) we obtain 
a similar behaviour: 
\eq{
 \begin{minipage}{80pt}\fbox{stable}\end{minipage}
 \renewcommand{\arraystretch}{1.1}
 \arraycolsep12pt
 \begin{array}{c||cc||c}
 \mathsf c & \mathsf N_{\mathsf c} & \mathsf N_{\mathsf c}/\mathsf N_0 & |\mathsf Q^A/64|_{\rm min} 
 \\
 \hline\hline
 0 & 1.51\cdot10^{6} & 1 & 1 
 \\
 1 & 2.89 \cdot 10^5 & 1.9 \cdot 10^{-1} & 17.7 
 \\
 2 & 1.04 \cdot 10^4 & 6.9 \cdot 10^{-3} & 42.1 
 \\
 3 & 82 & 5.4 \cdot 10^{-5} & 129.8 
 \\
 5 & 0 & 0 & 
 \end{array}
}
For $\mathsf c\geq 4$ we did not find any stable solutions in our data set, and
in table~\ref{table_examples_stu} we have collected some concrete examples for fully 
stabilized vacua  with imaginary parts greater than one.

\begin{sidewaystable}
\eq{\nonumber
\renewcommand{\arraystretch}{1.1}
\arraycolsep9pt
\begin{array}{c||c|c|c|c|c|c}
\mbox{data} & \mbox{example 1} &  \mbox{example 2}&  \mbox{example 3}&  \mbox{example 4}&  \mbox{example 5}&  \mbox{example 6}
\\ \hline \hline
f_0 &  -40 &224 &-336&-112&176 & 296\\
f_1 &  -88 & 24&96&104&24 & -368\\
f^0 &  -56 &8&32&40&48& 8\\
f^1 &  0& 0&24&0&24&-16\\
\hline
h_0 & -88& -64&-16&-96&72&200\\
h_1 &  0 &64&96&0&0&624\\
h^0 &  0& -16&-24&0&0&-192\\
h^1 &  0&0&-8&-8&0&-24\\
\hline
q^{01} & 0 & 8&8&0&0&72\\
q^{11} &  56&-104&112&24&-136&64\\
\tilde q^{11} & -16&48&-64&-16&48&0\\
q_0{}^1 & 72 &64&80&48&-112&-192\\
q_1{}^1 &  -64&8&96&-32&-40&488\\
\tilde q_1{}^1 & 32 &-56&-64&16&24&-608\\
\hline
\mathsf Q^0/64 & -77 &-48&-10&21&-54&-31\\
\mathsf Q^1/64 &  30&17&-94&-43&66&-193\\
\hline
\tau & +0.45+19.3\op i & -0.02+4.39\op i & -0.40+3.42\op i& +0.47 + 8.25\op i&+0.32 + 12.1\op i & -0.48 + 5.15\op i\\
U & \hphantom{+0.00}+3.53\op i& -0.33 + 3.52\op i&-0.10 + 3.13\op i& \hphantom{+0.47}+3.04 \op i&+0.10+3.13\op i&+0.32+3.14\op i\\
T & \hphantom{+0.00} +7.48\op i & +0.20+ 3.03\op i & +0.30+4.07\op i & -4.47 + 8.49 \op i &-0.48 + 3.10\op i & -0.26 + 4.57\op i
\end{array}
}
\caption{Examples for stable vacua (no tachyonic or flat directions) with $\tau_2,U_2,T_2>1$.\label{table_examples_stu}}
\end{sidewaystable}


\subsubsection*{Distribution of solutions in the $\mathsf Q^A$-plane}

For moduli stabilization of the axio-dilaton and complex-structure moduli studied in 
sections~\ref{sec_mod1} and \ref{sec_iso} we observed that the tadpole-contribution $\mathsf Q^0$ has
to be positive in order to obtain physical solutions with positive imaginary parts 
$\tau_2$ and $U_2$. 
However, when including non-geometric fluxes we see that 
positive as well as negative values of $\mathsf Q^0$ and $\mathsf Q^1$ can results in 
positive imaginary parts $\tau_2$, $U_2$ and $T_2$.

Having a large data set available, we have analyzed the distributions of vacua 
over $\mathsf Q^0$ and $\mathsf Q^1$. For the set of 
 stable vacua without flat or tachyonic directions we obtain
\eq{
 \begin{minipage}{110pt}\fbox{stable}\end{minipage}
 \renewcommand{\arraystretch}{1.1}
 \arraycolsep12pt
 \begin{array}{cc||c}
 \mathsf Q^0 & \mathsf Q^1 & \mbox{fraction of all vacua} \\
 \hline\hline
 \leq0 & \leq0 &0.51862
 \\
 \leq0 & >0 &0.15550
 \\
 >0 & \leq0 &0.32573
\\
>0 & >0 &0.00015
 \end{array}
}
It is somewhat surprising that the fraction of vacua with tadpole contributions
$\mathsf Q^0>0$ and $\mathsf Q^1>0$ 
is suppressed by three orders of magnitude compared to having at least one $\mathsf Q^A$ negative,
but we have no explanation for that. 
For our data set of solutions which include potentially tachyonic 
directions no such difference is found
\eq{
 \begin{minipage}{110pt}\fbox{all}\end{minipage}
 \renewcommand{\arraystretch}{1.1}
 \arraycolsep12pt
 \begin{array}{cc||c}
 \mathsf Q^0 & \mathsf Q^1 & \mbox{fraction of all vacua} \\
 \hline\hline
 \leq0 & \leq0 &0.36058
 \\
 \leq0 & >0 &0.25717
 \\
 >0 & \leq0 &0.25454
\\
>0 & >0 &0.12771
 \end{array}
}
We also note that both data sets do not contain any solution with $\mathsf Q^0=\mathsf Q^1=0$.


\subsubsection*{Distribution of solutions in moduli space}

We have also analyzed the distribution of solutions to the F-term equations \eqref{f_term_eq}
within the moduli space. Since the density of solutions is very small, we were not 
able to identify any patterns or structures.


\subsection{Summary}

We briefly summarize the main results obtained in this section 
for the combined moduli stabilization of the axio-dilaton, complex-structure
moduli and K\"ahler moduli by the fluxes shown in equation \eqref{IsotropyConditions_Nongeometric}:
\begin{itemize}

\item For fixed D3- and D7-brane tadpole contributions $\mathsf Q^0$ and $\mathsf Q^1$
the number of physically-distinct vacua is in general infinite \cite{Shelton:2006fd}. We therefore
were not able to generate a complete data set but used Monte-Carlo methods 
to randomly generate $6.8\cdot 10^6$ solutions to the F-term equations which fix all moduli.

\item We have shown that solutions at weak string coupling, large complex structure and 
large volume are only a small fraction of all vacua. 
For instance, stable solutions with $\tau_2,U_2,T_2\geq 5$ make only a fraction of $4\cdot 10^{-7}$
of all solutions. 
Requiring the solutions to be more reliable requires the tadpole contributions $\mathsf Q^A$ to be larger,
which is in tension with the tadpole-cancellation condition as discussed on page \pageref{page_ma}.

\item In table~\ref{table_examples_stu} we have  shown some concrete examples of stable vacua 
with axio-dilaton, complex-structure moduli and K\"ahler moduli fixed 
at imaginary parts greater than one.

\item Finally, we have pointed out that stable vacua with all tadpole contributions $\mathsf Q^0$ 
and $\mathsf Q^1$ positive 
are statistically disfavored. We do not have an explanation for this observation.

\end{itemize}


\clearpage
\section{Discussion}
\label{sec_disc}

In this work we have studied moduli stabilization with 
R-R and NS-NS fluxes in type IIB string theory for the example of the 
$\mathbb T^6/\mathbb Z_2\times \mathbb Z_2$ orientifold. 
We have analyzed the interplay between moduli stabilization 
and tadpole cancellation, in particular, we have shown
how properties of the vacua depend on 
the flux contribution to the tadpole-cancellation condition.


\subsubsection*{Summary of results}

More concretely, the axio-dilaton and complex-structure moduli are fixed by 
geometric fluxes while the K\"ahler moduli are fixed at tree-level using 
non-geometric $Q$-flux. 
In section~\ref{sec_mod1} we have focussed on the axio-dilaton only
and mainly ignored the complex-structure and K\"ahler moduli. 
In section~\ref{sec_iso} we included the complex-structure moduli, and 
in section~\ref{sec_stu}  we studied moduli stabilization for 
all closed-string moduli.
We analyzed the space of solutions to the F-term equations for these settings and found that it is not 
homogenous:
\begin{itemize}

\item For the axio-dilaton the space of solutions contains characteristic 
void structures (see  figure~\ref{DouglasPlot_Plane}) \cite{Denef:2004ze,DeWolfe:2004ns}. The radius of 
these voids depends on the flux contribution $\mathsf Q^0$ to the tadpole-cancellation
condition, and for larger $\mathsf Q^0$ the radii become smaller. 

When including the complex-structure moduli, we observe that vacua are 
accumulated on submanifolds within the space of solutions (see figure~\ref{IsotropicTorus_3DPlots}).
On these planes we again find void structures, which are connected by lines
between different planes. We therefore find cylindrical voids 
in (three-dimensional sections of) this four-dimensional space of solutions.

\end{itemize}
Furthermore, in section~\ref{sec_tadpole} we have argued that the flux contribution
to the tadpole-cancellation condition cannot be arbitrarily large. In particular, for 
many known examples this contribution is  small. 
We have then contrasted this observation with the requirement of having reliable solutions
at weak string-coupling, large complex structure  and large volume:
\begin{itemize}

\item We have seen that the fraction of vacua with small string coupling $\tau_2\gg1$, 
large complex structure $U_2\gg1$ and large volume $T_2\gg1$ is small. 
For instance, within the approach followed in this paper
around $20\%$ of the solutions satisfy $\tau_2\geq 5$,
around $0.4\%$ of the solutions satisfy $\tau_2,U_2\geq 5$,
and a fraction of around $10^{-7}$ of the solutions satisfy $\tau_2,U_2,T_2\geq 5$.
This suggests that for a large number of moduli, only a very small 
fraction of the solutions can be trusted (within the tree-level approach used in this work).

\item We have also observed that in order to find vacua at 
weak string-coupling, large complex structure  and large volume
the flux contribution to the tad\-pole-cancellation condition has to be large. 
For instance, within the approach followed in this paper 
for $\tau_2\geq 5$ one needs $\mathsf Q^0\geq 3840$, 
for $\tau_2,U_2\geq 5$ one needs $\mathsf Q^0\geq 38208$, 
and for $\tau_2,U_2,T_2\geq 5$ we have indications that one needs $|\mathsf Q^A|\gtrsim \mathcal O(10^4)$. 
This suggests that in order to stabilize a large number of moduli in a 
perturbatively-controlled regime a large flux contribution is needed. 
However, this conclusion is in stark contrast to the tadpole-cancellation condition 
which strongly disfavors large flux contributions.

\end{itemize}
To conclude, in order to stabilize moduli in a reliable way a large flux contribution is needed --- which
is however strongly restricted by the tadpole-cancellation condition. 
We therefore see that moduli stabilization and model building in string theory cannot be 
approached independently but have to be addressed simultaneously. This is a difficult task.


\subsubsection*{Limitations and future directions}

We now comment on the limitations of the analysis performed in this paper and on future directions:
\begin{itemize}

\item Our conclusions in this work are based on the study of a single compactification space. 
We believe that the $\mathbb T^6/\mathbb Z_2\times \mathbb Z_2$ orientifold captures
main features of the problem, but these have to be confirmed by other examples. 
We are planning to address this point in the future.

\item In this work we have stabilized moduli at tree-level.
Corrections to the effective theory can usually be ignored in the 
weak-coupling, large-complex-structure and large-volume regime, however,
many of the obtained solutions are not in this regime. 
We therefore should repeat our analysis and include 
various corrections from the start, which in turn will modify the space of 
solutions.

\item We have found that only a small fraction of solutions stabilize moduli in a perturbatively-controlled 
regime. This observation has implications for the landscape of string vacua, in particular, it suggests
that the landscape may be smaller than naively expected. It would be desirable to 
make this statement more precise. 

\item The $SL(2,\mathbb Z)$ duality of the axio-dilaton was broken by non-geometric 
$Q$-fluxes. Including so-called $P$-fluxes will restore this duality and may help
in showing that the corresponding  space of solutions is finite.

\item The contribution of orientifold planes to the tadpole-cancellation condition 
could only be estimated based on known examples. It would be desirable to 
have a criterium which can put a bound on the orientifold contribution 
for a particular compactification space.

\end{itemize}


\vskip2em
\subsubsection*{Acknowledgements}

The authors want to thank 
R.~Blumenhagen, 
M.~Haack,
D.~Klaewer, 
S.~Krippendorf,
C.~Long,
L.~McAllister and
L.~Schlechter 
for very helpful discussions.
EP thanks the 
Lorentz Institute for Theoretical Physics at the University in Leiden
for its hospitality, and 
he thanks the organizers of the 
BIRS-CMO workshop ``Geometrical Tools for String Cosmology'' in Oaxaca 
where preliminary results of this work have first been presented.


\newpage 
\appendix

\section{Finite number of solutions for setting II}
\label{app_proof_tu}

In this appendix we follow the proof of \cite{Kachru:2002he,DeWolfe:2004ns},
that for the setting of section~\ref{sec_tu_set} the number of physically-distinct solutions 
is finite for fixed $\mathsf Q^0$.
The important property for showing this are the 
$SL\left(2,\mathbb{Z}\right)$ dualities of the axio-dilaton and 
complex-structure moduli summarized in section~\ref{sec_dualities}.
Splitting the moduli into real and imaginary parts as
\eq{\label{ModuliSplit_IsotropicTorus}
\tau = \tau_{1}+ i\tau_{2}\,,\hspace{50pt} U= U_{1}+ iU_{2}\,,
}
we recall that the two equations \eqref{IsotropicTorus_EoMIsotropic1-2} define an overdetermined cubic system
for  $U$ and therefore do not allow for a closed-form solution in the
generic case. 
We will now follow the lines of \cite{Kachru:2002he,DeWolfe:2004ns} to demonstrate how a closed solution can still
be obtained for the physically relevant cases

In order for a physical solution to exist, 
both equations have to share a common root with non-vanishing
imaginary part. Since all coefficients  are real,
there  exists a second solution given by its complex conjugate
and the two equations share a common quadratic factor. 
The two cubic polynomials \eqref{IsotropicTorus_EoMIsotropic1-2} can then be factorized as 
\eq{\label{IsotropicTorus_FactorizedPolynomials}
\arraycolsep2pt
\begin{array}{c@{\op}c@{\op}ccc@{\op}c@{\op}lcl}
(&r&U&+&s&)&P(U) & =&0\,,
\\[4pt]
(&u&U&+&v&)&P(U) & =&0\,,
\end{array}
}
where $P(U)$ defines the common quadratic factor,
\eq{
P(U)=l\op (U)^{2}+m\op U+n\,,
}
and the seven new variables $l,m,n,r,s,u,v\in\mathbb{Z}$ are defined by an overdetermined system
of equations
\eq{
  \arraycolsep2pt
  \begin{array}{l@{\op}lcl@{\op}lcl@{\op}l@{\hspace{60pt}}l@{\op}lcc@{\op}l@{\op}l}
  r& m&+&s&l  &=&-3\op f^{1}&, & r&l  &=&&f^{0}&,
  \\[4pt]
  r&n&+&s&m  &=&-3\op f_{1}&, &s&n &=&-&f_{0}&,
  \\[4pt]
  u&m&+&v&l  &=&-3\op h^{1}&, &u&l &=&&h^{0}&,
  \\[4pt]
    u&n&+&v&m &=&-3\op h_{1}&, & v&n &=&-&h_{0}&.
  \end{array}
} 
The set of admissible septuples is
furthermore restricted by requiring the flux quanta to 
to satisfy the tadpole cancellation condition \eqref{IsotropicTorus_Tadpole2}, which
can be reformulated as 
\eq{\label{IsotropicTorus_Tadpole3}
\left(\vphantom{f^{1}}rv-su\right)\left(\vphantom{f^{1}}m^{2}-4ln\right)=-3\op\mathsf Q^{0}\,.
}
As  shown in \cite{Kachru:2002he}, this condition can only be satisfied if
$\mathsf Q^{0}$ is a multiple of three, yielding an overall factor of 192 when taking into account
the flux quantization conditions.
Since the prefactors appearing in
\eqref{IsotropicTorus_FactorizedPolynomials} are linear in $U$ with real coefficients,  the two solutions 
with non-vanishing imaginary part can be obtained by choosing $U$ such that
\eq{
P(U)=0\,.
}
Requiring furthermore the imaginary part of $U$ to be positive,
we  arrive at the physical solutions 
\eq{\label{IsotropicTorus_CSModulusSolution}
U  =\frac{-m+\sqrt{m^{2}-4ln}}{2l}\hspace{40pt}  \mbox{if }l>0\mbox{ and }n>0\,,
\\[4pt]
U  =\frac{-m-\sqrt{m^{2}-4ln}}{2l}\hspace{40pt}  \mbox{if }l<0\mbox{ and }n<0\,.
}
The F-term equation  \eqref{IsotropicTorus_EoMIsotropic2-2} is linear in $\tau$ and can
be solved analytically, leading to the stabilized value 
\eq{\label{IsotropicTorus_AxioDilatonModulusSolution}
\tau=\frac{s\left(m+2\op l\op U\right)+r\bigl[n+U\left(2\op m+3\op l\op U\right)\bigr]}{v\left(m+2\op l\op U\right)+u\bigl[n+U\left(2\op m+3\op l\op U\right)\bigr]}\,.
}
We will now proceed similarly to section~\ref{sec_t_mod_finite} to show that 
using the dualities for the axio-dilaton and complex-structure moduli, 
for fixed $\mathsf Q^0$ only a finite number of solutions can be found.
Without loss of generality we focus on the case $l>0\textrm{ and }n>0$, but the 
situation  $\textrm{ }l<0\textrm{ and }n<0$ is completely
analogous. 
\begin{itemize}
\item As can be read-off from the first line in \eqref{IsotropicTorus_CSModulusSolution}, the shift symmetry
\eqref{CS-transformations_T} of $U$ gives rise to an equivalence 
\eq{
m\sim m+2\op b\op l \,, \hspace{40pt} b\in\mathbb{Z}\,.
}
It therefore follows that all inequivalent values of $m$ are contained in the range
\eq{
m=-l,\ldots, l-1\,.
}

\item Considering the boundary $U_2=-\frac{1}{2}$, a minimal requirement for $U$ to be located in the fundamental
domain $\mathcal{F}_{U}$ is given by $U_2\geq\sqrt{3}/2$. This is equivalent to requiring
\eq{
m^{2}-4\op l\op n\leq-3\op  l^{2} \,.
}
On the other hand, both of the factors on the left-hand
side of the tadpole-cancellation condition \eqref{IsotropicTorus_Tadpole3} have to be integers, 
giving rise to a lower bound
\eq{
m^{2}-4\op l\op n\geq-3\op \mathsf Q^{0}\,.
} 
This restricts the inequivalent values of both $l$ and $n$ to finite ranges
\eq{
1\leq l\leq\sqrt{\mathsf Q^{0}}\,,\hspace{50pt}\frac{3l^{2}+m^{2}}{4l}\leq n\leq\frac{3\mathsf Q^{0}+m^{2}}{4l}\,.
}

\item Employing the same arguments for the axio-dilaton, one finds an additional equivalence 
\eq{
s\sim s+ b\op v\,, \hspace{50pt} b\in\mathbb{Z} \,,
}
as well as upper bounds for $u$ and $v$. 
\item The remaining degree of freedom $r$ is fixed by the tadpole cancellation condition  \eqref{IsotropicTorus_Tadpole3}.
\end{itemize}
 The above conditions leave only a finite number of inequivalent solutions for a fixed D3-tadpole contribution  $\mathsf Q^{0}$.


\clearpage
\nocite{*}
\bibliography{references}

\providecommand{\href}[2]{#2}\begingroup\raggedright\begin{thebibliography}{10}

\bibitem{Blumenhagen:2006ci}
R.~Blumenhagen, B.~K{\"o}rs, D.~L{\"u}st, and S.~Stieberger,
  ``{Four-dimensional String Compactifications with D-Branes, Orientifolds and
  Fluxes},'' {\em Phys. Rept.} {\bf 445} (2007) 1--193,
  \href{http://xxx.lanl.gov/abs/hep-th/0610327}{{\tt hep-th/0610327}}.

\bibitem{Freed:1999vc}
D.~S. Freed and E.~Witten, ``{Anomalies in string theory with D-branes},'' {\em
  Asian J. Math.} {\bf 3} (1999) 819,
  \href{http://xxx.lanl.gov/abs/hep-th/9907189}{{\tt hep-th/9907189}}.

\bibitem{Dasgupta:1999ss}
K.~Dasgupta, G.~Rajesh, and S.~Sethi, ``{M theory, orientifolds and G -
  flux},'' {\em JHEP} {\bf 08} (1999) 023,
  \href{http://xxx.lanl.gov/abs/hep-th/9908088}{{\tt hep-th/9908088}}.

\bibitem{Taylor:1999ii}
T.~R. Taylor and C.~Vafa, ``{R R flux on Calabi-Yau and partial supersymmetry
  breaking},'' {\em Phys. Lett.} {\bf B474} (2000) 130--137,
  \href{http://xxx.lanl.gov/abs/hep-th/9912152}{{\tt hep-th/9912152}}.

\bibitem{Giddings:2001yu}
S.~B. Giddings, S.~Kachru, and J.~Polchinski, ``{Hierarchies from fluxes in
  string compactifications},'' {\em Phys. Rev.} {\bf D66} (2002) 106006,
  \href{http://xxx.lanl.gov/abs/hep-th/0105097}{{\tt hep-th/0105097}}.

\bibitem{Derendinger:2004jn}
J.-P. Derendinger, C.~Kounnas, P.~M. Petropoulos, and F.~Zwirner,
  ``{Superpotentials in IIA compactifications with general fluxes},'' {\em
  Nucl. Phys.} {\bf B715} (2005) 211--233,
  \href{http://xxx.lanl.gov/abs/hep-th/0411276}{{\tt hep-th/0411276}}.

\bibitem{Villadoro:2005cu}
G.~Villadoro and F.~Zwirner, ``{N=1 effective potential from dual type-IIA
  D6/O6 orientifolds with general fluxes},'' {\em JHEP} {\bf 06} (2005) 047,
  \href{http://xxx.lanl.gov/abs/hep-th/0503169}{{\tt hep-th/0503169}}.

\bibitem{DeWolfe:2005uu}
O.~DeWolfe, A.~Giryavets, S.~Kachru, and W.~Taylor, ``{Type IIA moduli
  stabilization},'' {\em JHEP} {\bf 07} (2005) 066,
  \href{http://xxx.lanl.gov/abs/hep-th/0505160}{{\tt hep-th/0505160}}.

\bibitem{Kachru:2003aw}
S.~Kachru, R.~Kallosh, A.~D. Linde, and S.~P. Trivedi, ``{De Sitter vacua in
  string theory},'' {\em Phys. Rev.} {\bf D68} (2003) 046005,
  \href{http://xxx.lanl.gov/abs/hep-th/0301240}{{\tt hep-th/0301240}}.

\bibitem{Balasubramanian:2005zx}
V.~Balasubramanian, P.~Berglund, J.~P. Conlon, and F.~Quevedo, ``{Systematics
  of moduli stabilisation in Calabi-Yau flux compactifications},'' {\em JHEP}
  {\bf 03} (2005) 007, \href{http://xxx.lanl.gov/abs/hep-th/0502058}{{\tt
  hep-th/0502058}}.

\bibitem{Obied:2018sgi}
G.~Obied, H.~Ooguri, L.~Spodyneiko, and C.~Vafa, ``{De Sitter Space and the
  Swampland},'' \href{http://xxx.lanl.gov/abs/1806.08362}{{\tt 1806.08362}}.

\bibitem{Blumenhagen:2007sm}
R.~Blumenhagen, S.~Moster, and E.~Plauschinn, ``{Moduli Stabilisation versus
  Chirality for MSSM like Type IIB Orientifolds},'' {\em JHEP} {\bf 01} (2008)
  058, \href{http://xxx.lanl.gov/abs/0711.3389}{{\tt 0711.3389}}.

\bibitem{Ashok:2003gk}
S.~Ashok and M.~R. Douglas, ``{Counting flux vacua},'' {\em JHEP} {\bf 01}
  (2004) 060, \href{http://xxx.lanl.gov/abs/hep-th/0307049}{{\tt
  hep-th/0307049}}.

\bibitem{Douglas:2003um}
M.~R. Douglas, ``{The Statistics of string / M theory vacua},'' {\em JHEP} {\bf
  05} (2003) 046, \href{http://xxx.lanl.gov/abs/hep-th/0303194}{{\tt
  hep-th/0303194}}.

\bibitem{Douglas:2004kc}
M.~R. Douglas, B.~Shiffman, and S.~Zelditch, ``{Critical points and
  supersymmetric vacua, II: Asymptotics and extremal metrics},'' {\em J. Diff.
  Geom.} {\bf 72} (2006), no.~3 381--427,
  \href{http://xxx.lanl.gov/abs/math/0406089}{{\tt math/0406089}}.

\bibitem{Denef:2004ze}
F.~Denef and M.~R. Douglas, ``{Distributions of flux vacua},'' {\em JHEP} {\bf
  05} (2004) 072, \href{http://xxx.lanl.gov/abs/hep-th/0404116}{{\tt
  hep-th/0404116}}.

\bibitem{Giryavets:2004zr}
A.~Giryavets, S.~Kachru, and P.~K. Tripathy, ``{On the taxonomy of flux
  vacua},'' {\em JHEP} {\bf 08} (2004) 002,
  \href{http://xxx.lanl.gov/abs/hep-th/0404243}{{\tt hep-th/0404243}}.

\bibitem{Douglas:2004zg}
M.~R. Douglas, ``{Basic results in vacuum statistics},'' {\em Comptes Rendus
  Physique} {\bf 5} (2004) 965--977,
  \href{http://xxx.lanl.gov/abs/hep-th/0409207}{{\tt hep-th/0409207}}.

\bibitem{Conlon:2004ds}
J.~P. Conlon and F.~Quevedo, ``{On the explicit construction and statistics of
  Calabi-Yau flux vacua},'' {\em JHEP} {\bf 10} (2004) 039,
  \href{http://xxx.lanl.gov/abs/hep-th/0409215}{{\tt hep-th/0409215}}.

\bibitem{DeWolfe:2004ns}
O.~DeWolfe, A.~Giryavets, S.~Kachru, and W.~Taylor, ``{Enumerating flux vacua
  with enhanced symmetries},'' {\em JHEP} {\bf 02} (2005) 037,
  \href{http://xxx.lanl.gov/abs/hep-th/0411061}{{\tt hep-th/0411061}}.

\bibitem{Denef:2004cf}
F.~Denef and M.~R. Douglas, ``{Distributions of nonsupersymmetric flux
  vacua},'' {\em JHEP} {\bf 03} (2005) 061,
  \href{http://xxx.lanl.gov/abs/hep-th/0411183}{{\tt hep-th/0411183}}.

\bibitem{Eguchi:2005eh}
T.~Eguchi and Y.~Tachikawa, ``{Distribution of flux vacua around singular
  points in Calabi-Yau moduli space},'' {\em JHEP} {\bf 01} (2006) 100,
  \href{http://xxx.lanl.gov/abs/hep-th/0510061}{{\tt hep-th/0510061}}.

\bibitem{Shelton:2006fd}
J.~Shelton, W.~Taylor, and B.~Wecht, ``{Generalized Flux Vacua},'' {\em JHEP}
  {\bf 02} (2007) 095, \href{http://xxx.lanl.gov/abs/hep-th/0607015}{{\tt
  hep-th/0607015}}.

\bibitem{MartinezPedrera:2012rs}
D.~Martinez-Pedrera, D.~Mehta, M.~Rummel, and A.~Westphal, ``{Finding all flux
  vacua in an explicit example},'' {\em JHEP} {\bf 06} (2013) 110,
  \href{http://xxx.lanl.gov/abs/1212.4530}{{\tt 1212.4530}}.

\bibitem{Dibitetto:2011gm}
G.~Dibitetto, A.~Guarino, and D.~Roest, ``{Charting the landscape of N=4 flux
  compactifications},'' {\em JHEP} {\bf 03} (2011) 137,
  \href{http://xxx.lanl.gov/abs/1102.0239}{{\tt 1102.0239}}.

\bibitem{Acharya:2005ez}
B.~S. Acharya, F.~Denef, and R.~Valandro, ``{Statistics of M theory vacua},''
  {\em JHEP} {\bf 06} (2005) 056,
  \href{http://xxx.lanl.gov/abs/hep-th/0502060}{{\tt hep-th/0502060}}.

\bibitem{Watari:2015ysa}
T.~Watari, ``{Statistics of F-theory flux vacua for particle physics},'' {\em
  JHEP} {\bf 11} (2015) 065, \href{http://xxx.lanl.gov/abs/1506.08433}{{\tt
  1506.08433}}.

\bibitem{Cirafici:2015pky}
M.~Cirafici, ``{Persistent Homology and String Vacua},'' {\em JHEP} {\bf 03}
  (2016) 045, \href{http://xxx.lanl.gov/abs/1512.01170}{{\tt 1512.01170}}.

\bibitem{Cole:2018emh}
A.~Cole and G.~Shiu, ``{Topological Data Analysis for the String Landscape},''
  {\em JHEP} {\bf 03} (2019) 054,
  \href{http://xxx.lanl.gov/abs/1812.06960}{{\tt 1812.06960}}.

\bibitem{Halverson:2017vde}
J.~Halverson, C.~Long, and B.~Sung, ``{On the Scarcity of Weak Coupling in the
  String Landscape},'' {\em JHEP} {\bf 02} (2018) 113,
  \href{http://xxx.lanl.gov/abs/1710.09374}{{\tt 1710.09374}}.

\bibitem{Bena:2018fqc}
I.~Bena, E.~Dudas, M.~Gra\~na, and S.~L{\"u}st, ``{Uplifting Runaways},'' {\em
  Fortsch. Phys.} {\bf 67} (2019), no.~1-2 1800100,
  \href{http://xxx.lanl.gov/abs/1809.06861}{{\tt 1809.06861}}.

\bibitem{Blaback:2018hdo}
J.~Blaback, U.~Danielsson, and G.~Dibitetto, ``{A new light on the darkest
  corner of the landscape},'' \href{http://xxx.lanl.gov/abs/1810.11365}{{\tt
  1810.11365}}.

\bibitem{CaboBizet:2019sku}
N.~Cabo~Bizet, C.~Damian, O.~Loaiza-Brito, and D.~M. Peña, ``{Leaving the
  Swampland: Non-geometric fluxes and the Distance Conjecture},''
  \href{http://xxx.lanl.gov/abs/1904.11091}{{\tt 1904.11091}}.

\bibitem{Grimm:2004uq}
T.~W. Grimm and J.~Louis, ``{The Effective action of N = 1 Calabi-Yau
  orientifolds},'' {\em Nucl. Phys.} {\bf B699} (2004) 387--426,
  \href{http://xxx.lanl.gov/abs/hep-th/0403067}{{\tt hep-th/0403067}}.

\bibitem{Benmachiche:2006df}
I.~Benmachiche and T.~W. Grimm, ``{Generalized N=1 orientifold
  compactifications and the Hitchin functionals},'' {\em Nucl. Phys.} {\bf
  B748} (2006) 200--252, \href{http://xxx.lanl.gov/abs/hep-th/0602241}{{\tt
  hep-th/0602241}}.

\bibitem{Shelton:2005cf}
J.~Shelton, W.~Taylor, and B.~Wecht, ``{Nongeometric flux compactifications},''
  {\em JHEP} {\bf 10} (2005) 085,
  \href{http://xxx.lanl.gov/abs/hep-th/0508133}{{\tt hep-th/0508133}}.

\bibitem{Blumenhagen:2015lta}
R.~Blumenhagen, A.~Font, and E.~Plauschinn, ``{Relating double field theory to
  the scalar potential of N = 2 gauged supergravity},'' {\em JHEP} {\bf 12}
  (2015) 122, \href{http://xxx.lanl.gov/abs/1507.08059}{{\tt 1507.08059}}.

\bibitem{Grana:2005jc}
M.~Grana, ``{Flux compactifications in string theory: A Comprehensive
  review},'' {\em Phys. Rept.} {\bf 423} (2006) 91--158,
  \href{http://xxx.lanl.gov/abs/hep-th/0509003}{{\tt hep-th/0509003}}.

\bibitem{Villadoro:2006ia}
G.~Villadoro and F.~Zwirner, ``{D terms from D-branes, gauge invariance and
  moduli stabilization in flux compactifications},'' {\em JHEP} {\bf 03} (2006)
  087, \href{http://xxx.lanl.gov/abs/hep-th/0602120}{{\tt hep-th/0602120}}.

\bibitem{Betzler:2017kme}
P.~Betzler and E.~Plauschinn, ``{Dimensional reductions of DFT and mirror
  symmetry for Calabi-Yau three-folds and $K3\times T^{2}$},'' {\em Nucl.
  Phys.} {\bf B933} (2018) 384--432,
  \href{http://xxx.lanl.gov/abs/1712.08382}{{\tt 1712.08382}}.

\bibitem{Plauschinn:2018wbo}
E.~Plauschinn, ``{Non-geometric backgrounds in string theory},'' {\em Phys.
  Rept.} {\bf 798} (2019) 1--122,
  \href{http://xxx.lanl.gov/abs/1811.11203}{{\tt 1811.11203}}.

\bibitem{Kiritsis:2007zza}
E.~Kiritsis, {\em {String theory in a nutshell}}.
\newblock Princeton University Press, 2007.

\bibitem{Blumenhagen:2009zz}
R.~Blumenhagen and E.~Plauschinn, ``{Introduction to conformal field theory},''
  {\em Lect. Notes Phys.} {\bf 779} (2009) 1--256.

\bibitem{Sagnotti:1992qw}
A.~Sagnotti, ``{A Note on the Green-Schwarz mechanism in open string
  theories},'' {\em Phys. Lett.} {\bf B294} (1992) 196--203,
  \href{http://xxx.lanl.gov/abs/hep-th/9210127}{{\tt hep-th/9210127}}.

\bibitem{Minasian:1997mm}
R.~Minasian and G.~W. Moore, ``{K theory and Ramond-Ramond charge},'' {\em
  JHEP} {\bf 11} (1997) 002, \href{http://xxx.lanl.gov/abs/hep-th/9710230}{{\tt
  hep-th/9710230}}.

\bibitem{Aspinwall:2004jr}
P.~S. Aspinwall, ``{D-branes on Calabi-Yau manifolds},'' in {\em {Progress in
  string theory. Proceedings, Summer School, TASI 2003, Boulder, USA, June
  2-27, 2003}}, pp.~1--152, 2004.
\newblock \href{http://xxx.lanl.gov/abs/hep-th/0403166}{{\tt hep-th/0403166}}.

\bibitem{Blumenhagen:2015kja}
R.~Blumenhagen, A.~Font, M.~Fuchs, D.~Herschmann, E.~Plauschinn, Y.~Sekiguchi,
  and F.~Wolf, ``{A Flux-Scaling Scenario for High-Scale Moduli Stabilization
  in String Theory},'' {\em Nucl. Phys.} {\bf B897} (2015) 500--554,
  \href{http://xxx.lanl.gov/abs/1503.07634}{{\tt 1503.07634}}.

\bibitem{Plauschinn:2008yd}
E.~Plauschinn, ``{The Generalized Green-Schwarz Mechanism for Type IIB
  Orientifolds with D3- and D7-Branes},'' {\em JHEP} {\bf 05} (2009) 062,
  \href{http://xxx.lanl.gov/abs/0811.2804}{{\tt 0811.2804}}.

\bibitem{Lust:2006zh}
D.~L{\"u}st, S.~Reffert, E.~Scheidegger, and S.~Stieberger, ``{Resolved
  Toroidal Orbifolds and their Orientifolds},'' {\em Adv. Theor. Math. Phys.}
  {\bf 12} (2008), no.~1 67--183,
  \href{http://xxx.lanl.gov/abs/hep-th/0609014}{{\tt hep-th/0609014}}.

\bibitem{Blumenhagen:2008zz}
R.~Blumenhagen, V.~Braun, T.~W. Grimm, and T.~Weigand, ``{GUTs in Type IIB
  Orientifold Compactifications},'' {\em Nucl. Phys.} {\bf B815} (2009) 1--94,
  \href{http://xxx.lanl.gov/abs/0811.2936}{{\tt 0811.2936}}.

\bibitem{Candelas:1997eh}
P.~Candelas, E.~Perevalov, and G.~Rajesh, ``{Toric geometry and enhanced gauge
  symmetry of F theory / heterotic vacua},'' {\em Nucl. Phys.} {\bf B507}
  (1997) 445--474, \href{http://xxx.lanl.gov/abs/hep-th/9704097}{{\tt
  hep-th/9704097}}.

\bibitem{Lynker:1998pb}
M.~Lynker, R.~Schimmrigk, and A.~Wisskirchen, ``{Landau-Ginzburg vacua of
  string, M theory and F theory at c = 12},'' {\em Nucl. Phys.} {\bf B550}
  (1999) 123--150, \href{http://xxx.lanl.gov/abs/hep-th/9812195}{{\tt
  hep-th/9812195}}.

\bibitem{Taylor:2015xtz}
W.~Taylor and Y.-N. Wang, ``{The F-theory geometry with most flux vacua},''
  {\em JHEP} {\bf 12} (2015) 164,
  \href{http://xxx.lanl.gov/abs/1511.03209}{{\tt 1511.03209}}.

\bibitem{Junghans:2013xza}
D.~Junghans, {\em {Backreaction of Localised Sources in String
  Compactifications}}.
\newblock PhD thesis, Leibniz U., Hannover, 2013.
\newblock \href{http://xxx.lanl.gov/abs/1309.5990}{{\tt 1309.5990}}.

\bibitem{Font:1988mk}
A.~Font, L.~E. Ibanez, and F.~Quevedo, ``{$Z(N$) X $Z$(m) Orbifolds and
  Discrete Torsion},'' {\em Phys. Lett.} {\bf B217} (1989) 272--276.

\bibitem{Berkooz:1996dw}
M.~Berkooz and R.~G. Leigh, ``{A D = 4 N=1 orbifold of type I strings},'' {\em
  Nucl. Phys.} {\bf B483} (1997) 187--208,
  \href{http://xxx.lanl.gov/abs/hep-th/9605049}{{\tt hep-th/9605049}}.

\bibitem{Antoniadis:1996vw}
I.~Antoniadis, C.~Bachas, C.~Fabre, H.~Partouche, and T.~R. Taylor, ``{Aspects
  of type I - type II - heterotic triality in four-dimensions},'' {\em Nucl.
  Phys.} {\bf B489} (1997) 160--178,
  \href{http://xxx.lanl.gov/abs/hep-th/9608012}{{\tt hep-th/9608012}}.

\bibitem{Lust:2005dy}
D.~L{\"u}st, S.~Reffert, W.~Schulgin, and S.~Stieberger, ``{Moduli
  stabilization in type IIB orientifolds (I): Orbifold limits},'' {\em Nucl.
  Phys.} {\bf B766} (2007) 68--149,
  \href{http://xxx.lanl.gov/abs/hep-th/0506090}{{\tt hep-th/0506090}}.

\bibitem{Frey:2002hf}
A.~R. Frey and J.~Polchinski, ``{N=3 warped compactifications},'' {\em Phys.
  Rev.} {\bf D65} (2002) 126009,
  \href{http://xxx.lanl.gov/abs/hep-th/0201029}{{\tt hep-th/0201029}}.

\bibitem{Kachru:2002he}
S.~Kachru, M.~B. Schulz, and S.~Trivedi, ``{Moduli stabilization from fluxes in
  a simple IIB orientifold},'' {\em JHEP} {\bf 10} (2003) 007,
  \href{http://xxx.lanl.gov/abs/hep-th/0201028}{{\tt hep-th/0201028}}.

\bibitem{Blumenhagen:2003vr}
R.~Blumenhagen, D.~L{\"u}st, and T.~R. Taylor, ``{Moduli stabilization in
  chiral type IIB orientifold models with fluxes},'' {\em Nucl. Phys.} {\bf
  B663} (2003) 319--342, \href{http://xxx.lanl.gov/abs/hep-th/0303016}{{\tt
  hep-th/0303016}}.

\bibitem{Cascales:2003zp}
J.~F.~G. Cascales and A.~M. Uranga, ``{Chiral 4d string vacua with D branes and
  NSNS and RR fluxes},'' {\em JHEP} {\bf 05} (2003) 011,
  \href{http://xxx.lanl.gov/abs/hep-th/0303024}{{\tt hep-th/0303024}}.

\bibitem{Vafa:1986wx}
C.~Vafa, ``{Modular Invariance and Discrete Torsion on Orbifolds},'' {\em Nucl.
  Phys.} {\bf B273} (1986) 592--606.

\bibitem{Buscher:1987sk}
T.~H. Buscher, ``{A Symmetry of the String Background Field Equations},'' {\em
  Phys. Lett.} {\bf B194} (1987) 59--62.

\bibitem{Buscher:1987qj}
T.~H. Buscher, ``{Path Integral Derivation of Quantum Duality in Nonlinear
  Sigma Models},'' {\em Phys. Lett.} {\bf B201} (1988) 466--472.

\bibitem{Plauschinn:2014nha}
E.~Plauschinn, ``{On T-duality transformations for the three-sphere},'' {\em
  Nucl. Phys.} {\bf B893} (2015) 257--286,
  \href{http://xxx.lanl.gov/abs/1408.1715}{{\tt 1408.1715}}.

\bibitem{Hassan:1999bv}
S.~F. Hassan, ``{T duality, space-time spinors and RR fields in curved
  backgrounds},'' {\em Nucl. Phys.} {\bf B568} (2000) 145--161,
  \href{http://xxx.lanl.gov/abs/hep-th/9907152}{{\tt hep-th/9907152}}.

\bibitem{Aldazabal:2006up}
G.~Aldazabal, P.~G. Camara, A.~Font, and L.~E. Ibanez, ``{More dual fluxes and
  moduli fixing},'' {\em JHEP} {\bf 05} (2006) 070,
  \href{http://xxx.lanl.gov/abs/hep-th/0602089}{{\tt hep-th/0602089}}.

\bibitem{Aldazabal:2008zza}
G.~Aldazabal, P.~G. Camara, and J.~A. Rosabal, ``{Flux algebra, Bianchi
  identities and Freed-Witten anomalies in F-theory compactifications},'' {\em
  Nucl. Phys.} {\bf B814} (2009) 21--52,
  \href{http://xxx.lanl.gov/abs/0811.2900}{{\tt 0811.2900}}.

\bibitem{Guarino:2008ik}
A.~Guarino and G.~J. Weatherill, ``{Non-geometric flux vacua, S-duality and
  algebraic geometry},'' {\em JHEP} {\bf 02} (2009) 042,
  \href{http://xxx.lanl.gov/abs/0811.2190}{{\tt 0811.2190}}.

\bibitem{Chatzistavrakidis:2013jqa}
A.~Chatzistavrakidis, F.~F. Gautason, G.~Moutsopoulos, and M.~Zagermann,
  ``{Effective actions of nongeometric five-branes},'' {\em Phys. Rev.} {\bf
  D89} (2014), no.~6 066004, \href{http://xxx.lanl.gov/abs/1309.2653}{{\tt
  1309.2653}}.

\bibitem{Sakatani:2014hba}
Y.~Sakatani, ``{Exotic branes and non-geometric fluxes},'' {\em JHEP} {\bf 03}
  (2015) 135, \href{http://xxx.lanl.gov/abs/1412.8769}{{\tt 1412.8769}}.

\bibitem{Bergshoeff:2015cba}
E.~A. Bergshoeff, V.~A. Penas, F.~Riccioni, and S.~Risoli, ``{Non-geometric
  fluxes and mixed-symmetry potentials},'' {\em JHEP} {\bf 11} (2015) 020,
  \href{http://xxx.lanl.gov/abs/1508.00780}{{\tt 1508.00780}}.

\bibitem{Shukla:2015rua}
P.~Shukla, ``{On modular completion of generalized flux orbits},'' {\em JHEP}
  {\bf 11} (2015) 075, \href{http://xxx.lanl.gov/abs/1505.00544}{{\tt
  1505.00544}}.

\bibitem{Lombardo:2016swq}
D.~M. Lombardo, F.~Riccioni, and S.~Risoli, ``{$P$ fluxes and exotic branes},''
  {\em JHEP} {\bf 12} (2016) 114,
  \href{http://xxx.lanl.gov/abs/1610.07975}{{\tt 1610.07975}}.

\end{thebibliography}\endgroup
\bibliographystyle{utphys}


\end{document}